\documentclass[12pt]{scrartcl}    

\usepackage{silence}

\usepackage{a4} 
\usepackage{amsmath}    
\usepackage{paralist}
\usepackage{latexsym} 
\usepackage{amssymb}  
\usepackage{amsfonts}  
\usepackage{mathrsfs}   
\usepackage{mathtools}   
\usepackage{dsfont}
\usepackage{xcolor}
\usepackage{bbm,exscale}
\definecolor{Myblue}{rgb}{0,0,0.6}  
\usepackage[colorlinks,citecolor=Myblue,linkcolor=Myblue,urlcolor=Myblue,pdfpagemode=None]{hyperref}
\usepackage{amsthm}
\usepackage{accents}
\usepackage[square,numbers,sort&compress]{natbib} 
\usepackage[all,cmtip]{xy}
\usepackage{ifthen} 
\usepackage{bbding}
\usepackage{stmaryrd}  
\usepackage{verbatim}
\usepackage{bbding} 
\usepackage{wasysym}  
\usepackage{soul}  
	\setuldepth{Berlin}
\usepackage[yyyymmdd,hhmmss]{datetime}
\usepackage{booktabs}
\usepackage{enumitem}
\usepackage{color}
\usepackage{textcomp}
\usepackage{gensymb}
\usepackage{pdfpages}
\usepackage{tcolorbox}
\usepackage[bb=boondox]{mathalfa} 
\usepackage{tikz}
\usepackage{tikz-cd}
\usetikzlibrary{calc}
\usetikzlibrary{decorations.markings}
\usetikzlibrary{fadings,decorations.pathreplacing}
\usetikzlibrary{matrix,arrows}
\usetikzlibrary{patterns}
\usetikzlibrary{arrows,calc,decorations.pathreplacing,decorations.markings,shapes.geometric,shadows}

\DeclareSymbolFont{bbold}{U}{bbold}{m}{n}
\DeclareSymbolFontAlphabet{\mathbbold}{bbold}

\colorlet{SurfaceColor}{orange!30!white}
\colorlet{BackSurfaceColor}{orange!20!white}
\colorlet{IdentityColor}{red!80!black}
\colorlet{IdentityLineColor}{gray}
\colorlet{IdentityColorMid}{red!50!white}
\colorlet{DefectColor}{blue!50!black}
\colorlet{GreenColor}{green!80!black}
\colorlet{SidelineColor}{black}
\tikzset{
	Surface/.style={ SurfaceColor, opacity=0.8 },
	BackSurface/.style={ BackSurfaceColor, opacity=0.8 },
	LineDefect/.style = {very thick, DefectColor },
	Sideline/.style = {	very thin , black },
	IdentitySlice/.style = { very thick , IdentityColor },
	IdentitySliceMid/.style = { very thin , IdentityColorMid },
	Multiplication/.style = {thick, BlueColor },
	IdentityLine/.style = {thin, IdentityLineColor,dashed},
	DefaultSettings/.style = {very thick,scale=0.7,color=blue!50!black, baseline=0.5cm}
}
\tikzset{
	string/.style={draw=#1, postaction={decorate}, decoration={markings,mark=at position .51 with {\arrow[draw=#1]{>}}}},
	costring/.style={draw=#1, postaction={decorate}, decoration={markings,mark=at position .51 with {\arrow[draw=#1]{<}}}},
	ostring/.style={draw=#1, postaction={decorate}, decoration={markings,mark=at position .47 with {\arrow[draw=#1]{>}}}},
	ustring/.style={draw=#1, postaction={decorate}, decoration={markings,mark=at position .56 with {\arrow[draw=#1]{>}}}},
	oostring/.style={draw=#1, postaction={decorate}, decoration={markings,mark=at position .43 with {\arrow[draw=#1]{>}}}},
	uustring/.style={draw=#1, postaction={decorate}, decoration={markings,mark=at position .59 with {\arrow[draw=#1]{>}}}},
	directed/.style={string=blue!50!black}, 
	odirected/.style={ostring=blue!50!black}, 
	udirected/.style={ustring=blue!50!black}, 
	oodirected/.style={oostring=blue!50!black}, 
	uudirected/.style={uustring=blue!50!black},     
	redirected/.style={costring= blue!50!black},
	redirectedgreen/.style={costring= green!50!black},
	directedgreen/.style={string= green!50!black},
}

\tikzset{-dot-/.style={decoration={
			markings,
			mark=at position 0.5 with {\fill circle (2pt);}},postaction={decorate}}}

\tikzset{
	Fdot/.style={circle, draw, fill, inner sep=0pt}, 
	Odot/.style={circle, draw, inner sep=0.1pt, minimum size=0.1cm}
}

\newcommand\tikzzbox[1]
{#1}

\vfuzz \hfuzz
\makeatletter
\newcommand{\raisemath}[1]{\mathpalette{\raisem@th{#1}}}
\newcommand{\raisem@th}[3]{\raisebox{#1}{$#2#3$}}
\makeatother

\newcommand{\A}{\mathcal{A}}
\newcommand{\Aca}{\mathcal{A}^{\mathcal{S}}}
\newcommand{\orb}{\mathcal{A}}
\newcommand{\sgn}{\mathrm{sgn}}
\renewcommand{\leq}{\leqslant}
\renewcommand{\geq}{\geqslant}
\newcommand{\rp}[1]{\widetilde{#1}}
\newcommand{\btimes}{\mathbin{\square}} 
\newcommand{\bigbtimes}{\mathop{\vcenter{\hbox{\scalebox{2.00}{$\btimes$}}}}\limits}
\newcommand*{\longhookleftarrow}{\ensuremath{\leftarrow\joinrel\relbar\joinrel\rhook}}
\newcommand*{\longhookrightarrow}{\ensuremath{\lhook\joinrel\relbar\joinrel\lra}}
\newcommand*{\twoheadlongrightarrow}{\ensuremath{\relbar\joinrel\twoheadrightarrow}}

\newcommand{\E}{\text{e}} 
\newcommand{\I}{\text{i}}
\newcommand{\B}{\mathcal{B}}
\newcommand{\Bfd}{\mathcal{B}^{\textrm{fd}}}

\newcommand{\Bs}{\mathds{B}}
\newcommand{\Borb}{\B_{\mathrm{orb}}}
\newcommand{\Beq}{\B_{\mathrm{eq}}}
\newcommand{\C}{\mathds{C}}
\newcommand{\Cc}{\mathcal{C}}
\newcommand{\D}{\mathds{D}}
\newcommand{\Ee}{\mathds{E}}
\newcommand{\K}{\mathds{K}}
\newcommand{\Kc}{\mathcal{K}}
\newcommand{\M}{\mathds{M}}
\newcommand{\N}{\mathds{N}}
\newcommand{\Q}{\mathds{Q}}
\newcommand{\R}{\mathds{R}}
\newcommand{\Z}{\mathds{Z}}
\newcommand{\Zb}{\mathbb{Z}}
\newcommand{\Zc}{\mathcal{Z}}
\newcommand{\IP}{\mathds{P}}
\def\1{\ifmmode\mathrm{1\!l}\else\mbox{\(\mathrm{1\!l}\)}\fi}
\newcommand{\one}{\mathbbm{1}}
\newcommand{\be}{\begin{equation}}
  \newcommand{\ee}{\end{equation}}
\newcommand{\bes}{\begin{equation*}}
  \newcommand{\ees}{\end{equation*}}
\newcommand{\cc}[1] {\overline{#1}}
\newcommand{\inv}[0]{{-1}}
\newcommand{\DC}{\mathds{D}^{\mathcal C}}
\newcommand{\cat}{{\mathcal C}}

\newcommand{\MF}{\operatorname{MF}_{\operatorname{bi}}}
\newcommand{\MFW}{\operatorname{MF}_{\operatorname{bi}}(W)}
\newcommand{\MFR}{\operatorname{MF}^\text{R}_{\operatorname{bi}}}
\newcommand{\DG}{\operatorname{DG}_{\operatorname{bi}}}
\newcommand{\DGW}{\operatorname{DG}_{\text{bi}}(W)}
\newcommand{\DGR}{\operatorname{DG}^\text{R}_{\text{bi}}}
\newcommand{\id}{\operatorname{id}}
\newcommand{\Id}{\operatorname{Id}}
\newcommand{\KMF}{K_{0}(\operatorname{MF}_{\text{bi}}}
\newcommand{\Ext}{\operatorname{Ext}}
\newcommand{\Hom}{\operatorname{Hom}}
\newcommand{\Aut}{\operatorname{Aut}}
\newcommand{\End}{\operatorname{End}}
\newcommand{\modu}{\operatorname{mod}}
\def\LG{\mathcal{LG}}
\def\LGgr{\mathcal{LG}^{\mathrm{gr}}}
\def\LGhuge{\mathcal{LG}^{\mathrm{GR}}}
\def\LGgrs{\mathcal{LG}'^{\mathrm{gr}}}
\def\LGgrso{\mathcal{LG}'^{\mathrm{gr}}_{\mathrm{orb}}}
\def\LGs{\mathcal{LG}'}
\def\LGsorb{\mathcal{LG}'_{\mathrm{orb}}}
\def\LGorb{\mathcal{LG}_{\mathrm{orb}}}
\def\LGeq{\mathcal{LG}_{\mathrm{eq}}}
\def\LGs{\mathcal{LG}^{\textrm{{\tiny $\bullet/2$}}}}
\def\RW{\mathcal{RW}}
\newcommand{\hmf}{\operatorname{hmf}}
\newcommand{\HMF}{\operatorname{HMF}}
\newcommand{\ev}{\operatorname{ev}}
\newcommand{\eval}{\operatorname{eval}}
\newcommand{\tev}{\widetilde{\operatorname{ev}}}
\newcommand{\coev}{\operatorname{coev}}
\newcommand{\tcoev}{\widetilde{\operatorname{coev}}}
\def\lra{\longrightarrow}

\newcommand\leftidx[3]{%
  {\vphantom{#2}}#1#2#3%
}

\def\lmt{\longmapsto}
\DeclareMathOperator{\tr}{tr}
\DeclareMathOperator{\str}{str}
\DeclareMathOperator{\Jac}{Jac}
\def\Re{R^{\operatorname{e}}}
\DeclareMathOperator{\Res}{Res}
\newcommand{\Ga}[1]{\Gamma_{\hspace{-2pt}#1}}
\newcommand{\HIA}{\Hom(I,A)}
\newcommand{\ZA}{Z_A(\Hom(I,A))}
\newcommand{\gZA}{\!Z_A^\gamma(\Hom(I,A))}
\newcommand{\gA}{{}_{\gamma_A}A}
\newcommand{\Aginv}{A_{\gamma_A^{-1}}}
\newcommand{\picc}{\pi^{(\text{c,c})}_A}
\newcommand{\pirr}{\pi^{\text{RR}}_A}
\newcommand{\tpirr}{{\widetilde\pi}^{\text{RR}}} 
\newcommand{\im}{\operatorname{im}}
\DeclareMathOperator*{\eq}{=}
\DeclareMathOperator*{\congscript}{\cong}
\newcommand{\specflow}{\mathcal U_{-\frac{1}{2},-\frac{1}{2}}}
\newcommand{\Hil}{\mathcal{H}}
\newcommand{\Hpcc}{\mathcal{H}'_{\text{(c,c)}}}
\newcommand{\Hprr}{\mathcal{H}'_{\text{RR}}}
\newcommand{\Hcc}{\mathcal{H}_{\text{(c,c)}}^A}
\newcommand{\Hrr}{\mathcal{H}_{\text{RR}}^A}
\newcommand{\HccnoA}{\mathcal{H}_{\text{(c,c)}}}
\newcommand{\HrrnoA}{\mathcal{H}_{\text{RR}}}
\newcommand{\Hrrbo}{\Hom_A(X,{}_{\gamma_A}X)}
\newcommand{\buboorb}{\beta_X^{\text{orb}}}
\newcommand{\bobuorb}{\beta^X_{\text{orb}}} 
\def\Cong{C_g}
\def\Centg{N_g}                 
\def\alphaKK{\alpha^{\{K\}}}
\def\alphagKg{\alpha_g^{K(g)}}
\newcommand{\AGC}{A_G^c}
\DeclareMathOperator{\colim}{colim}

\newcommand{\FD}[1]{\textrm{2D}^{#1}}
\newcommand{\Fun}{\textrm{Fun}^{\mathrm{sm}}}
\newcommand{\Nat}{\textrm{Nat}}
\newcommand{\SMBicat}{\textrm{SMBicat}}
\newcommand{\Spin}{\textrm{Spin}}
\newcommand{\GL}{\textrm{GL}}
\newcommand{\SO}{\textrm{SO}}
\renewcommand{\O}{\textrm{O}}
\newcommand{\Bord}{\textrm{Bord}}
\newcommand{\BBord}{\mathds{B}\mathbbold{o}\mathbb{r}\mathbbold{d}}
\newcommand{\Bordfr}{\Bord_{2,1,0}^{\textrm{fr}}}
\newcommand{\Bordor}{\Bord_{2,1,0}^{\textrm{or}}}
\newcommand{\Bordrspin}{\Bord_{2,1,0}^{r\textrm{-spin}}}
\newcommand{\zzwfr}{\zz_W^{\textrm{fr}}}
\newcommand{\zzwor}{\zz_W^{\textrm{or}}}
\newcommand{\sotwo}{\textrm{SO}(2)}
\newcommand{\kbso}{\mathscr{K}(\Bfd)^{\sotwo}}
\newcommand{\zz}{\mathcal{Z}}

\newcommand{\unit}{\operatorname{\mathbf{1}}}  
\newcommand{\Vect}{\operatorname{Vect}}
\newcommand{\SVect}{\operatorname{sVect}_\C}
\newcommand{\Vectk}{\operatorname{Vect}_\Bbbk}

\newcommand{\eps}{\varepsilon}
\newcommand{\al}{\alpha}
\newcommand{\alb}{\overline{\alpha}}
\newcommand{\T}{\mathcal{T}}
\newcommand{\Sb}{\mathbb{S}^1}
\newcommand{\Ss}{\mathcal{S}}
\newcommand{\X}{\mathcal{X}}
\newcommand{\Y}{\mathcal{Y}}
\newcommand{\fus}{\otimes}
\newcommand{\sd}{^{\star}}
\newcommand{\dagg}{^{\dagger}}
\newcommand{\hash}{^{\#}}
\newcommand{\dual}{\#}
\newcommand{\Ae}{A^{\textrm{e}}}

\newcommand{\dX}{{}^\dagger\hspace{-1.8pt}X}
\newcommand{\Xd}{X^\dagger}
\newcommand{\dM}{{}^\dagger\hspace{-1.8pt}M}
\newcommand{\Md}{M^\dagger}
\newcommand{\dA}{{}^\dagger\hspace{-1.8pt}A}
\newcommand{\dsX}{{}^\dagger\hspace{-1.8pt}\mathcal{X}}
\newcommand{\deqX}{{}^\star\hspace{-1.8pt}X} 
\newcommand{\dseqX}{{}^\star\hspace{-1.8pt}\mathcal{X}} 
\newcommand{\dY}{{}^\dagger\hspace{-0.3pt}Y}
\newcommand{\dphi}{{}^\dagger\hspace{-0.9pt}\phi}
\newcommand{\dPhi}{{}^\dagger\hspace{-0.9pt}\Phi}
\newcommand{\FEnd}{\mathcal{E}\hspace*{-.7pt}nd}

\newcommand{\bi}{\mathcal C}
\newcommand{\bigra}{{\mathcal C}}
\newcommand{\idlg}{{\textrm{Id}_{\bi}}}
\newcommand{\sta}{\btimes}
\newcommand{\au}{\text{\ul{$a$}}}
\newcommand{\bu}{\text{\ul{$b$}}}
\newcommand{\cu}{\text{\ul{$c$}}}
\newcommand{\du}{\text{\ul{$d$}}}
\newcommand{\eu}{\text{\ul{$e$}}}
\newcommand{\gu}{\text{\ul{$g$}}}
\newcommand{\alphau}{\text{\ul{$\alpha$}}}
\newcommand{\betau}{\text{\ul{$\beta$}}}
\newcommand{\gammau}{\text{\ul{$\gamma$}}}
\newcommand{\deltau}{\text{\ul{$\delta$}}}
\newcommand{\epsilonu}{\text{\ul{$\epsilon$}}}
\newcommand{\zetau}{\text{\ul{$\zeta$}}}
\newcommand{\etau}{\text{\ul{$\eta$}}}
\newcommand{\lambdau}{\text{\ul{$\lambda$}}}
\newcommand{\muu}{\text{\ul{$\mu$}}}
\newcommand{\rhou}{\text{\ul{$\rho$}}}
\newcommand{\sigmau}{\text{\ul{$\sigma$}}}
\newcommand{\tauu}{\text{\ul{$\tau$}}}
\newcommand{\chiu}{\text{\ul{$\chi$}}}
\newcommand{\psiu}{\text{\ul{$\psi$}}}
\newcommand{\xiu}{\text{\ul{$\xi$}}}
\newcommand{\omegau}{\text{\ul{$\omega$}}}
\newcommand{\df}{\text{\ul{$f$}}}
\newcommand{\pu}{\text{\ul{$p$}}}
\newcommand{\qu}{\text{\ul{$q$}}}
\newcommand{\wu}{\text{\ul{$w$}}}
\newcommand{\xu}{\text{\ul{$x$}}}
\newcommand{\yu}{\text{\ul{$y$}}}
\newcommand{\zu}{\text{\ul{$z$}}}
\newcommand{\uu}{\text{\ul{$u$}}}
\newcommand{\vu}{\text{\ul{$v$}}}
\newcommand{\zeru}{\text{\ul{$0$}}}
\newcommand{\Deltau}{\text{\ul{$\Delta$}}}
\newcommand{\Bhfp}[1]{{#1}^{\circlearrowleft}}
\newcommand{\vect}[2]{{\tiny \begin{pmatrix} #1\\ #2\end{pmatrix}}}
\newcommand{\Borddef}{\Bord_{2,1,0}^{\textrm{def}}(\mathds{D})}
\newcommand{\Bordoc}{\Bord_{2,1,0}^{\textrm{oc}}}

\newcommand\nxt{\noindent\raisebox{.08em}{\rule{.44em}{.44em}}\hspace{.4em}}
\newcommand\arxiv[2]      {\href{https://arXiv.org/abs/#1}{#2}}
\newcommand\doi[2]        {\href{https://dx.doi.org/#1}{#2}}
\newcommand\httpurl[2]    {\href{https://#1}{#2}}

\renewcommand{\labelenumi}{(\roman{enumi})}

\allowdisplaybreaks

\deffootnote[1em]{1em}{1em}{\textsuperscript{\thefootnotemark}}

\theoremstyle{definition} 
\newtheorem{definition}{Definition}

\newtheorem{theorem}[definition]{Theorem}

\numberwithin{equation}{section}
\numberwithin{definition}{section}
\numberwithin{figure}{section}

\newcommand\void[1]{}

\begin{document}

\title{%
\vskip -1cm
Truncated affine Rozansky--Witten models as extended defect TQFTs%
}

\author{%
	Ilka Brunner$^*$ 
	\\ 
	Nils Carqueville$^\#$ 
	\\ 
	Pantelis Fragkos$^*$ 
	\\ 
	Daniel Roggenkamp$^{\vee,\vee\vee}$
	\\[0.5cm]
	\normalsize{\texttt{\href{mailto:ilka.brunner@physik.uni-muenchen.de}{ilka.brunner@physik.uni-muenchen.de}}} \\ %
	\normalsize{\texttt{\href{mailto:fragkos.pantelis@physik.uni-muenchen.de}{fragkos.pantelis@physik.uni-muenchen.de}}} 
	\\
	\normalsize{\texttt{\href{mailto:nils.carqueville@univie.ac.at}{nils.carqueville@univie.ac.at}}} \\  %
	\normalsize{\texttt{\href{mailto:daniel.roggenkamp@mis.mpg.de}{daniel.roggenkamp@mis.mpg.de}}}
	\\[0.5cm]  %
	{\normalsize\slshape $^*$Arnold Sommerfeld Center,  LMU M\"unchen,} \\[-0.1cm]  {\normalsize\slshape \phantom{$^*$}Theresienstra\ss e 37, 80333 M\"unchen, Deutschland}\\
	{\normalsize\slshape $^\#$Universit\"at Wien, Fakult\"at f\"ur Physik,}
	\\[-0.1cm] 
	{\normalsize\slshape Boltzmanngasse 5, 1090 Wien, \"{O}sterreich}\\
	{\normalsize\slshape $^\vee$Max Planck Institute for Mathematics in the Sciences,} 
	\\[-0.1cm]
	{\normalsize\slshape Inselstra\ss e 22, 04103 Leipzig, Deutschland}\\
	{\normalsize\slshape $^{\vee\vee}$Institute for Theoretical Physics, University of Leipzig,} 
	\\[-0.1cm]
	{\normalsize\slshape Br\"uderstra\ss e 16, 04103 Leipzig, Deutschland} 
}

\date{}
\maketitle

\vspace{-0.8cm}
\vspace{-18.8cm}
\hfill {\scriptsize LMU-ASC 23/23}
\vspace{18cm}

\begin{abstract} 
	We apply the cobordism hypothesis with singularities to the case of affine Rozansky--Witten models, providing a construction of extended TQFTs that includes all line and surface defects. 
	On a technical level, this amounts to proving that the associated homotopy 2-category is pivotal, and to systematically employing its 3-dimensional graphical calculus. 
	This in particular allows us to explicitly calculate state spaces for surfaces with arbitrary defect networks. 
	As specific examples we discuss symmetry defects which can be used to model non-trivial background gauge fields, as well as boundary conditions.
\end{abstract}

\thispagestyle{empty}

\newpage 

\tableofcontents

\section{Introduction and summary}
\label{sec:IntroductionSummary}

Rozansky--Witten models are topological twists of 3-dimensional $\mathcal N=4$ supersymmetric sigma models with holomorphic symplectic target manifolds \cite{RW1996, Kapranov, KontsevichRW}. 
Such targets arise as Coulomb moduli spaces of supersymmetric gauge theories, and Rozansky--Witten models are thought to describe their infrared fixed points under renormalisation group flow. 
They also participate in 3-dimensional mirror symmetry and  relate Casson invariants to Atiyah--Hitchin moduli spaces.

Despite widely held expectations and significant progress \cite{Sawon, r0112209, RobWill, KRS, KR0909.3643}, Rozansky--Witten models have not yet been rigorously constructed as 3-dimensional topological quantum field theories (TQFTs) in the functorial sense. 
Based on the path integral analysis of \cite{KRS}, an extended definition sketch of the 3-category $\mathcal{RW}$ which would encode all bulk theories as well as surface, line and point defects was presented in \cite{KR0909.3643}. Indeed, it was shown in \cite{BCR} that restriction to the 
3-category $\RW^{\textrm{aff}}$ of affine Rozansky--Witten models whose objects are target manifolds of the form $T^*\C^n$, and subsequent truncation at the 2-dimensional level yields a symmetric monoidal 2-category 
\be 
\bigra := \operatorname{Ho}_2\big(\RW^{\textrm{aff}}\big) \, .
\ee 
Objects of this 2-category are lists of variables $\xu = (x_1,\dots,x_n)$, 1-morphisms are polynomials~$W$, and 2-morphisms are isomorphism classes of matrix factorisations~$X$, see Section~\ref{subsec:targetcategory} for a detailed review. 
The interpretation of~$\xu$ is in terms of the base coordinates of $T^*\C^n$, while~$W$ and~$X$ describe potentials for surface defects and isomorphism classes of line defect, respectively. 
Point defects are truncated away in the homotopy 2-category~$\bigra$ of $\RW^{\textrm{aff}}$. 

The main result in \cite{BCR} was to explicitly construct fully extended TQFTs valued in~$\bigra$, thus rigorously exhibiting truncated affine Rozansky--Witten models as functorial field theories. 
This was done by proving that every object in~$\bigra$ is fully dualisable, by computing all trivialisations of their Serre automorphisms, and then applying the cobordism hypothesis of \cite{BDpaper, l0905.0465}, which in the 2-dimensional oriented case is a theorem \cite{spthesis, PstragowskiJournalVersion, Hesse} as reviewed in some detail in Section~\ref{sec:2dTQFTsWithoutDefects} below. 
In particular this allowed for the computation of state spaces for arbitrary surfaces from first principles, recovering some results of \cite{RW1996} which were based on a path integral analysis. 

\medskip 

In the present paper we extend the work \cite{BCR} on bulk theories by refining it to include all surface and line defects (from the 3-dimensional perspective). 
Our main technical result (proved in Section~\ref{subsec:adjunctions}) is: 

\begin{theorem}
	\label{thm:CPivotal}
	The symmetric monoidal 2-category~$\bigra$ has a natural pivotal structure (and every object is fully dualisable). 
\end{theorem}

This essentially means that all 1-morphisms have coherently isomorphic left and right adjoints (see Section~\ref{subsec:GraphicalCalculus} for details), which is a necessary and sufficient condition for them to describe surface defects. (It formalises the statement that an orientation-reversed surface defect is ``dual'' to the original one.)

Our main technical tools and inspirations are the ``cobordism hypothesis with singularities'' of \cite{l0905.0465} and the 3-dimensional graphical calculus of \cite{BMS}. 
In particular, we restrict the contents of \cite[Sect.\,4.3]{l0905.0465} to the case at hand to introduce a symmetric monoidal 2-category $\Borddef$ of ``extended defect bordisms'', where~$\mathds{D}$ is a chosen collection of defect data (which in our applications are obtained from Rozansky--Witten models). 
This is a generalisation of the non-extended defect bordisms of \cite{dkr1107.0495, CRS1} as we explain in detail in Section~\ref{subsec:ExtendedDefectBordisms}. 
Roughly, objects in $\Borddef$ are points that are labelled by bulk theories, 1-morphisms are stratified lines whose strata are labelled by bulk theories and line defects (from the 2-dimensional perspective), and 2-morphisms are $\mathds{D}$-labelled stratified surfaces with corners such as 
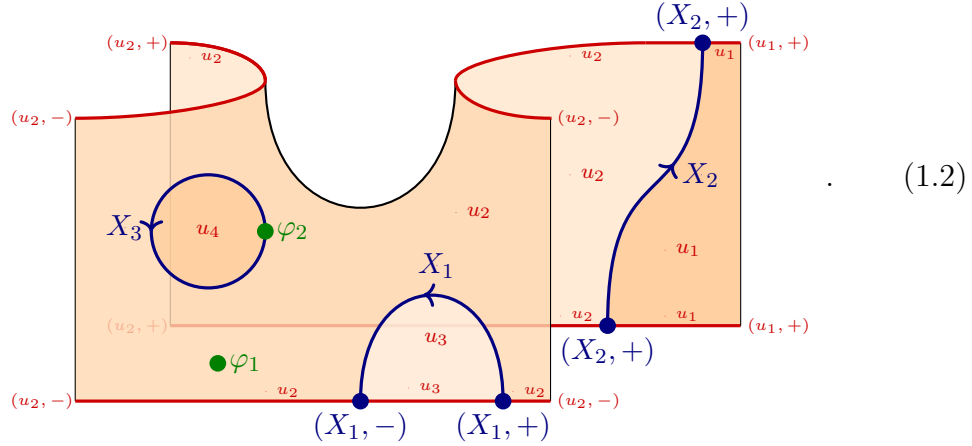
\be 
\tikzzbox{%

}%
\, . 
\ee 
Here the labels~$u_i$ encode bulk theories, the labels $X_j$ correspond to line defects, and~$\varphi_k$ to point defects. 

In Section~\ref{subsec:ExtendedDefectTQFTs} we extract from \cite[Sect.\,4.3]{l0905.0465} what we call the (2-dimensional oriented) ``cobordism hypothesis with defects'', i.\,e.\ a quasi-algorithmic instruction how to construct fully extended defect TQFTs as symmetric monoidal 2-functors
\be 
\zz \colon \Borddef \lra \B 
\ee 
for some appropriate target 2-category~$\B$. 
The essence of the cobordism hypothesis from this perspective is that \textsl{fully extended TQFTs are evaluated on bordisms by interpreting them in the graphical calculus of the target}. 
In particular, only 1-morphisms in~$\B$ which admit adjoints can describe line defects. 
Hence Theorem~\ref{thm:CPivotal} ensures that we can apply the cobordism hypothesis with defects to the truncated Rozansky--Witten case, i.\,e.\ the choice $\B=\bigra$. 
Of course one can equally well consider other examples, such as the 2-categories of B-twisted sigma models or Landau--Ginzburg models. 

Since~$\bigra$ is under very explicit computational control, by applying the cobordism hypothesis with defects we can evaluate the affine Rozansky--Witten TQFT on surfaces with arbitrary configurations of defects. 
This is explained in Section~\ref{subsec:ExtendedRWTQFT} and illustrated in two classes of concrete examples in Section~\ref{sec:Examples}. 

In Section~\ref{subsec:gldefects} we discuss symmetry defects. 
We identify the defects (1-morphisms) incorporating the $\text{Sp}_{2n}$ symmetry of Rozansky--Witten models with target space $T^*\C^n$. 
We then focus on the subgroup $\text{GL}_n\subset\text{Sp}_{2n}$ and determine the state spaces for surfaces with arbitrary networks of the respective symmetry defects. 
Insertion of such defect networks can be interpreted as ``turning on a flat background gauge connection''. 
We reproduce the state spaces of free hypermultiplets with non-trivial background gauge fields as well as categories of line operators in twisted sectors obtained by different methods in \cite{CDGG}. 

In Section~\ref{subsec:boundaries} we treat boundary conditions, which can be regarded as defects separating a given bulk theory from the trivial theory on the other side. 
We calculate the state spaces of affine Rozansky--Witten theories on surfaces with boundaries, such as the one associated to the disc with arbitrary boundary conditions. 
The state space associated to a cylinder yields a pairing on the category of boundary conditions which can be regarded as a categorification of the open string Witten index. 
Indeed, we explain that for boundary conditions associated to Lagrangian submanifolds of the target geometry $T^* \C^n$ it categorifies the geometric intersection pairing. 

Finally, we determine all the generators of the respective open-closed TQFT. 
Since our general defect TQFT is a functor by construction, all constraints including the Cardy condition are automatically satisfied. 
In a geometric context, the latter implies a novel example of a Hirzebruch--Riemann--Roch-type theorem that we expect to hold also for more interesting target geometries.

\medskip 

We conclude this introduction with a few comments on potential future directions. 
Firstly, from a physics perspective, gauge theories are  more interesting than the free theories of hypermultiplets discussed in the present paper, and we expect that the formalism adopted here can be applied to describe the Higgs branch of abelian gauge theories. 
Indeed, it might even provide a way to obtain information about the Coulomb branch, 
even though the latter is not directly accessible after the Rozansky--Witten twist. 
The idea here is to use a strategy similar to the case of 2-dimensional Landau--Ginzburg orbifolds, where sectors compatible with the A-twist can be realised by twisted sectors in the B-twisted models, see e.\,g.~\cite{BCP1}. 
Adapting this strategy to the 3-dimensional case using techniques developed in 
\cite{cr1210.6363, CaMue} might for instance yield information about 
``line operators in the twisted sector''. 
This could have applications in the context of 3-dimensional mirror symmetry.

Secondly, in this paper we treat defects and boundary conditions compatible with the Rozansky--Witten topological twist. 
They can be characterised by the supersymmetry they preserve in the initial untwisted theory, namely $\mathcal N = (2,2)$ in two dimensions. 
However, as explained in \cite{CG}, there are additional very interesting boundary conditions that in the untwisted theory preserve $\mathcal N = (0,4)$ supersymmetry in two dimensions. 
While they are not directly compatible with a topological twist, some of them allow for deformations which are. 
They give rise to holomorphic boundary conditions for topological theories that can be described in terms of non-semisimple vertex operator algebras. 
It is expected that the category of line operators is equivalent to a derived category of modules over the boundary VOA.
In the specific case of free hypermultiplets, compatible boundary conditions are ``Dirichlet in all directions'', and the VOA is that of symplectic fermions. 

It would be interesting to explore whether one can bring the methods developed in this paper to bear in the description of the holomorphic boundary conditions just mentioned.
Since our formalism gives complete control over line and surface defects of the topological 3-dimensional bulk theory, a possible realisation of the holomorphic boundary conditions could potentially lead to new results on the representation theory of logarithmic VOAs.

\subsubsection*{Acknowledgements} 

We are grateful to 
	Tudor Dimofte, 
	Lukas Müller, 
	David Reutter, 
	Ingmar Saberi, 
	and 
	Yang Yang
for helpful discussions. 
I.\,B.\ and P.\,F.\ are supported by the Deutsche Forschungsgemeinschaft (DFG) under Germany's Excellence Strategy EXC-2094 390783311, I.\,B.\ is further supported by the DFG grant ID 17448, 
N.\,C.\ is supported by the DFG Heisenberg Programme, 
and 
D.\,R.\ was supported by the Heidelberg Institute for Theoretical Studies. 
D.\,R.\ thanks the Institute for Advanced Study Princeton for its hospitality, where part of this work was done.

\section{2-dimensional extended TQFTs without defects}
\label{sec:2dTQFTsWithoutDefects}

In this review section, we start with a short exposition of the graphical calculus for monoidal 2-categories with duals and adjoints in Section~\ref{subsec:GraphicalCalculus}. 
In Section~\ref{subsec:ExtendedBordisms} we briefly discuss the oriented bordism 2-category, and in Section~\ref{subsec:CHwithoutDfects} we review how 2-dimensional extended oriented TQFTs are described by the cobordism hypothesis. 
In particular, we systematically employ the graphical calculus, and stress the explicit, quasi-algorithmic character of the cobordism hypothesis.

\subsection{Graphical calculus for duals in 2-categories}
\label{subsec:GraphicalCalculus}

Non-extended topological quantum field theories with defects in two dimensions naturally give rise to 2-categories whose objects, 1- and 2-morphisms correspond to bulk theories, line and surface defects. 
Moreover, these 2-categories come with (canonically identified left and right) adjoints for 1-morphisms, encoding orientation reversal of line defects. 
This was established in \cite{dkr1107.0495}, see \cite{2dDefectTQFTLectureNotes} for a review. 
Examples of such 2-categories are those of 
\begin{itemize}
	\item 
	state sum models: objects are separable symmetric Frobenius algebras over~$\C$, Hom categories are those of bimodules and bimodule maps \cite{lp0602047, spthesis}; 
	\item 
	B-twisted sigma models: Calabi--Yau manifolds and their derived categories \cite{cw1007.2679, BanksOnRozanskyWitten}; 
	\item 
	affine Landau--Ginzburg models: isolated singularities and homotopy categories of matrix factorisations \cite{cm1208.1481, CMM}; 
	\item 
	truncated affine Rozansky--Witten models \cite{BCR}: discussed in Section~\ref{sec:DefectsRW}. 
\end{itemize} 
All the above examples are further endowed with a natural symmetric monidal structure in which every object is fully dualisable (as proven in the above references). 
Hence they can be taken as codomains for extended TQFTs as discussed in Section~\ref{subsec:CHwithoutDfects} below. 

In the present section we review some aspects of the graphical calculus for symmetric monoidal 2-categories with duals established in \cite{BMS}, and we recall some of the notions relevant for our applications.\footnote{In fact the graphical calculus of \cite{BMS} applies more generally to generically maximally strict 3-categories with coherent adjoints (called ``Gray categories with duals''), of which (generically maximally strict) monoidal 2-categories with duals provide special cases. Every 3-category is equivalent to a generically maximally strict one, cf.\ \cite{Gurskibook}; we do not spell out such strictifications explicitly.} 
For a detailed discussion of symmetric monoidal 2-categories and dualisability we refer to \cite{spthesis} and \cite{DouglasSchommerPriesSnyder2013, PstragowskiJournalVersion}, respectively. 

\medskip 

Given two objects $u,v$ and two parallel 1-morphisms $X,Y\colon u\lra v$ in some 2-category~$\B$, the graphical presentation of a 2-morphism $\varphi\colon X\lra Y$ is
\be 
\tikzzbox{%

}
\, . 
\ee 
Our convention is to read vertical composition (denoted~$\cdot$, or simply as concatenation) from bottom to top, and horizontal composition (denoted~$\circ$) from right to left. 
Hence for further 1-morphisms $Z\colon u\lra v$ and $A,B\colon v\lra w$ as well as 2-morphisms $\psi\colon Y\lra Z$ and $\zeta\colon A\lra B$ we have
\be 
\tikzzbox{%
	%
} 
\, .
\ee 

A 1-morphism $X\colon u\lra v$ has a \textsl{left adjoint} if there is a 1-morphism $\dX\colon v\lra u$ together with 2-morphisms 
\be 
\label{eq:Adjunction1Morphisms}
\ev_X  = 
\tikzzbox{%
	%
}
\ee 
such that the \textsl{Zorro moves}
\be 
\label{eq:ZorroMoves}
\tikzzbox{%
	%
}
\ee 
hold.\footnote{We usually suppress identity and structure morphisms in the graphical calculus. For example, the 1-morphism~$1_u$ is not displayed for $\ev_X\colon \dX\circ X \lra 1_u$, and the left-hand side of~\eqref{eq:ZorroMoves} implicitly involves the (inverse) unitor 2-isomorphisms $\lambda_X^{-1}\colon X \lra 1_v \circ X$ and $\rho_X\colon X \circ 1_u \lra X$ as well as an associator.} 
The \textsl{adjunction data} $(\dX, \ev_X, \coev_X)$ are unique up to unique isomorphism in the sense that for any other triple $(\dX', \ev'_X, \coev'_X)$ as above, there exists a unique 2-isomorphism $\varphi\colon \dX \lra \dX'$ such that $\ev'_X=\ev_X \cdot (\varphi^{-1} \circ 1_X)$ and $\coev'_X = (1_X \circ \varphi) \cdot \coev_X$. 
Similarly, a \textsl{right adjoint} consists of $\Xd\colon v\lra u$ and 2-morphisms 
\be 
\label{eq:RightAdjunction1Morphisms}
\tev_X  = 
\tikzzbox{%

}
\ee 
subject to analogous Zorro moves (and analogous uniqueness). 
If~$\dX$ and~$\Xd$ can be canonically identified (which is true in all of the examples of 2-categories mentioned above), for $\chi\colon X \lra X$ the \textsl{left} and \textsl{right traces} 
\be \label{eq:traces}
\tr_{\textrm{l}}(\chi) := 
\tikzzbox
{
	%
 
}
\in \End(1_v)
\ee 
make sense. 
In particular, we have the \textsl{quantum dimensions} 
$\dim_{\textrm{l}}(X) := \tr_{\textrm{l}}(1_X)$ 
and 
$\dim_{\textrm{r}}(X) := \tr_{\textrm{r}}(1_X)$. 
A 2-category with coherently isomorphic adjoints for all 1-morphisms is called \textsl{pivotal}. 
If $\dX = \Xd$ for every 1-morphism~$X$, then the pivotality coherence conditions read
\be
\label{eq:pivotality}
\tikzzbox{
	%
}
\ee 
for all 2-morphisms $\xi\colon Z\lra X$ and all composable 1-morphisms $X,Y$. 
Note that by definition, the two sides of the first equation are the right and left adjoints~$\xi^\dagger$ and~${}^\dagger\xi$, respectively. 

\medskip 

The categories~$\B$ that we want to consider also come with a monoidal structure. 
This means that there is not only vertical composition for 2-morphisms and horizontal composition for 1- and 2-morphisms, but also ``monoidal composition'' (denoted~$\btimes$) for objects, 1- and 2-morphisms. 
In our interpretation, the composition~$\cdot$ corresponds to the operator product of point defects, $\circ$ corresponds to the fusion of line defects (or the fusion of surface defects from a truncated 3-dimensional perspective as in Section~\ref{sec:DefectsRW}), and~$\btimes$ corresponds to stacking of bulk theories. 

In the graphical calculus, the ``third direction'' from front to back\footnote{Our conventions for reading diagrams from bottom to top, from right to left, and from front to back is as in, e.\,g., \cite{CMS, CRS1, BCR}, but differs from the conventions in \cite{BMS}.} is used to present~$\btimes$: for 1-morphisms $X,Y\colon u\lra v$ and $X',Y'\colon u'\lra v'$ the \textsl{monoidal product} $\varphi' \btimes \varphi$ of $\varphi\colon X\lra X$ and $\varphi'\colon X'\lra Y'$ is presented as 
\begin{align}
&
\tikzzbox{%

}
\, . 
\end{align}

The monoidal product~$\btimes$ on~$\B$ has a \textsl{symmetric} structure if for all $u,v\in\B$ there is (among other data) a \textsl{braiding} 1-morphism 
\be 
\label{eq:Braiding1Morphism}
%
= 
b_{u,v} \colon u\btimes v \lra v\btimes u 
\ee 
as well as \textsl{braiding 2-morphisms} $b_{X,Y} \colon X\btimes Y \lra Y \btimes X$ for all 1-morphisms $X,Y$, subject to certain coherence axioms, see \cite[App.\,C]{spthesis}. 

\medskip 

Let~$\B$ from now on be a symmetric monoidal 2-category. 
An object $u\in\B$ is (right) \textsl{dualisable} if there exists $u^\# \in \B$ together with \textsl{adjunction 1-morphisms} 
\be 
\label{eq:TildeCoEv}
\tikzzbox{%

}   
=
\tcoev_u \colon \one \lra u^\# \btimes u 
\, , 
\ee 
where $\one\in\B$ is the unit object (corresponding to the trivial bulk theory). 
The 1-morphisms $\tev_u, \tcoev_u$ are subject to a categorification of the Zorro moves in the sense that there must exist so-called \textsl{cusp 2-isomorphisms} 
\be 
\label{eq:CuspIsomorphisms}
\tikzzbox{%
	%
}     
\, . 
\ee 
The uniqueness of the \textsl{duality data} $(u,u^\#,\tev_u,\tcoev_u,c^u_{\textrm{l}},c^u_{\textrm{r}})$ is described in detail in \cite{PstragowskiJournalVersion}; below in Section~\ref{subsec:CHwithoutDfects} we will review some aspects relevant for us. 

Using the symmetric braiding~$b$, one finds that the adjunction maps 
\be 
\label{eq:LeftDuals}
\ev_u := \tev_u \circ b_{u^\#,u} 
	\, , \quad 
	\coev_u := b_{u^\#,u} \circ \tcoev_u 
\ee 
also witness~$u^\#$ as a left dual of~$u$. 
Since~$\B$ has a symmetric monoidal structure, one finds that $u^{\dual\dual} \cong u$, and we can and will assume that $u^{\dual\dual} = u$ for all dualisable objects~$u$. 

A dualisable object $u\in\B$ is \textsl{fully dualisable} if its adjunction 1-morphisms $\tev_u, \tcoev_u$ have both left and right adjoints, in the sense explained for 1-morphisms~$X$ around \eqref{eq:Adjunction1Morphisms}--\eqref{eq:RightAdjunction1Morphisms} above. 
In particular, every dualisable object in a symmetric monoidal pivotal 2-category is fully dualisable. 
If they exist, we denote the (left and right) adjunction 2-morphisms for $\tev_u$ as follows: 
\begin{align}
\ev_{\tev_u} = 
\tikzzbox{%

}%
\, . 
\end{align}
Note that in general ${}^\dagger \tev_u$ and $\tev_u^\dagger$ need not be isomorphic. 
However, both are related to $\coev_u$ via the \textsl{Serre automorphism}\footnote{The displayed formula depends on the choice of evaluation 1-morphism $\tev_u$ and its right adjoint $\tev_u^\dagger$, but every other choice leads to something that is 2-isomorphic to~$S_u$ as in~\eqref{eq:SerreAutomorphism}, see \cite[Cor.\,4.11]{PstragowskiJournalVersion}.} 
\be 
\label{eq:SerreAutomorphism}
S_u := (1_u \btimes \tev_u) \circ (b_{u,u} \btimes 1_{u^\#}) \circ (1_u \btimes \tev_u^\dagger) 
\ee 
as 
\be 
\label{eq:AdjointsViaSerre}
\tev_u^\dagger 
	\cong 
	\big( S_u \btimes 1_{u^\#} \big) \circ \coev_u
\, , \quad 
{}^\dagger\tev_u 
	\cong
	\big( S^{-1}_u \btimes 1_{u^\#} \big) \circ \coev_u 
\, . 
\ee 
This is a special case of \cite[Thm.\,4.13]{PstragowskiJournalVersion}, where it is shown that arbitrary multiple adjoints of $\tev_u, \tcoev_u$ like $\tev_u^{\dagger\dagger}$ or ${}^{\dagger\dagger\dagger}\tcoev_u$ are given only in terms of $\tev_u, \tcoev_u$, $b_{u^\#,u}$ and an appropriate power of~$S_u$. 

It follows that if the Serre automorphism is \textsl{trivialisable}, i.\,e.\ if there exists a 2-isomorphism $\lambda_u\colon S_u\lra 1_u$, then 
$ 
{}^\dagger \tev_u \cong \coev_u \cong \tev_u^\dagger 
$ 
and 
$ 
{}^\dagger \tcoev_u \cong \ev_u \cong \tcoev_u^\dagger 
$. 
In detail, the isomorphism between ${}^\dagger \tev_u$ and $\tev_u^\dagger$ induced by~$\lambda_u$ is 
\be 
\label{eq:Lambda}
\Lambda_u 
	:= 
	\big( \lambda_u^\dagger \cdot \lambda_u \btimes 1_{1_{u^\#}} \big) \circ 1_{\coev_u} 
	\colon 
	{}^\dagger \tev_u \stackrel{\cong}{\lra} \tev_u^\dagger 
\ee 
and similarly for ${}^\dagger \tcoev_u \cong \tcoev_u^\dagger$. 
This is the case relevant for oriented TQFT (see Section~\ref{subsec:CHwithoutDfects} below), so that $\tev_u$ and $\tcoev_u$ are mutually adjoint up to the braiding~$b_{u^\#,u}$, and 2-morphisms like $\ev_{\tev_u}$ and $\tev_{\tcoev_u}$ are identified up to the trivialisation~$\lambda_u$. 
As an illustration we can interpret a $u$-labelled torus as a 2-morphism $1_\one \lra 1_\one$ as follows: 
\begin{align}
\hspace{-0.6cm}
\tikzzbox{%

}%
\nonumber
\\ 
& = 
\tev_{\tev_u} 
\cdot \Big[ 1_{\tev_u} \circ \Big( \ev_{\tev_u} \cdot \big[ \Lambda_u^{-1} \circ 1_{\tev_u} \big] \cdot \tcoev_{\tev_u} \Big) \circ \Lambda_u \Big] \cdot \coev_{\tev_u} 
\label{eq:Torus1}
\end{align}
where as usual we suppress associator and unitor 2-morphisms, as we already did in~\eqref{eq:pivotality}. 
Note that a $u$-labelled torus represents the same 2-morphism as a $u^\#$-labelled torus; if one label is on the ``front'', the dual label is on the ``back''. 

\medskip 

Given dualisable objects $u,v\in\B$, the \textsl{right dual} of a 1-morphism $X\colon u\lra v$ is the 1-morphism $X^\# \colon v^\# \lra u^\# $ given by 
\be 
\label{eq:LeftDual1Morphism}
\tikzzbox{%

}     
\widehat{=} \, 
X^\# 
	:= 
	\big( 1_{u^\#} \btimes \tev_{v} \big) 
		\circ 
		\big( 1_{u^\#} \btimes X \btimes 1_{v^\#} \big) 
		\circ 
		\big( \tcoev_u \btimes 1_{v^\#} \big) \, . 
\,  
\ee 
Note that~$X^\#$ is different from the right adjoint $\Xd\colon v\lra u$ (if it exists). 
Moreover, for another 1-morphism $Y\colon u\lra v$ and a 2-morphism $\varphi\colon X\lra Y$, the \textsl{right dual} of the latter is 
\be 
\label{eq:varphiDual}
\varphi^\dual 
:= 
\tikzzbox{%
	%
}    
\, .
\ee 
Analogously one defines left duals ${}^\dual\varphi\colon {}^\dual\!X \lra {}^\dual Y$ in terms of the non-tilded adjunction 1-morphisms as in~\eqref{eq:LeftDuals}. 
Using the relation~\eqref{eq:LeftDuals} between left and right adjunction maps, one finds that the braiding induces canonical isomorphisms 
\be 
\label{eq:LeftRightDualX}
\omega_X\colon X^\# \stackrel{\cong}{\lra} {}^\#\!X \, . 
\ee 

\medskip 

From now on we assume that~$\B$ has a pivotal structure, and that there are coherent 2-isomorphisms 
\be 
\label{eq:CoherenceDaggerHash}
X^{\#\#} \cong X 
	\, , \quad 
	(X^\dagger)^\# \cong (X^\#)^\dagger
\ee 
for all 1-morphisms~$X$, where the first isomorphism is used to make the second compile. 
This coherence is studied in detail in the case of generically maximally strict 3-categories in \cite[Sect.\,4.3]{BMS}. 
Then one finds that $\#$-duals of 2-morphisms are compatible with the adjunction 1-morphisms in a sense that is expressed (in~\eqref{eq:Omegaphi} below) in terms of 2-isomorphisms 
\begin{align}
\tikzzbox{%

} 
& = 
	\Omega'_X
	\colon 
	\big( 1_{u^\#} \btimes X \big) \circ \tcoev_u
	\stackrel{\cong}{\lra} 
	\big( X^\# \btimes 1_v \big)\circ \tcoev_v 
	\label{eq:OmegaTilde}
\end{align}
for all $X\colon u\lra v$, which allow $X$-labelled lines to ``move to the other side'' of adjunction 1-morphisms. 
As explained in \cite[Sect.\,4.4]{BMS}, $\Omega_X$ is built from a cusp isomorphism, 
\be 
\label{eq:OmegaX}
\Omega_X = 
\tikzzbox{%

}
\, , 
\ee 
while~$\Omega'_X$ can either be taken to be $(\Omega_{X^\dagger}^\dagger)^{-1}$ (for an appropriate choice of pivotal structure, and using~\eqref{eq:CoherenceDaggerHash}), or it can be defined analogously to~\eqref{eq:OmegaX} in terms of $(c_{\textrm{l}}^v)^{-1}$; these two definitions of~$\Omega'_X$ are precisely related by the uniqueness result on ``coherent full duality data'' reviewed below in Section~\ref{subsec:CHwithoutDfects}. 
In the setting of \cite{BMS} one then has compatibility with ($\#$-duals of) 2-morphisms in the sense that 
\be 
\label{eq:Omegaphi}
\tikzzbox{%

} 
\ee 
for all $\varphi\colon X \lra Y$ as above. 

As an illustration, we may consider any number of parallel 1-morphisms $X_1,X_2,\dots, X_n\colon u\lra v$ and a sequence of 2-morphisms 
\be 
X_1 
	\stackrel{\varphi_1}{\lra}
	X_2 
	\stackrel{\varphi_2}{\lra}
	X_3 
	\stackrel{\varphi_3}{\lra}
	\dots 
	\stackrel{\varphi_{n-1}}{\lra}
	X_n
	\stackrel{\varphi_n}{\lra}
	X_1 
\, . 
\ee
If we assume~$X_1$ to have adjoints and a canonical isomorphism $\dX\cong\Xd$, then we can interpret a ``cylinder with a $\varphi_1$-$\dots$-$\varphi_n$-necklace'' as the 2-morphism 
\be 
N_{\varphi_1,\dots,\varphi_n} 
	:= 
	\tikzzbox{%
 
	}  
\colon \ev_v \circ \,\ev_v^\dagger \lra \ev_u \circ \,\ev_u^\dagger \, ,
\ee 
where we also used the canonical isomorphisms \eqref{eq:LeftRightDualX} and~\eqref{eq:CoherenceDaggerHash} for~$X_1$. 
Since by pivotality all~$X_i$ have canonically isomorphic adjoints, it follows from the compatibility condition~\eqref{eq:Omegaphi} that the $\varphi_i$-labelled ``beads'' can be moved around cyclically without changing the 2-morphism. 
In particular, we have 
$N_{\varphi_1,\dots,\varphi_n}  
	= 
	N_{\varphi_i,\dots,\varphi_n,\varphi_1,\dots,\varphi_{i-1}}$ 
for all $i\in\{1,\dots,n\}$.

\subsection{Extended bordism 2-category}
\label{subsec:ExtendedBordisms}

We now briefly recall the symmetric monoidal 2-category $\Bordor$ of oriented points, lines, and surfaces with corners, which is constructed in detail in \cite[Sect.\,3.1--3.2]{spthesis}. 

Objects of $\Bordor$ are 2-haloed\footnote{A 2-haloed point is a point~$p$ together with an embedding of~$p$ into an open arc~$a$, and an embedding of~$a$ into an oriented open disc. Similarly, a 2-halo for a 1-dimensional manifold~$\ell$ is an embedding of~$\ell$ into an oriented 2-dimensional neighbourhood. Bordisms between haloed manifolds must be compatible with the haloes. In light of the generator-and-relations description of $\Bordor$ recalled below, we need not consider haloes after all.} 0-dimensional compact oriented manifolds, i.\,e.\ finite disjoint unions of positively ($+$) and negatively ($-$) oriented points. 
1-morphisms are (2-haloed) 1-dimensional oriented bordisms, i.\,e.\ compact oriented lines whose orientation is compatible with their source and target, and similarly 2-morphisms are diffeomorphism classes of 2-dimensional compact oriented bordisms with corners. 
In line with our diagrammatic conventions for monoidal, horizontal and vertical composition in any monoidal 2-category (see Section~\ref{subsec:GraphicalCalculus}), we thus have the following examples of objects, 1- and 2-morphisms of $\Bordor$: 
\begin{align}
\textrm{objects: } \;  
	& 
	+ \, , -
\\ 
\textrm{1-morphisms: }
	& 
	\tikzzbox{%

	}   
	\, . 
	\label{eq:BordCusps}
\end{align}

In fact it is shown in \cite{spthesis} that the above generate $\Bordor$ as a symmetric monoidal pivotal 2-category with duals, with relations expressing that 
\begin{itemize}
	\item 
	$+$ and~$-$ are mutually dual, 
	\item 
	$
	\coev_+ := 
		%
	$ 
	(where the left-hand side involves the braiding~$b_{-,+}$ which by definition is the mapping cylinder over the swap diffeomorphism $-\sqcup + \lra +\sqcup -$) 
	is left and right adjoint to $\tev_+$, 
	\item 
	$
	\ev_+ := 
		%
	$ 
	is left and right adjoint to $\tcoev_+$, 
	\item 
	the cusps~\eqref{eq:BordCusps} are invertible, and compatible with the adjunction 1-morphisms \eqref{eq:BordAdj1}--\eqref{eq:BordAdj4}. 
\end{itemize}
In practice this simply means that every 1-morphism can be (non-uniquely) decomposed into straight lines $1_+, 1_-$ as well as the ``elbows'' \eqref{eq:Elbow1}, \eqref{eq:Elbow2} and their flipped versions (using the symmetric braiding), and every 2-morphism can be (non-uniquely) decomposed into cylinders, saddles, caps, and their upside-down versions. 
For example, a closed surface of genus~$g$, viewed as a 2-morphism $1_\varnothing \lra 1_\varnothing$, can be decomposed into~$g$ saddles, $g$ upside-down saddles, two cylinders over semicircles, one cup and one cap (compare~\eqref{eq:Torus1} for the case $g=1$). 

Put differently, the generators-and-relations description of $\Bordor$ together with the 3-dimensional graphical calculus makes intuition about decomposing oriented surfaces rigorous. 
Since every object in $\Bordor$ is a disjoint union of~+ and $+^\#=-$, we suppress the object labels for 2-dimensional regions in the graphical calculus for $\Bordor$; this translates between the algebraic reasoning for general symmetric monoidal 2-categories and the geometric content of $\Bordor$.

\subsection{Cobordism hypothesis without defects}
\label{subsec:CHwithoutDfects}

Let~$\B$ be a symmetric monoidal 2-category. 
A \textsl{2-dimensional extended oriented TQFT with values in~$\B$} is a symmetric monoidal 2-functor 
$
\zz \colon \Bordor \lra \B 
$. 
In this section we review the 2-dimensional cobordism hypothesis for such TQFTs, i.\,e.\ how to equivalently describe them in terms of data internal to~$\B$, based on \cite{spthesis, PstragowskiJournalVersion, HSV, HV, Hesse}. 

\medskip 

The idea of the cobordism hypothesis is to give a generators-and-relations description which is equivalent to the bordism category. 
The same generators-and-relations conditions in the target category~$\B$ then equivalently describe extended TQFTs. 
This leads to the concise formulation of the cobordism hypothesis that the 2-groupoid of extended framed TQFTs is equivalent to the maximal sub-2-groupoid of fully dualisable objects in~$\B$, while oriented TQFTs need the additional structure of an $\SO(2)$-homotopy fixed point, i.\,e.\ a trivialisation of the Serre automorphism. 
In the remainder of this section, we explain what this means explicitly. 

\medskip 

We start by observing that the uniqueness property of adjoints of 1-morphisms in~$\B$ stated after~\eqref{eq:ZorroMoves} corresponds to the following equivalence of categories. 
For a given monoidal category~$\mathcal M$, there is a category $\textrm{DuDa}(\mathcal M)$ whose objects are duality data $(X,\Xd,\tev_X,\tcoev_X)$ satisfying the Zorro moves in~$\mathcal M$, and whose morphisms are given by morphisms in~$\mathcal M$ that are compatible with the duality maps. 
On the other hand, there is a category~$\mathcal M^{\textrm{d}}$ which consists of dualisable objects in~$\mathcal M$ and their morphisms; no choices of duality data are involved in the definition of~$\mathcal M^{\textrm{d}}$. 
The uniqueness of duals is then precisely that the forgetful functor
\begin{align}
\textrm{DuDa}(\mathcal M) & \lra \big( \mathcal M^{\textrm{d}} \big)\!^\times \nonumber
	\\
	\big( X,\Xd,\tev_X,\tcoev_X \big) & \lmt X
\end{align}
is an equivalence, where $(\mathcal M^{\textrm{d}})^\times$ is the maximal subgroupoid of~$\mathcal M^{\textrm{d}}$ (whose morphisms by definition are the isomorphisms in~$\mathcal M^{\textrm{d}}$). 

\medskip 

The analogue  of the above for symmetric monoidal 2-categories~$\B$ is more involved, as explained in basically two steps in \cite{PstragowskiJournalVersion}. 
The first step is to consider the maximal sub-2-groupoid $(\mathcal B^{\textrm{d}})^\times$ of dualisable objects in~$\B$, and then ask for an equivalent 2-category whose objects explicitly witness dualisability. 
The subtlety here is that the naive 2-category whose objects are duality data $(u,u^\#,\tev_u,\tcoev_u,c^u_{\textrm{l}},c^u_{\textrm{r}})$ as in \eqref{eq:TildeCoEv}--\eqref{eq:CuspIsomorphisms} is \textsl{not} equivalent to $(\mathcal B^{\textrm{d}})^\times$. 
Instead one finds that \textsl{coherent} duality data are needed, in which by definition the cusp isomorphisms have to satisfy the \textsl{swallowtail identities} of \cite[Fig.\,2\,\&\,3]{PstragowskiJournalVersion}, which in the graphical calculus (where we suppress structure morphisms of the underlying symmetric monoidal 2-category) are equivalent to 
\be 
\label{eq:Swallowtail}
\tikzzbox{%

}
\ee 
which is also the second relation in \cite[Fig.\,3.13]{spthesis}. 

Coherent duality data give rise to a 2-category $\textrm{DuDa}^{\textrm{coh}}(\mathcal B)$, and by \cite[Thm.\,3.14]{PstragowskiJournalVersion} the forgetful functor 
\begin{align}
\textrm{DuDa}^{\textrm{coh}}(\mathcal B) & \lra \big( \mathcal B^{\textrm{d}} \big)\!^\times \nonumber
\\
\big( u,u^\#,\tev_u,\tcoev_u,c^u_{\textrm{l}},c^u_{\textrm{r}} \big) & \lmt u
\end{align}
is an equivalence. 
As explained in loc.\ cit.\ every object in $\textrm{DuDa}(\mathcal B)$ can be made into one in $\textrm{DuDa}^{\textrm{coh}}(\mathcal B)$ by (uniquely) changing only one of the cusp isomorphisms. 

\medskip 

The second step is to consider the full sub-2-groupoid $(\mathcal B^{\textrm{fd}})^\times$ of \textsl{fully} dualisable objects in~$\B$, and again ask for an equivalent 2-category whose objects are \textsl{full duality data} 
\be 
\label{eq:FDD}
\Big( 
 	u,u^\#,\tev_u,\tcoev_u,S_u, S_u^{-1}, c^u_{\textrm{l}},c^u_{\textrm{r}}, \ev_{\tev_u}, \coev_{\tev_u}, \ev_{\tcoev_u}, \coev_{\tcoev_u}, \phi, \psi 
 	\Big) 
\ee 
where a choice of Serre automorphism~$S_u$ and its weak inverse~$S_u^{-1}$ gives us the adjoints 
$
{}^\dagger\tev_u 
	= 
	( S^{-1}_u \btimes 1_{u^\#} ) \circ b_{u^\#,u} \circ  \tcoev_u 
$ 
and 
$
{}^\dagger\tcoev_u 
	= 
	\tev_u \circ b_{u^\#,u} \circ (1_{u^\#} \btimes S_u)
$ 
(recall \eqref{eq:AdjointsViaSerre}), 
while $\phi\colon S_u^{-1} \circ S_u \lra 1_u$ and $\psi\colon S_u \circ S_u^{-1} \lra 1_u$ are chosen 2-isomorphisms. 
As shown in \cite{PstragowskiJournalVersion}, the naive 2-category of all such data is \textsl{not} equivalent to $(\mathcal B^{\textrm{fd}})^\times$. 
One instead has to impose an additional compatibility condition between adjunction 2-morphisms and cusp isomorphisms: full duality data as in~\eqref{eq:FDD} are \textsl{coherent} if in addition to the swallowtail identity~\eqref{eq:Swallowtail} they satisfy the \textsl{cusp-counit identity} 
\be 
\label{eq:CuspCounit}
\tikzzbox{%

}
\ee 
is defined in terms of~$c^u_{\textrm{l}}$ and~$\psi$ as well as the 2-isomorphism marked ``$\cong$'' which is obtained from the braided monoidal structure on~$\B$. 
As observed in \cite[Rem.\,4.24]{PstragowskiJournalVersion}, one checks that the cusp-counit identity corresponds to the ``cusp flip identity'', i.\,e.\ the third relation in \cite[Fig.\,3.13]{spthesis}. 
Moreover, it follows from~\eqref{eq:CuspCounit} and the standard ``sliding relations'' between morphisms and their adjoints that 
\be 
\label{eq:crTildeFromcl}
\widetilde c^u_{\textrm{r}} = \big( {}^\dagger\! c^u_{\textrm{l}} \big)^{-1} 
	\quad 
	\textrm{(up to coherence 2-isomorphisms of~$\B$)} \, . 
\ee 

Coherent full duality data give rise to a 2-category $\textrm{FuDuDa}^{\textrm{coh}}(\mathcal B)$, and by \cite[Thm.\,4.27]{PstragowskiJournalVersion} the forgetful functor
\begin{align}
\textrm{FuDuDa}^{\textrm{coh}}(\mathcal B) & \lra \big( \mathcal B^{\textrm{fd}} \big)\!^\times \nonumber
\\
\big( u,u^\#,\tev_u,\dots \big) \textrm{ as in \eqref{eq:FDD}} & \lmt u
\end{align}
is an equivalence. 
In particular, it is shown in loc.\ cit.\ that any full duality data~\eqref{eq:FDD} for some $u\in\mathcal B^{\textrm{fd}}$ can be made coherent by (uniquely) changing only one of the cusp isomorphisms as well as $\ev_{\tev_u}$ and $\coev_{\tev_u}$. 

\medskip 

We are now ready to explicitly re-state the  

\medskip 

\noindent
\textbf{2-dimensional oriented cobordism hypothesis without defects: }

\noindent 
An extended TQFT $\zz \colon \Bordor \lra \B$ is equivalently described by a pair $(u,\lambda_u)$ of a fully dualisable object $u \in \B$ together with chosen coherent full duality data~\eqref{eq:FDD} and a 2-isomorphism $\lambda_u\colon S_u \lra 1_u$. 
Evaluating~$\zz$ on any 2-morphisms $[\Sigma]$ in $\Borddef$ then amounts to 
\begin{itemize}
	\item 
	choosing a representative bordism~$\Sigma$, 
	\item 
	choosing a generic embedding of~$\Sigma$ into the cube $[0,1]^{\times 3}$ compatible with the graphical calculus of $\Bordor$, and 
	\item 
	interpreting the resulting diagram in the graphical calculus for~$\B$ (applying the maps~$\lambda_u$ as needed, recall~\eqref{eq:Lambda} and~\eqref{eq:Torus1}). 
\end{itemize}
By restricting this procedure to boundaries one also obtains the action of~$\zz$ on objects and 1-morphisms. 

\medskip 

As an illustration, we may consider the formula in~\eqref{eq:Torus1} as the torus partition function $\zz(T^2)$ of a TQFT~$\zz$ described by the pair $(u,\lambda_u)$ with chosen coherent full duality data. 
More generally, we can express a closed surface~$\Sigma_g$ of genus~$g$ by replacing the middle ``handle operator'' in~\eqref{eq:Torus1} by its $g$-th power, so we arrive at  
\be 
\label{eq:ZStateSpacesGeneral}
\zz(\Sigma_g) 
= 
\tev_{\tev_u} 
\cdot \Big[ 1_{\tev_u} \circ \Big( \ev_{\tev_u} \cdot \big[ \Lambda_u^{-1} \circ 1_{\tev_u} \big] \cdot \tcoev_{\tev_u} \Big)^g \circ \Lambda_u \Big] \cdot \coev_{\tev_u} 
. 
\ee 

\medskip 

As a final comment before moving on to defects, we recall that extended \textsl{framed} TQFTs -- where all bordisms in the domain category by definition are endowed with a trivialisation of the (stabilised) tangent bundle -- are equivalently described solely by a fully dualisable objects along with chosen coherent fully duality data. 
Note that not every orientable bordism admits a framing, e.\,g.\ the sphere~$S^2$ does not.

\section{2-dimensional extended TQFTs with defects}
\label{sec:ExtendedDefects}

In Section~\ref{subsec:ExtendedDefectBordisms} we first describe extended defect bordism 2-categories $\Borddef$ that encode arbitrary oriented line and point defects on surfaces with corners. 
Our presentation builds on \cite[Sect.\,4.3]{l0905.0465} and is formulated as a natural generalisation of the non-extended defect bordisms of \cite[Sect.\,2]{dkr1107.0495} and \cite[Sect.\,2]{CRS1}, to which we refer for further details. 
Then in Section~\ref{subsec:ExtendedDefectTQFTs} we introduce extended defect TQFTs as functors on $\Borddef$ and phrase a special case of the ``cobordism hypothesis with singularities'' of \cite{l0905.0465} as a natural instruction how to ``compute'' extended defect TQFTs.

\subsection{Extended defect bordism 2-categories}
\label{subsec:ExtendedDefectBordisms}

We describe topological defects in two dimensions in terms of \textsl{defect data} $\mathds{D} =(D_2,D_1,D_0,s,t,j)$ as in \cite{dkr1107.0495, CRS1}: 
the sets~$D_j$ have $j$-dimensional defect labels as their elements (where 2-dimensional defects are bulk theories), while the source and target maps 
\be 
s,t\colon D_1 \lra D_2 
\ee 
and the junction map 
\be 
\label{eq:JunctionMap}
j\colon D_0 \lra D_2 \sqcup \bigsqcup_{m\geqslant 1} \big( (D_1 \times \{\pm\}) \times_{D_2} \dots \times_{D_2} (D_1 \times \{\pm\}) \big)/\Z_m
\ee  
(where the quotient by~$\Z_m$ expresses the cyclic symmetry around junction points, see the last picture in~\eqref{eq:STJExamples} below)
encode how defects are allowed to meet locally. 
This is illustrated by the following examples of local patches on defect bordisms, where elements of~$D_j$ are used to label $j$-dimensional strata: 
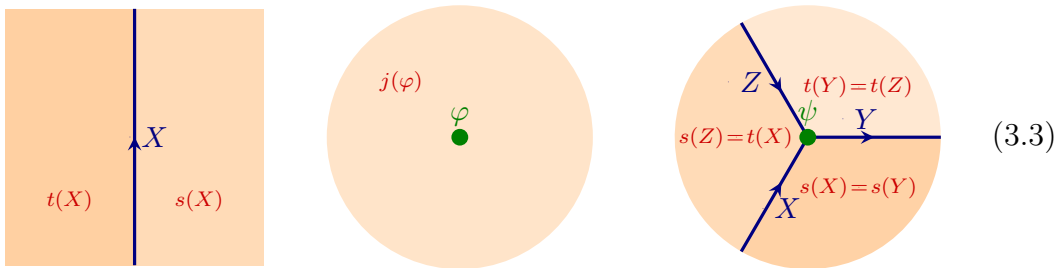
\be 
\label{eq:STJExamples}
\tikzzbox{
}
\ee 
Here, $X,Y,Z \in D_1$, $\varphi\in D_0$ is such that $j(\varphi) \in D_2$, and $\psi\in D_0$ is such that $j(\psi) \in [(X,+), (Y,-), (Z,+)]$. 
Note that with this our conventions regarding orientations can be read off of the above pictures; in particular, the signs~$\pm$ in~\eqref{eq:JunctionMap} encode the orientations of line defects incident on point defects. 

To give some examples of defect data, recall that in the case of Landau--Ginzburg models we can choose~$\mathds{D}$ such that~$D_2$ consists of all potentials, while~$D_1$ and~$D_0$ consist of matrix factorisations and their maps up to homotopy, respectively. 
As another example, in subsequent sections we will consider the defect data whose sets $D_2, D_1, D_0$ are given by the objects, 1- and 2-morphisms of the 2-category of truncated affine Rozansky--Witten models~$\bi$. 
Note that also the maps $s,t,j$ are naturally obtained from the structure of~$\bi$. 
More generally, in all examples known to us the defect data~$\mathds{D}$ for a given type of model can be extracted from the structure of the associated pivotal 2-category, which in turn is the codomain of the associated extended TQFT, cf.\ the discussion in \cite[Sect.\,1]{CMM}. 
However, by definition defect data are just three sets together with adjacency maps $s,t,j$. 

\medskip 

From now on we fix a set of defect data~$\mathds{D}$. 
The idea of the 2-category $\Borddef$ is to add $\mathds{D}$-labelled stratifications to the surfaces with corners representing 2-morphisms in $\Bordor$, as well as compatibly labelled stratifications to 1-morphisms and objects. 
For example, we will see below in~\eqref{eq:ExampleDefectBordismEvaluated} how 
\begin{align}
\label{eq:StratBordExample}
\tikzzbox{%

}%
\end{align}
represents a 2-morphism in $\Borddef$, whose underlying 2-morphism in $\Bordor$ is the saddle in~\eqref{eq:BordAdj4}. 

By definition, objects in $\Borddef$ are objects in $\Bordor$ together with a label from~$D_2$ for every individual oriented point. 
Hence 
\be 
\varnothing  \, , 
\quad 
(u,+) \equiv 
\tikzzbox{%
	%
}
\ee 
are three distinct objects, where $u,v\in D_2$. 
Similarly, 1-morphisms are those in $\Bordor$ together with stratifications whose $(j-1)$-strata are labelled by~$D_j$ in a way compatible with the maps $s,t$ as well as the source and target objects. 
All 1-strata inherit their orientation from the underlying 1-morphism in $\Bordor$ (which we  suppress notationally), and 0-strata can have arbitrary orientations as long as they are compatible with the maps $s,t$. 
This is illustrated in the following examples: 
\begin{align}
\tikzzbox{%

}
&
\lra 
\varnothing \, . 
\end{align}
Finally, 2-morphisms in $\Borddef$ are represented by $\mathds{D}$-labelled stratified surfaces with corners such that 1-strata may only end (transversally) at compatibly labelled 0-strata in the bulk or boundary of the surface. 
Hence 
\be 
\tikzzbox{
	%
}
\, . 
\ee 
Two such defect bordisms represent the same 2-morphism if there is a homeomorphism~$f$ between them that is compatible with the boundary parametrisation and restricts to diffeomorphisms on 2-, 1- and 0-strata, respectively. 
In particular all representatives have the same number of defects, whose labels must match under~$f$. 

Horizontal and vertical composition in $\Borddef$ are given by horizontal and vertical glueing, and taking disjoint unions leads to a symmetric monoidal structure in a standard fashion. 
To describe the relation to $\Bordor$ more precisely, for every $u\in D_2$ we write $\Borddef|_u$ for the non-full sub-2-category of trivially stratified 1- and 2-morphisms where everything (on the level of objects, 1- and 2-morphisms) is labelled only by~$u$. 
Then simply forgetting these labels gives rise to equivalences of symmetric monoidal 2-categories 
\be 
\label{eq:RestrictBordDef}
\Borddef\Big|_u 
	\stackrel{\cong}{\lra} 
	\Bordor
	\quad \textrm{ for all } u \in D_2 \, . 
\ee 

Similarly, the relation to the non-extended defect bordism categories of \cite{dkr1107.0495, CRS1} is given by restricting to closed 1-morphisms, or more precisely to the endomorphism category of the unit object~$\varnothing$: 
\be 
\label{eq:NonExtendedDefectBordisms}
\Bord_{2,1}^{\textrm{def}}(\mathds{D}) 
	\cong 
	\Omega \big( \Borddef \big) 
	:= 
	\textrm{End}_{\Borddef}(\varnothing) \, . 
\ee 
A special case is the one without defects, $\Bord_{2,1}^{\textrm{or}} \cong \Omega(\Bordor)$. 

\medskip 

Every object in $\Borddef$ has a dual, with $(u,\pm)^\# = (u,\mp)$ for all $u\in D_2$. 
The adjunction 1-morphisms witnessing the dualities are simply those of $\Bordor$ dressed with the appropriate labels, for example 
\be 
\tev_{(u,+)} = 
\tikzzbox{%

}  
. 
\ee
Moreover, every 1-morphism has a left and right adjoint obtained from the adjoint of the underlying 1-morphism in $\Bordor$ by reversing the order of strata and the orientation of all 0-strata (while the orientation of 1-strata does not change). 
We illustrate this with two examples: 
\begin{align}
\left(
\tikzzbox{%
	%
}
\!\!\!\! .
\label{eq:AdjEx2}
\end{align}
The adjunction 2-morphisms witnessing these adjunctions are those of the underlying data in $\Bordor$ together with appropriately labelled U-shaped 1-strata. 
For example, the evaluation 2-morphisms witnessing~\eqref{eq:AdjEx} and~\eqref{eq:AdjEx2} as right adjoints are, respectively, 
\vspace{-2.5em}
\be 
\tikzzbox{%
	%
}%
\, . 
\ee 
By pre-composing left/right evaluations with right/left coevaluations, we obtain left/right quantum dimensions of 1-morphisms. 
For example, the defect cylinder 
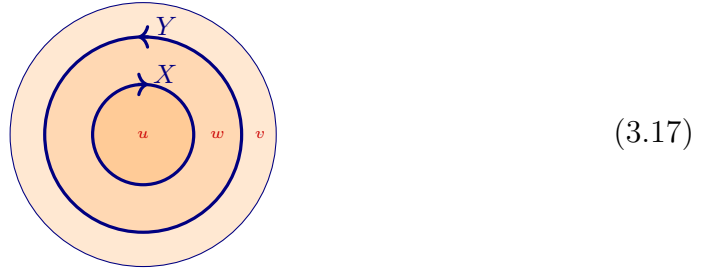
\be 
\tikzzbox
{
	%
 
} 
\ee 
viewed as a 2-morphism $1_{(v,+)} \lra 1_{(v,+)}$, is the right quantum dimension of 
\be 
\tikzzbox{%
	%
}
\ee 
in $\Borddef$. 

\medskip 

The special case of a defect datum~$\mathds{D}^\partial$ with 
\be \label{eq:boundarydatum}
D^\partial_0 := \varnothing 
	\, , \quad 
	D^\partial_2 := \{\bullet, \circ\} 
	\, , \quad 
	s(D^\partial_1) := \{\circ\} 
	\, , \quad 
	t(D^\partial_1) := \{\bullet\}
\ee 
and~$D^\partial_1$ an arbitrary set can be used to describe bordisms with (only) boundary conditions for a single bulk theory. 
Indeed, by forgetting all $\circ$-labelled strata and the $\bullet$-label of the remaining strata, we obtain from $\Bord_{2,1,0}^{\textrm{def}}(\mathds{D}^\partial)$ a symmetric monoidal 2-category $\Bordoc(D_1^\partial)$ which extends the non-extended open/closed bordism category with boundary conditions, compare e.\,g.\ \cite[Sect.\,2.1.1]{2dDefectTQFTLectureNotes}. 
Here we think of~$\circ$ as the trivial bulk theory, of~$\bullet$ as \textsl{the} bulk theory, and elements of~$D^\partial_1$ are boundary conditions. 
Note that as recalled below, we typically extract defect data from associated symmetric monoidal 2-categories~$\B$; then~$\circ$ corresponds to the monoidal unit of~$\B$. 
Note also that $D_1^\partial$-labelled ``endpoints'' of 1-morphisms in $\Bordoc(D_1^\partial)$ should not be confused with objects; for example, by setting $u=\circ$ and $v=\bullet$ in~\eqref{eq:1MorBordEx}, we obtain
\be 
\tikzzbox{%
	\begin{tikzpicture}[thick,scale=1,color=black, baseline=-0.08cm]
	\coordinate (l) at (0,0);
	\coordinate (r) at (4,0);
	\coordinate (m) at ($(l)-0.5*(l)+0.5*(r)$);
	\coordinate (lm) at ($(l)-0.2*(l)+0.2*(r)$);
	\coordinate (rm) at ($(l)-0.8*(l)+0.8*(r)$);
	\coordinate (l2) at (0,-1);
	\coordinate (r2) at (4,-1);
	\coordinate (m2) at ($(l2)-0.5*(l2)+0.5*(r2)$);
	\coordinate (ml2) at ($(l2)-0.3*(l2)+0.3*(r2)$);
	\coordinate (mr2) at ($(l2)-0.7*(l2)+0.7*(r2)$);
	\coordinate (lm2) at ($(l2)-0.15*(l2)+0.15*(r2)$);
	\coordinate (rm2) at ($(l2)-0.85*(l2)+0.85*(r2)$);
	\draw[very thick, red!80!black] (l) -- (m); 
	\fill[red!80!black] (l) circle (0pt) node[left] {{\tiny$+$}};
	\fill[color=blue!50!black] (m) circle (2.5pt) node[above] {{\footnotesize $(X,+)$}};
	\end{tikzpicture}
}
\colon 
\varnothing \lra + \, . 
\ee 
Versions of $\Bordoc(\{*\})$ for various tangential structures are described in more detail in \cite[\S2.1.6\,\&\,App.\,A.3]{FreedTeleman2020}.

\subsection{Cobordism hypothesis with defects}
\label{subsec:ExtendedDefectTQFTs}

We fix defect data~$\mathds{D}$ as well as a symmetric monoidal 2-category~$\B$, for example one of those listed at the beginning of Section~\ref{subsec:GraphicalCalculus}. 
A \textsl{2-dimensional extended defect TQFT with defect data~$\mathds{D}$ and with values in~$\B$} is a symmetric mononoidal 2-functor 
\be 
\zz \colon \Borddef \lra \B 
\, . 
\ee 

Combining the ordinary cobordism hypothesis as reviewed in Section~\ref{subsec:CHwithoutDfects} with the equivalence~\eqref{eq:RestrictBordDef}, we see that an extended defect TQFT restricted to $\Borddef|_u$ for some $u\in D_2$ is equivalently described by a fully dualisable object $\zz(u) := \zz(u,+) \in \B$ together with a trivialisation $\lambda_u \colon S_{\zz(u)} \lra 1_{\zz(u)}$ of its Serre automorphism. 
We are thus left with the question of how to describe (internally to~$\B$) the action of~$\zz$ on nontrivially stratified bordisms. 
An answer was put forward in the context of $(\infty,n)$-categories in \cite[Sect.\,4.3]{l0905.0465} as the ``cobordism hypothesis with singularities'', which restricts to the standard cobordism hypothesis in the case of trivial stratifications. 
Lurie also provides strong evidence for the validity of the cobordism hypothesis with singularities/defects. 
To our knowledge a detailed, complete proof has however not been published -- neither for arbitrary $(\infty,n)$-categories nor in a setting of (weak) 2-categories relevant for our purposes here. 
(Clearly though, \cite{l0905.0465} as well as other important work such as \cite{AyalaFrancis2017CH} has come very close to a complete rigorous proof.)

Absent an established theorem, we base our applications in subsequent sections on the assumption that the following version of the cobordism hypothesis with singularities holds: 

\medskip 

\noindent
\textbf{2-dimensional cobordism hypothesis with defects: }

\noindent 
An extended defect TQFT $\zz \colon \Borddef \lra \B$ is equivalently described by 
\begin{itemize}
	\item[(0)] 
	pairs $(\zz(u),\lambda_u)$ of fully dualisable objects $\zz(u) \in \B$ together with chosen coherent full duality data as in Section~\ref{subsec:CHwithoutDfects} and 2-isomorphisms $\lambda_u \colon S_{\zz(u)} \lra 1_{\zz(u)}$ for all $u\in D_2$, 
	\item[(1)] 
	1-morphisms $\zz(X) \colon \zz(s(X)) \lra \zz(t(X))$ for all $X\in D_1$ that have coherently isomorphic left and right adjoints ${}^\dagger\zz(X) \cong \zz(X)^\dagger$ with chosen adjunction data, 
	\item[(2)] 
	2-morphisms $\zz(\varphi) \colon 1_{\zz(s(X_1))} \lra \bigotimes_{i=1}^m \zz(X_i)^{\varepsilon_i}$ (where $Z^+ := Z$ and $Z^- := Z^\dagger$) if $j(\varphi) = [(X_1,\varepsilon_1), \dots, (X_m, \varepsilon_m)]$, and $\zz(\varphi) \in \End_\B(1_{\zz(j(\varphi))})$ if $j(\varphi) \in D_2$, for all $\varphi \in D_0$.\footnote{Here we arbitrarily single out~$X_1$, but there is a canonical isomorphism for every choice of linear order on $[(X_1,\varepsilon_1), \dots, (X_m, \varepsilon_m)]$, compare~\eqref{eq:JunctionMap}.}
\end{itemize}
Evaluating~$\zz$ on any 2-morphisms $[\Sigma]$ in $\Borddef$ then amounts to 
\begin{itemize}
	\item 
	choosing a representative defect bordism~$\Sigma$, 
	\item 
	choosing a generic embedding of~$\Sigma$ into the cube $[0,1]^{\times 3}$ compatible with the graphical calculus of $\Borddef$, 
	\item 
	replacing every label $\xi\in D_j$ of $j$-strata by $\zz(\xi)$ for all $j\in \{0,1,2\}$, and  
	\item 
	interpreting the resulting diagram in the graphical calculus for~$\B$ (applying the maps~$\lambda_u$ as needed, recall~\eqref{eq:Lambda} and~\eqref{eq:Torus1}). 
\end{itemize}
By restricting this procedure to boundaries one also obtains the action of~$\zz$ on objects and 1-morphisms. 

\medskip 

In short, the 2-dimensional cobordism hypothesis with defects precisely instructs us to interpret defect bordisms as 2-morphisms in the target~$\B$, and analogously in higher dimensions. 
That their diagrammatic evaluation is independent of the choices made is a non-trivial statement. 

We also observe that the natural isomorphism ${}^\dagger\zz(X) \cong \zz(X)^\dagger$ follows from the fact that every extended defect TQFT~$\zz$ gives rise to a non-extended defect TQFT~$\zz'$ by restricting to surfaces without corners, cf.~\eqref{eq:NonExtendedDefectBordisms}. 
Indeed, the 2-category associated to~$\zz'$ in \cite{dkr1107.0495} has a pivotal structure, and it is equivalent to the image of~$\zz$ in the above formulation of the cobordism hypothesis with defects. 

\medskip 

As a first illustration, a 2-sphere stratified by a single $X$-labelled loop with $s(X) = u$ and $t(X) = v$, we have 
\be 
\label{eq:ZonLoopSphere}
\mathcal Z \Bigg( 
\tikzzbox{\begin{tikzpicture}[very thick,scale=1.2,color=blue!50!black, >=stealth, baseline=0]
	\fill[ball color=orange!40!white] (0,0) circle (0.95 cm);
	\draw (0,0) circle (0.65);
	\fill (90:0.42) circle (0pt) node {{\small$X$}};
	\fill (90:0) circle (0pt) node[red!80!black] {{\scriptsize$v$}};
	\fill (180:0.8) circle (0pt) node[red!80!black] {{\scriptsize$u$}};
	\draw[<-, very thick] (-0.100,0.65) -- (0.10001,0.65) node[below] {}; 
	\end{tikzpicture}}
\Bigg) 
	= 
	\tev_{\tev_{\zz(u)}} 
		\cdot \Big[ 1_{1_{\zz(u)}} \btimes \textrm{dim}_{\textrm{l}} \big(\zz(X)\big)\Big] 
		\cdot 
		\coev_{\tev_{\zz(u)}}
\, . 
\ee 
Another example is that~$\zz$ applied to the bordism in~\eqref{eq:StratBordExample} produces the 2-morphism 
\begin{align}
& \Big[ 1_{1_{\zz(v)}}  \Big( \btimes \tr_{\textrm{l}}\big( \zz(\varphi) \big) \cdot {\zz(\xi)} \Big) \Big] 
	\circ \tcoev_{\tev_{\zz(v)}} \circ \Big[ 1_{\zz(Y)} \btimes \ev_{\zz(X)} \Big]
	\nonumber
	\\
	& 
	\qquad \qquad \qquad \qquad 
	\in 
	\B \Big( Y \btimes \big[ X^\dagger \circ X\big] , \, \ev_{\zz(v)}^\dagger \circ \ev_{\zz(v)} \circ \big[ Y \btimes 1_v \big] \Big)
	\label{eq:ExampleDefectBordismEvaluated}
\end{align}

If in the example of~\eqref{eq:ZonLoopSphere} we have that $\zz(u) \cong \one$ is equivalent to the unit object of~$\B$, then~$X$ is a boundary condition for the $v$-labelled bulk theory of the TQFT~$\zz$. 
Hence the ``disc partition function'' is equal to the left quantum dimension of $\zz(X)$ in~$\B$, 
\be 
\mathcal Z \Bigg(
\tikzzbox
{
 
} 
= 
\dim_{\textrm{l}} \big( \mathcal Z(X) \big)
\, . 
\ee 
This is of course well-known also from non-extended TQFTs, recall e.\,g.\ \cite[Sect.\,3.3.1]{2dDefectTQFTLectureNotes} and~\eqref{eq:RestrictBordDef}. 
In particular we recover the full structure of 2-dimensional open/closed TQFT of \cite{l0010269, MooreSegal2dTQFT, lp0510664}, including the bulk-boundary map
\be
\label{eq:Whistle}
\mathcal Z \left(
\tikzzbox{%
	%
}
\ee
from the $v$-circle state space to the interval state space for the boundary condition $\zz(X)$. 
Below in Section~\ref{subsubsec:open-closed} we discuss this in more detail in the context of truncated affine Rozansky--Witten models.

\section{Defects in truncated affine Rozansky--Witten models}
\label{sec:DefectsRW}

In \cite{BCR} extended TQFTs were associated to 3-dimensional Rozansky--Witten models with target spaces $T^*\C^n$. 
Due to the non-compactness of the targets, these models exhibit infinite-dimensional state spaces and hence do not give rise to well-defined 3-dimensional TQFTs valued in $\Vect_\C$. 
In particular, such affine Rozansky--Witten models cannot be extended, over~$\C$, all the way down to the point.\footnote{Individual affine Rozansky--Witten models can very likely be extended to the point, but with values in a target 3-category which is a looping not of $\Vect_\C$, but a category of modules over a polynomial ring.} 
One can however consider a truncation, essentially by forgetting the 3-dimensional part of the theory. Such a truncation was constructed in \cite{BCR}. 
It is a 
2-dimensional extended TQFT whose target 2-category~$\bigra$ is given by a truncation of the 3-category of defects in affine Rozansky--Witten models. 
Since the latter is under very good control, the construction is very explicit.

In this section we discuss how to incorporate line and point defects in this truncated extended TQFT, which are surface and line defects from the point of view of the original 3-dimensional TQFT.
We start in Section~\ref{subsec:targetcategory} with a review of the target 2-category $\bigra$. In Section~\ref{subsec:adjunctions} we review that all objects are fully dualisable, and prove that~$\bigra$ has a pivotal structure, so in particular every 1-morphism has left and right adjoints. 
Having assembled all necessary ingredients, we spell out the explicit construction of the extended defect TQFT in Section~\ref{subsubsec:extendedtqft}. 
Finally in Section~\ref{subsec:samplecalculations} we compute some concrete examples of correlation functions involving line defects.

\subsection{The target 2-category}
\label{subsec:targetcategory}

The starting point of our construction is the (slightly conjectural\footnote{Rigorous proofs of the respective axioms of 3-categories have not been published.}) 3-category of defects in Rozansky--Witten models which was first described in \cite{KRS,KR0909.3643}. 
Restricted to affine Rozansky--Witten models, i.\,e.\ those with target spaces $T^*\C^n$, it takes a rather manageable form, and by truncation one obtains a 2-category~$\bigra$ that was shown to have a symmetric monoidal structure in \cite{BCR}.\footnote{In \cite{BCR} we distinguished between 2-categories with and without extra gradings to encode R- and flavour charges, so as to avoid additional technical details in the initial presentation. 
Here we consider such extra degrees from the start (though they mostly only appear in shifts~\eqref{eq:defPhi} as described below) and simply write~$\bigra$ for the 2-category denoted~$\bi^{\textrm{gr}}$ in \cite{BCR}.} 
In this section we review the definition of~$\bigra$. 

Objects of~$\bigra$ are ordered sets of variables $\xu=(x_1,\ldots,x_n)$ with $n\geq 0$, corresponding to Rozansky--Witten models with target spaces $T^*\C^n$. 
The case $n=0$ is the trivial theory. 
The variables $x_i$ have an interpretation in terms of complex scalar free fields and geometrically parametrise the base of $T^*\C^n$.
As such they carry R- and $\textrm{U}(1)$-flavour charges, which we take to be $1$ and $-1$ respectively. 
Hence the polynomial ring $\C[x_1,\ldots,x_n]$ in these variables is $(\Q\times\Q)$-graded, with~$x_i$ having bidegree $(1,-1)$. 
We think of different lists of the same length as just different names for one and the same object.  

A 1-morphism $(x_1,\ldots,x_n)\lra (y_1,\ldots,y_m)$ in $\bigra$ is a pair $D=(\alphau; W)$,
where $\alphau=(\alpha_1,\ldots,\alpha_k)$ is yet another (possibly empty) list of bigraded variables, and~$W$ is a homogeneous polynomial in $\xu,\yu,\alphau$ of bidegree $(2,0)$. 
(Even though the specification of the bidegrees of~$\alphau$ is part of the data of 1-morphisms, we omit it from the notation.)
Here again, the names of the variables $\alphau$ are not relevant. 
1-morphisms represent surface defects between Rozansky--Witten models, which in turn are specified by additional fields on the defect, the $\alpha_t$, and a \textsl{superpotential}~$W$ that couples the bulk variables $x_r,y_s$ on either side of the defect to the defect fields~$\alpha_t$.\footnote{The hypermultiplets of the 3-dimensional theory contain two complex scalars of opposite flavour charge and are thus described by a pair $(\xu,\alphau)$, where geometrically $\xu$ are base and $\alphau$ are fibre coordinates of $T^*\C^n$.}

The horizontal composition of 1-morphisms $(\alphau;W(\alphau,\xu,\yu))\colon \xu\lra\yu$ and $(\betau;W(\betau,\yu,\zu))\colon\yu\lra\zu$ is given by
\be
\big(\betau;W(\betau,\yu,\zu)\big)\circ\big(\alphau;W(\alphau,\xu,\yu)\big)
:=
\big(\alphau,\betau,\yu; W(\alphau,\xu,\yu)+V(\betau,\yu,\zu)\big)\,,
\ee
i.\,e.\ the potentials are added and intermediate bulk variables become new defect fields. 

Because of the truncation, a 2-morphism $(\alphau;W(\alphau,\xu,\yu))\lra (\betau;V(\betau,\xu,\yu))$ is an isomorphism class of line defects separating the respective surface defects. 
They are given by isomorphism classes of graded matrix factorisations of the difference $V(\betau,\xu,\yu)-W(\alphau,\xu,\yu)$ of the respective potentials over the polynomial ring $\C[\alphau,\betau,\xu,\yu]$.

A \textsl{matrix factorisation} (see e.\,g.~\cite{kr0401268, cm1208.1481} for more details) of a given polynomial $f$ in a polynomial ring~$R$ is a pair $(X,d_X)$, where $X=X_0 \oplus X_1$ is a free $\Z_2$-graded $R$-module and $d_X\colon X\lra X$ is an odd $R$-linear module map such that $d_X^2 = f\cdot \id_X$. 
In the case at hand the polynomial ring $R=\C[\alphau,\betau,\xu,\yu]$ is bigraded, and the polynomial $f=W$ has bidegree $(2,0)$. 
We call a matrix factorisation \textsl{graded} if the modules $X_i$ are bigraded modules such that the map~$d_X$ is homogeneous of bidegree $(1,0)$. 
Thus, in addition to the homological $\Z_2$-grading, the matrix factorisation carries a $(\Q\times\Q)$-grading, where the first factor corresponds to R-charge, and the second factor to flavour charge. 
There is a natural notion of degree-preserving map of matrix factorisations up to homotopy, in terms of which isomorphism classes are defined in a standard way, see e.\,g.\ \cite[Sect.\,2.1\,\&\,2.5]{BCR}.

The $(\Z_2\times\Q\times\Q)$-grading of a matrix factorisation $(X,d_X)$ can be shifted:  
denoting by $X_{n,r,s}$ the degree $(n,r,s)$-part of~$X$, we define the shifted module $X\{(m,t,u)\}$ by
$
(X\{(m,t,u)\})_{n,r,s}=X_{m+n,t+r,u+s}
$. 
The shifted matrix factorisation is then given by
$
(X,d_X)\{(m,t,u)\} := (X\{m,t,u\}, (-1)^m d_X)
$. 
Below we will repeatedly encounter shifts by the degrees
\be
\label{eq:defPhi}
\Phi_{\alphau}
: =
\Big(n,\, \sum_{i}(r_{\alpha_i}-1),\, \sum_{i} q_{\alpha_i}\Big) 
	\;\;\in\;\; \Z_2\times\Q\times\Q
\, , 
\ee
associated to lists of variables $\alphau=(\alpha_1,\ldots,\alpha_n)$ of degrees 
$\textrm{deg}(\alpha_i)=(r_{\alpha_i},q_{\alpha_i})$.
Hence in particular for objects $\xu \in \bigra$ we have $\Phi_\xu=(n,0,-n)$.

Horizontal as well as vertical composition of two appropriately composable 2-morphisms $(X,d_X)$ and $(Y,d_Y)$ in $\bigra$ are both given by  the tensor product
\be
\label{eq:XtensorY} 
(X,d_X)\circ(Y,d_Y) 
:= 
\big(X\otimes Y,\, d_{X\otimes Y}=d_X\otimes 1_Y+1_X\otimes d_Y\big)
\ee
of matrix factorisations, taken however over different rings. 
Here, ``$\otimes$'' refers to the tensor product over the respective intermediate\footnote{For horizontal composition we take the tensor product over the ring of intermediate bulk variables, while for vertical composition it is the ring of intermediate defect variables.} polynomial ring, see \cite[Eq.\,(2.11)]{BCR} for details. 
We often abbreviate~\eqref{eq:XtensorY} to $X\circ Y$, or to $X\otimes Y$ if we want to stress that it is computed as a tensor product. 

\medskip 

A class of matrix factorisations which will be very important in our construction is that of Koszul type: 
for $k\in\Z_{\geqslant 1}$ let $\pu=(p_1,\ldots,p_k)$ and $\qu=(q_1,\ldots,q_k)$ be lists of homogeneous polynomials in $\C[\xu]$ such that $f=\sum_{i=1}^k p_iq_i$ has bidegree $(2,0)$.
Then the \textsl{Koszul matrix factorisation} 
$
[\pu,\,\qu] := ( K(\pu,\qu), d_{K(\pu,\qu)} )
$ 
is given by 
\be 
\label{eq:Koszul2}
K(\pu,\qu) 
= 
\bigwedge\Big( \bigoplus_{i=1}^k \C[\xu] \cdot \theta_i \Big) \, , 
\quad 
d_{K(\pu,\qu)} 
= 
\sum_{i=1}^k \big( p_i\cdot \theta_i + q_i \cdot \theta_i^* \big) \, ,  
\ee 
where $\{\theta_i\}$ is a chosen $\C[\xu]$-basis of $\C[\xu]^{\oplus k}$. 
It is straightforward to check that $[\pu,\qu]$ is a matrix factorisation of $f=\sum_{i=1}^k p_i q_i$. 
The $\Z_2$-grading is given by the parity of the wedge degree in $K(\pu,\qu)$, and the  $(\Q\times\Q)$-grading is fixed by the grading on the wedge degree~0 component, which we take to be $(0,0)$. 
Note that 
$
K(\pu,\qu)\cong\bigotimes_i K(p_i,q_i)
$.

The following facts (which are well-known and/or established in \cite[App.\,A]{BCR}) about Koszul matrix factorisations will be repeatedly used for manipulations of 2-morphisms in this paper: 
\begin{enumerate}
	\item
	The lists $\pu$ and $\qu$ can be permuted: 
	if $\pu'$ is a permutation of the list~$\pu$ and~$\qu'$ is a permutation of $\qu$, then
	$
	[\pu',\qu']\cong [\pu,\qu]
	$.
	\item 
	Interchanging $\pu$ and $\qu$ leads to a shift (recall~\eqref{eq:defPhi}), 
	\be
	\label{eq:Koszulshift}
	\big[\qu,\pu\big]\cong \big[\pu,\qu\big]\{-\Phi_\pu\} \, . 
	\ee 
	\item
	For $z\in\C^\times$, we have
	$
	[z\pu,z^{-1}\qu]\cong[\pu,\qu]
	$. 
	\item
	For two matrix factorisations $[\pu,\qu]$ and $[\pu',\qu']$ of the same size, we have
	\be
	\label{eq:mfmanipulation}
	\big[\pu,\qu\big]\otimes\big[\pu',\qu'\big]
	\cong
	\big[\pu+\pu',\qu\big]\otimes\big[\pu',\qu'-\qu\big]\,.
	\ee
	\item
	Let $P=P(\au,\bu,\xu)$ be a matrix factorisation of $W\in\C[\au,\bu,\xu]$, where~$\au$ and~$\bu$ are lists of the same length. 
	Then for a list $\pu$ of polynomials in $\C[\au,\bu,\xu]$ such that $(\bu-\au)\cdot\pu+W$ does not depend on the~$b_i$, we have an isomorphism of matrix factorisations over $\C[\au,\xu]$ (i.\,e.\ $\bu$ are ``internal''{} variables)
	\be
	\label{eq:varelimination}
	\big[\bu-\au,\pu\big]\otimes P(\au,\bu,\xu)\cong P(\au,\au,\xu)\,,
	\ee
	where the right-hand side is obtained from the matrix factorisation $P$ by setting $\bu=\au$. 
	Thus internal variables can be eliminated using Koszul matrix factorisations.
\end{enumerate}

To describe more of the structure of the 2-category $\bigra$, the following notation turns out to be convenient: 
for a polynomial $f=f(\cdots,\xu,\cdots)$ in variables $\xu=(x_1,\ldots,x_n)$ (and possibly more, indicated by the centred ellipses), we set
\be
\Delta_i f(\cdots,\vect{\xu}{\yu},\cdots)
=
\tfrac{f(\cdots,y_1,\ldots,y_{i-1},x_i,x_{i+1},\ldots,x_n,\cdots) - f(\cdots,y_1,\ldots,y_{i-1},y_i,x_{i+1},\ldots,x_n,\cdots)}{x_i-y_i}
\ee
and denote the list of these polynomials as 
\be
\Deltau f(\cdots,\vect{\xu}{\yu},\cdots) 
=
\big(\Delta_i f(\cdots,\vect{\xu}{\yu},\cdots)\big)_{i=1}^n \, . 
\ee

For example, the unitors in~$\bigra$ can concisely be expressed in this notation: 
for $\xu = (x_1,\dots,x_n) \in\bigra$ and $\yu,\gammau$ two other lists of variables of the same length, the 1-morphism 
\be
1_\xu 		
:=
\big(\gammau;\, \gammau\cdot(\yu-\xu)\big)\colon \xu\lra\yu 
\, , \quad \textrm{with} \quad \gammau\cdot(\yu-\xu) := \sum_{i=1}^n\gamma_i(y_i-x_i) \, ,
\ee
is a weak unit, as for every 1-morphism $D=(\alphau;W(\alphau,\xu,\zu))\colon\xu\lra\zu$ we have 2-isomorphisms\footnote{%
	In these string diagrams, the blue lines represent a 1-morphism $D$, while the dashed grey lines represent the identity 1-morphism. 
	For easier translation into equations, we display the names of the respective variables/fields (in the bulk and on the 1-morphisms).}
\begin{eqnarray}
\lambda_D&=&
\tikzzbox{%

}
\;\;
=
\begin{array}{l}
\big[\zu-\wu,-\gammau+\Deltau W(\alphau,\xu,\vect{\zu}{\wu})\big]\\
\qquad\otimes\big[\betau-\alphau,\Deltau W(\vect{\betau}{\alphau},\xu,\zu)\big]\{-\Phi_{\wu}\}\, ,
\end{array}%
\label{eq:defrho}
\end{eqnarray}
One verifies that the respective inverses are given by
\begin{eqnarray}
\lambda^{-1}_D&=&
\tikzzbox{%
	
}
\;\;
=
\begin{array}{l}
\big[\wu-\zu,-\gammau+\Deltau W(\alphau,\xu,\vect{\zu}{\wu})\big]\\
\qquad\otimes\big[\betau-\alphau,\Deltau W(\vect{\betau}{\alphau},\xu,\wu)\big]  \, ,
\end{array}%
\label{eq:rhoinv}
\end{eqnarray}
Finally, the unit 2-morphism on $D=(\alphau;W(\alphau,\xu,\zu))\colon\xu\lra\zu$ is given by 
\be 
1_D = \big[\betau-\alphau,\Deltau W(\vect{\betau}{\alphau},\xu,\zu)\big]
\,.
\ee 
It follows from the variable elimination formula~\eqref{eq:varelimination} that this is indeed a strict unit. 
This concludes the presentation of the 2-category structure on~$\bigra$. 

\medskip 

As explained in \cite[Sect.\,2.2\,\&\,2.3]{BCR}, the 2-category~$\bigra$ carries a symmetric monoidal structure. 
The monoidal product~$\btimes$ corresponds to the stacking of 3-dimensional Rozansky--Witten models and is basically given by concatenation of lists of variables, addition of potentials, and the tensor product of matrix factorisations over~$\C$:
\begin{align}
	(x_1,\ldots,x_m)\btimes(x'_1,\ldots,x'_n) & := (x_1,\ldots,x_m,x'_1,\ldots,x'_n) \,,
	\nonumber
	\\
	\big(\alphau;W(\alphau,\xu,\yu)\big)\btimes\big(\alphau';W'(\alphau',\xu',\yu')\big) & := \big(\alphau\btimes\alphau'; W(\alphau,\xu,\yu)+W'(\alphau',\xu',\yu')\big) \,,
	\nonumber
	\\
	(X,d_X) \btimes (Y,d_Y) & := \big( X\otimes_\C Y, d_X \otimes 1_Y + 1_X \otimes d_Y \big) \, . 
\end{align}
For details on structure morphisms (which will not feature prominently in the present paper) we refer to \cite{BCR}. 
Here we only mention that~$\bigra$ also has a symmetric braiding~$b$, whose 1-morphism components 
\be 
\label{eq:Braiding1morphismComponents}
b_{\xu,\yu} := \big(\cu,\du; \du \cdot (\yu'-\yu) + \cu \cdot (\xu'-\xu)\big) \colon \xu\btimes\yu 
\lra 
\yu \btimes \xu 
\equiv \yu' \btimes \xu'\,.
\ee 
are basically the unit 1-morphisms $1_{\xu\btimes\yu}$ with the order of the lists $\xu,\yu$ swapped in the codomain. 
Under this identification the 2-morphism components of~$b$ are trivial as well.

\subsection{Dualisability}
\label{subsec:adjunctions}

As discussed in Sections~\ref{subsec:CHwithoutDfects} and~\ref{subsec:ExtendedDefectTQFTs}, dualisability is a key feature of extended TQFTs. 
On the one hand, objects describing bulk theories have to be fully dualisable, and 1-morphisms describing line defects have to have adjoints.
On the other hand, the respective duality data are central to the evaluation of correlation functions. 
In this section we show that all 1-morphisms in~$\bigra$ have adjoints, and that~$\bigra$ indeed has a pivotal structure. 
We also review that all objects are fully dualisable, as already established in \cite{BCR}.
Along the way, we explicitly construct the respective duality data.

\subsubsection{Dualisability of 1-morphisms}
\label{sectdualisability}

Let $D=(\alphau;W(\alphau,\xu,\zu))\colon\xu\lra \zu$ be a 1-morphism in $\bigra$. 
The left and right adjoints of~$D$ can be chosen to be 
\be\label{eq:defdual}
D^\dagger 
:=
\big(\alphau;-W(\alphau,\xu,\zu)\big)
=:
{}^\dagger\! D\colon\zu\lra \xu\,,
\ee
where besides the sign of the potential, only the order of source and target bulk variables in $W$ change in comparison with~$D$. 
It follows that 
$
(D^\dagger)^\dagger=D
$,
and that the unit 1-morphism
$
1_\xu=(\alphau;\alphau\cdot(\xu^\prime-\xu))\colon\xu\lra\xu^\prime
$
is self-dual, $1_\xu^\dagger=1_\xu={}^\dagger\! 1_\xu$. 

The adjunction 2-morphisms witnessing~\eqref{eq:defdual} as a left adjoint are 
\begin{eqnarray}
\ev_D&=&
\tikzzbox{%

}
\;\;
=
\begin{array}{l}\big[\yu-\xu,\gammau+\Deltau W(\betau,\vect{\yu}{\xu},\zu)\big]\\
\qquad \otimes\big[\betau-\alphau,\Deltau W(\vect{\betau}{\alphau},\xu,\zu\big)]\{-\Phi_{\zu}\} \, ,
\end{array}%
\label{eq:defcoev}
\end{eqnarray}
For the right adjunction we use $D^{\dagger\dagger}=D$ and set
\be
\label{eq:deftildeadj}
\tev_D:=\ev_{D^\dagger}\,,\quad
\tcoev_D:=\coev_{D^\dagger}\,.
\ee

We explicitly check the Zorro move (recall~\eqref{eq:ZorroMoves})
\be
\label{eq:zorro_string}
\tikzzbox{%
	%
}
\, . 
\ee
The other moves are verified analogously. 
Cutting the string diagram on the left-hand side along the red lines as indicated corresponds to a decomposition of the 2-morphism into elementary building blocks. 
Treating the associator as an identity, this decomposition translates~\eqref{eq:zorro_string} into 
\begin{equation}
	\lambda_D\cdot\left(\tev_D\circ1_D\right)\cdot\left(1_D\circ \tcoev_D\right)\cdot\rho^{-1}_D
	= 
	1_{D} \, . 
\end{equation}
The left-hand side can be evaluated explicitly by inserting the definitions \eqref{eq:defev}--\eqref{eq:deftildeadj} as well as the formulas \eqref{eq:deflambda}--\eqref{eq:rhoinv} for unitors and their inverses. 
To wit, with the variable assignments in \eqref{eq:zorro_string}, we obtain the following expression for the matrix factorisation representing the left-hand side: 
\begin{equation}
	\begin{array}{ll}
		&\big[\zu-\wu,-\psiu+\Deltau W(\zetau,\xu,\vect{\zu}{\wu})\big]\otimes
		\big[\etau-\zetau,\Deltau W(\vect{\etau}{\zetau},\xu,\zu)\big]\{-\Phi_\wu\}\\[0.05cm]
		\otimes&\big[\wu-\zu,-\psiu+\Deltau W(\gammau,\yu,\vect{\zu}{\wu})\big]\otimes
		\big[\deltau-\gammau,\Deltau W(\vect{\gammau}{\deltau},\yu,\wu)\big]\{-\Phi_\yu\}\\[0.05cm]
		\otimes&\big[\zetau-\epsilonu,\Deltau W(\vect{\zetau}{\epsilonu},\xu,\wu)\big]\\[0.05cm]
		\otimes & \big[\gammau-\betau,\Deltau W(\vect{\gammau}{\betau},\yu,\zu)\big]\\[0.05cm]
		\otimes & \big[\yu-\xu,-\chiu+\Deltau W(\deltau,\vect{\yu}{\xu},\wu)\big]\otimes
		\big[\epsilonu-\deltau,\Deltau W(\vect{\deltau}{\epsilonu},\xu,\wu)\big]\\[0.05cm]
		\otimes & \big[\yu-\xu,\chiu+\Deltau W(\alphau,\vect{\yu}{\xu},\zu)\big]\otimes \big[\betau-\alphau,\Deltau W(\vect{\betau}{\alphau},\yu,\zu)\big]	\, . 
	\end{array}
\end{equation}
By means of the variable elimination identity~\eqref{eq:varelimination} and the shift relation~\eqref{eq:Koszulshift} this matrix factorisation can be simplified, and one finds that it is isomorphic to 
\begin{equation}
	\big[\etau-\alphau,\Deltau W(\vect{\etau}{\alphau},\xu,\zu)\big] \, . 
\end{equation}
But this precisely represents the unit~$1_D$ on the right-hand side of~\eqref{eq:zorro_string}. 

Recall that in general the right and left adjoint of a 2-morphism are defined as the two sides of the first identity in~\eqref{eq:pivotality}, respectively. 
With the above adjunction data it is now straightforward to show that in~$\bigra$ all 2-morphisms are self-dual in the following sense: 
if $\xi\colon (\alphau;W(\alphau,\xu,\yu))\lra (\alphau';W'(\alphau',\xu,\yu))$ is represented by a matrix factorisation~$P$ of $W'(\alphau',\xu,\yu)-W(\alphau,\xu,\yu)$, then both~$\xi^\dagger$ and~${}^\dagger\!\xi$ are represented by~$P$ as well. 
In particular, we have 
\be 
\label{eq:2MorphismsSelfDual}
\xi^\dagger = {}^\dagger\!\xi
\ee 
for all 2-morphisms~$\xi$.

\subsubsection{Pivotality}
\label{subsubsec:Pivotality}

The adjunction data of Section~\ref{sectdualisability} admit a trivial pivotal structure on~$\bigra$. 
To show this, we have to verify the two relations in~\eqref{eq:pivotality} for composable 1-morphisms and arbitrary 2-morphisms. 
The first relation is simply~\eqref{eq:2MorphismsSelfDual}, so we are left to consider the second relation in~\eqref{eq:pivotality}. 
For $X=(\alphau;W(\alphau,\yu,\xu))\colon\yu\lra\xu$ and 
$Y=(\betau;V(\betau,\zu,\yu))\colon\zu\lra\yu$ it reads as follows: 
\begin{equation}\label{eq:pivotality2}
	\tikzzbox{

	}
\end{equation}
Inserting the definitions (\ref{eq:defev}) and (\ref{eq:defcoev})
of evaluation and coevaluation and the formulas (\ref{eq:deflambda})--(\ref{eq:rhoinv}) for unitors and their inverses, the left-hand side evaluates to 
\begin{equation}
	\begin{array}{cl}
		&\big[\xu-\xu_1,-\gammau_1+\Deltau W(\alphau_1,\yu,\vect{\xu_1}{\xu|})\big]\otimes
		\big[\alphau_2-\alphau_1,-\Deltau W(\vect{\alphau_2}{\alphau_1},\yu,\xu)\big]\\[0.05cm]
		\otimes & \big[\zu-\zu_1,\gammau_3+\Deltau V(\betau_1,\vect{\zu}{\zu_1},\yu)\big]\otimes
		\big[\betau_1-\betau_2,\Deltau V(\vect{\betau_1}{\betau_2},\zu_1,\yu)\big]\{-\Phi_\yu\}\\[0.05cm]
		\otimes & \big[\yu-\yu_1,\gammau_2+\Deltau W(\alphau_2,\vect{\yu}{\yu_1},\xu_1)\big]\otimes
		\big[\alphau_2-\alphau_3,\Deltau W(\vect{\alphau_2}{\alphau_3},\yu_1,\xu_1)\big]\{-\Phi_\xu\}\\[0.05cm]
		\otimes & \big[\yu-\yu_1,-\gammau_2+\Deltau V(\betau_3,\zu_1,\vect{\yu}{\yu_1})\big]\otimes
		\big[\betau_2-\betau_3,\Deltau V(\vect{\betau_2}{\betau_3},\zu_1,\yu)\big]\{-\Phi_\yu\}\\[0.05cm]
		\otimes & \big[\xu_1-\xu,-\gammau_1+\Deltau (W+V)((\alpha_3,\yu_1,\betau_3),\zu_1,\vect{\xu_1}{\xu})\big]\\[0.05cm]
		\otimes&
		\big[(\alphau_3,\yu_1,\betau_3)-(\alphau_4,\yu_4,\betau_4),
		\Deltau (W+V)(\vect{(\alphau_3,\yu_1,\betau_3)}{(\alphau_4,\yu_4,\betau_4)},\zu_1,\xu)\big]\\[0.05cm]
		\otimes & \big[\zu_1-\zu,\gammau_3+\Deltau (W+V)((\alphau_4,\yu_4,\betau_4),\vect{\zu}{\zu_1},\xu)\big]\\[0.05cm]
		\otimes&
		\big[(\alphau_5,\yu_5,\betau_5)-(\alphau_4,\yu_4,\betau_4),-\Deltau(W+V)(\vect{(\alphau_5,\yu_5,\betau_5)}{(\alphau_4,\yu_4,\betau_4)},\zu_1,\xu)\big]\{-\Phi_\zu\} \, .
	\end{array}
\end{equation}
This expression can be simplified using the formulas (\ref{eq:varelimination}) and~(\ref{eq:Koszulshift}) to obtain
\begin{equation}
	\big[(\alphau_5,\yu_5,\betau_5)-(\alphau_1,\yu,\betau_1),-\Deltau (W+V)
	(\vect{(\alphau_5,\yu_5,\betau_5)}{(\alphau_4,\yu_4,\betau_4)},\zu_1,\xu)\big]\{-\Phi_\yu\}
	= 1_{(X\circ Y)^\dagger}\{-\Phi_\yu\} 
\end{equation}
where we used the fact that the adjunction as defined in (\ref{eq:defdual}) satisfies
\be
(X\circ Y)^\dagger = Y^\dagger \circ X^\dagger\,,\quad
{}^\dagger(X\circ Y) = {}^\dagger Y \circ {}^\dagger X
\ee

In a similar fashion one shows that the right-hand side of~(\ref{eq:pivotality2}) evaluates to the same result. 
Hence we have 
\begin{equation}
\label{eq:pivshift1}
	\tikzzbox{

	} 
\, . 
\end{equation}
for the inverse. These expressions define the isomorphisms
\be
\phi_{X,Y}\colon Y^\dagger\circ X^\dagger = {}^\dagger Y\circ {}^\dagger X
	\lra (X\circ Y)^\dagger = {}^\dagger(X\circ Y)
\ee
and their inverses compatible with our choice of pivotal structure. 
Note that they differ from the identity on $Y^\dagger\circ X^\dagger = {}^\dagger Y\circ {}^\dagger X$, but only by a shift $\{-\Phi_\yu\}$ by minus the degree associated to the bulk variables $\yu$ between the two parallel defects, cf.~(\ref{eq:defPhi}): 
\be\label{eq:phi}
\phi_{X,Y}=1_{(X\circ Y)^\dagger}\{-\Phi_\yu\}\,,\quad
\phi_{X,Y}^{-1}=1_{(X\circ Y)^\dagger}\{\Phi_\yu\}\,.
\ee
These shifts are crucial for the behaviour under adjunction of 2-morphisms between composite 1-morphisms.
Indeed, consider a 2-morphism
\be
\varphi=
\tikzzbox{
	\begin{tikzpicture}[DefaultSettings, baseline = -0.05cm]
	\coordinate (f) at (0,0);
	\coordinate (d) at (-0.5,-2);
	\coordinate (e) at (0.5,-2);
	\coordinate (d') at (-0.5,2);
	\coordinate (e') at (0.5,2);
	\coordinate (y) at (0,-1);
	\coordinate (y') at (0,1);
	\filldraw[color = black, fill = BackSurfaceColor, thin] ($(d)+(-1,0)$) -- ($(d)+(-1,4)$) -- ($(e')+(1,0)$) -- ($(e')+(1,-4)$) -- ($(d)+(-1,0)$);
	\draw[LineDefect] (d') -- (d);
	\draw[LineDefect] (e') -- (e);
	\node[black, fill=BackSurfaceColor, inner sep=2pt, draw, opacity = 1] (f) {{\small$\quad \varphi \quad\vphantom{I}$}}; 
	\fill[IdentityColor] (y) circle (0pt) node {\scriptsize $\yu_1$};
	\fill[IdentityColor] (y') circle (0pt) node {\scriptsize $\yu_2$};
	\fill[DefectColor] (d) circle (0pt) node[below] {\small $X_1$};
	\fill[DefectColor] (d') circle (0pt) node[above] {\small $X_2$};
	\fill[DefectColor] (e) circle (0pt) node[below] {\small $Y_1$};
	\fill[DefectColor] (e') circle (0pt) node[above] {\small $Y_2$};
	\end{tikzpicture}
}
\colon X_1\circ Y_1 \lra X_2\circ Y_2
\ee
between composite 1-morphisms $X_1\circ Y_1$ and $X_2\circ Y_2$. 
If~$\varphi$ is represented by a matrix factorisation $Q$, then its adjoint $\varphi^\dagger\colon(X_2\circ Y_2)^\dagger\lra(X_1\circ Y_1)^\dagger$ is represented by the same matrix factorisation $Q$, cf.~the discussion at the end of Section~\ref{sectdualisability}. However, if one wants to view it as a 2-morphism $Y_2^\dagger\circ X_2^\dagger\lra Y_1^\dagger\circ X_1^\dagger$ it has to be composed with the (inverse of the) appropriate isomorphisms $\phi_{X_i,Y_i}$:
\be
\phi_{X_1,Y_1}^{-1}\cdot\varphi^\dagger\cdot\phi_{X_2,Y_2}=
\tikzzbox{
	\begin{tikzpicture}[DefaultSettings,baseline = 0cm,xscale=-1]
	\coordinate (d) at (-0.5,0);
	\coordinate (e) at (0.5,0);
	\coordinate (dd) at ($(d)+(3,0)$);
	\coordinate (ed) at ($(e)+(1,0)$);
	\coordinate (d') at ($(d)+(-1,0)$);
	\coordinate (e') at ($(e)+(-3,0)$);
	\coordinate (h) at (0,2);
	\coordinate (y) at (2,0);
	\coordinate (y') at (-2,0);
	\filldraw[color = black, fill = BackSurfaceColor, thin] ($(e')-(h)+(-1,0)$) -- ($(e')-(h)+(-1,4)$) -- ($(dd)+(h)+(1,0)$) -- ($(dd)+(h)+(1,-4)$) -- ($(e')-(h)+(-1,0)$);
	\draw[costring, LineDefect] (e) .. controls +(0,-1) and +(0,-1) .. (ed);
	\draw[costring, LineDefect] (d) .. controls +(0,-2) and +(0,-2) .. (dd);
	\draw[costring, LineDefect] (e') .. controls +(0,2) and +(0,2) .. (e);
	\draw[costring, LineDefect] (d') .. controls +(0,1) and +(0,1) .. (d);
	\draw[LineDefect] (ed) -- ($(ed)+(h)$);
	\draw[LineDefect] (dd) -- ($(dd)+(h)$);
	\draw[LineDefect] (e') -- ($(e')-(h)$);
	\draw[LineDefect] (d') -- ($(d')-(h)$);
	\node[black, fill=BackSurfaceColor, inner sep=2pt, draw, opacity = 1] (0,0) {{\small$\quad \varphi \quad\vphantom{I}$}};
	\fill[IdentityColor] (y) circle (0pt) node {\scriptsize $\yu_1$};
	\fill[IdentityColor] (y') circle (0pt) node {\scriptsize $\yu_2$};
	\fill[DefectColor] ($(e')-(h)$) circle (0pt) node[below] {\small $X_2^\dagger$};
	\fill[DefectColor] ($(d')-(h)$) circle (0pt) node[below] {\small $Y_2^\dagger$};
	\fill[DefectColor] ($(ed)+(h)$) circle (0pt) node[above] {\small $X_1^\dagger$};
	\fill[DefectColor] ($(dd)+(h)$) circle (0pt) node[above] {\small $Y_1^\dagger$};
	\end{tikzpicture}
}
\,.
\ee
This 2-morphism is represented by the shifted matrix factorisation $Q\{\Phi_{\yu_1}-\Phi_{\yu_2}\}$. 
Thus adjunction acts on 2-morphisms between composite 1-morphisms by shifting the respective matrix factorisation by the difference of degrees of the intermediate incoming and outgoing bulk variables.

The above applies for instance to the adjunction 2-morphisms 
\be
\ev_D\colon{}^\dagger\!D\circ D\lra 1_\xu\,,\quad
\coev_D\colon 1_\xu\lra D\circ{}^\dagger\!D
\ee
defined in (\ref{eq:defev}) and~(\ref{eq:defcoev}). 
For example one finds $\phi^{-1}_{{}^\dagger\!D, D}\cdot \ev_D^\dagger=\coev_{D^\dagger}=\tcoev_D$. 
It is straightforward to verify that indeed the matrix factorisations associated to $\ev_D$ and $\coev_{D^\dagger}$ differ by the shift $\Phi_\zu$. Similarly $\coev_D^\dagger\cdot\phi_{D,{}^\dagger\!D}=\ev_{D^\dagger}=\tev_D$.

In the following we will suppress the isomorphisms $\phi_{X,Y}$, since it should always be clear where they have to be inserted. For instance, we will just write
\be
\label{eq:daggeradj}
\ev_D^\dagger=\tcoev_D={}^\dagger\!\ev_D\,,\quad
\coev_D^\dagger=\tev_D={}^\dagger\!\coev_D \,. 
\ee
Note however that one needs to insert the respective shifts when translating these relations into matrix factorisations.

\subsubsection{Full dualisability of objects}
\label{subsubsec:fulldualisability}

Next we explain how all objects $\xu$ in $\bigra$ are fully dualisable. This was already shown in \cite{BCR}, but here we make slightly different choices for the adjunction 1- and 2-morphisms that are more convenient in our applications.\footnote{For one thing we use $\ev_\xu=\widetilde{\ev}_\xu$, which is isomorphic but not equal to the choice made in \cite{BCR}. (Since adjunctions are unique up to unique isomorphism, cf.\ the discussion after~\eqref{eq:ZorroMoves}, the self-duality $\xu^\#=\xu$ implies that all choices of $\ev_\xu$ and $\tev_\xu$ are isomorphic.)
Also our choices for adjunction 2-morphisms for $\ev_\xu$ and $\coev_\xu$ differ from the ones in \cite{BCR} by a grade shift.}

All objects in~$\bigra$ are self-dual, 
\be
\xu^\#=\xu={}^\#\xu\,, 
\ee
and we choose the respective adjunction 1-morphisms to be 
\be\label{eq:adj1mor}
\begin{array}{ll}
	\ev_\xu&=
	\tikzzbox{%
		\begin{tikzpicture}[thick,scale=1.5,color=black, baseline=0.2cm]
			\coordinate (p1) at (0,0);
			\coordinate (p2) at (0,0.4);
			\coordinate (g1) at (0,-0.05);
			\coordinate (g2) at (0,0.3);
			\draw[costring, IdentitySlice] (p2) .. controls +(-0.4,0) and +(-0.4,0) ..  (p1); 

			\fill[red!80!black] (0.257,0) circle (0pt) node[left] {{\tiny$\xu$}};
			\fill[red!80!black] (0.8,0.4) circle (0pt) node[left] {{\tiny$\xu^\dual = x^\prime$}};
		\end{tikzpicture}
	}
	:=\big(\alphau;\,\alphau(\xu-\xu^\prime)\big)\colon \xu^\prime\btimes\xu\lra\varnothing\,,
	\\
	\coev_\xu&=
	\tikzzbox{%
		\begin{tikzpicture}[thick,scale=1.5,color=black, baseline=0.2cm]
			\coordinate (p1) at (0,0);
			\coordinate (p2) at (0,0.4);
			\coordinate (g1) at (0,-0.05);
			\coordinate (g2) at (0,0.3);
			\draw[costring, IdentitySlice] (p2) .. controls +(0.4,0) and +(0.4,0) ..  (p1); 

			\fill[red!80!black] (0.08,0) circle (0pt) node[left] {{\tiny$\xu^\dual=x^\prime$}};
			\fill[red!80!black] (0.03,0.4) circle (0pt) node[left] {{\tiny$\xu\vphantom{\xu^\dual}$}};
		\end{tikzpicture}
	}
	:=\big(\alphau;\,\alphau(\xu-\xu^\prime)\big)\colon \varnothing\lra \xu\btimes\xu^\prime\,,
	\\
	\tev_\xu&=
	\tikzzbox{%
		\begin{tikzpicture}[thick,scale=1.5,color=black, baseline=0.2cm]
			\coordinate (p1) at (0,0);
			\coordinate (p2) at (0,0.4);
			\coordinate (g1) at (0,-0.05);
			\coordinate (g2) at (0,0.3);
			\draw[costring, IdentitySlice] (p1) .. controls +(-0.4,0) and +(-0.4,0) ..  (p2); 

			\fill[red!80!black] (0.257,0.4) circle (0pt) node[left] {{\tiny$\xu$}};
			\fill[red!80!black] (0.8,0) circle (0pt) node[left] {{\tiny$\xu^\dual = x^\prime$}};
		\end{tikzpicture}
	}
	:=\big(\alphau;\,\alphau(\xu^\prime-\xu)\big)\colon \xu\btimes\xu^\prime\lra\varnothing \,,
	\\
\end{array}
\ee
Note that in particular 
\be\label{eq:ev=tev}
\ev_\xu=\tev_\xu\,,\quad
\coev_\xu=\tcoev_\xu\,.
\ee
Together with the cusp 2-isomorphisms \begin{align}\label{eq:defcl}
	c_{\textrm{l}}^\xu&=
	\tikzzbox{%
		%
	}     
	\,
	=
	\big[\alphau+\betau,\yu-\xu\big]\otimes\big[\xu^\prime-\yu,\alphau-\gammau\big]\{-\Phi_\xu\} 
	\label{eq:defcr}
\end{align}
these are indeed duality data for~$\xu$ (as reviewed in Section~\ref{subsec:GraphicalCalculus}). 
Moreover, as established in Section~\ref{sectdualisability}, all 1-morphisms in $\bigra$ have adjoints, so this holds in particular for adjunction 1-morphisms. 
Thus all objects are fully dualisable.

\medskip 

By the general formula (\ref{eq:defdual}) for adjoints of 1-morphisms in~$\bigra$, evaluation and coevaluation 1-morphisms are adjoint to each other: 
\begin{align}
	\ev_\xu^\dagger& =\tcoev_\xu={}^\dagger\!\ev_\xu \,, &
	\coev_\xu^\dagger&=\tev_\xu={}^\dagger\!\coev_\xu\,, \\
	\tev_\xu^\dagger&=\coev_\xu={}^\dagger\tev_\xu\,, &
	\tcoev_\xu^\dagger&=\ev_\xu={}^\dagger\tcoev_\xu \,.
\end{align}
This is in line with the general relation~\eqref{eq:AdjointsViaSerre} and the fact that the Serre automorphism~$S_\xu$ in~$\bigra$ is trivialisable, cf.\ \cite[Prop.\,2.5]{BCR}. 
The adjunction 2-morphisms can be read off from~(\ref{eq:defev}) and~(\ref{eq:defcoev}). 
This yields 
\be\label{eq:adjevcoev}

\end{array}
\ee
By inserting the appropriate isomorphisms~(\ref{eq:phi}), it becomes apparent that these 2-morphisms indeed satisfy~(\ref{eq:daggeradj}), which specialises to the relations
$
\ev_{\coev_\xu}^\dagger=\tcoev_{\coev_\xu}
$
and 
$
\ev_{\ev_\xu}^\dagger=\tcoev_{\ev_\xu}
$.

Note also that our choices for left and right cusp isomorphisms are related by adjunction:
\be
\label{eq:CuspInverse}
\big(c_{\textrm{l}}^\xu\big)^\dagger
=
\big(c_{\textrm{r}}^\xu\big)^{-1} \,.
\ee
Hence, because of the self-duality of~$\xu$, relation~\eqref{eq:crTildeFromcl} holds on the nose for our choices. 
This in turn implies that the cusp-counit identity~\eqref{eq:CuspCounit} holds. 

\medskip 

For later use, we spell out how $\#$-duality acts on 1- and 2-morphisms in $\bigra$.
Given a 1-morphism $D=(\alphau;W(\alphau,\xu,\zu))\colon\xu\lra\zu$, the definition~\eqref{eq:LeftDual1Morphism} yields
\be
\label{eq:ComplicatedDdual}
D^\#
=
\big(\alphau,\betau,\xu^\prime,\zu^\prime;\;\betau\cdot(\zu^\prime-\zu)+W(\alphau,\xu^\prime,\zu^\prime)+\gammau\cdot(\xu^\prime-\xu)\big)\,.
\ee
The left dual ${}^\#\!D$ is isomorphic but not equal to~$D^\#$. 
The formula for~${}^\#\!D$ can be obtained from the expression for $D^\#$ by changing the signs of $\betau$ and $\gammau$. 
This sign flip is implemented by the 2-isomorphisms
\be
\label{eq:omegaD}
\omega_D 
:= \begin{array}{l}
	\phantom{\otimes}\big[\alphau'-\alphau,\Deltau W(\vect{\alphau'}{\alphau},\xu',\zu')\big]
	\\[0.05cm]
	\otimes \big[\betau'+\betau,\zu-\zu'\big]\otimes\big[\gammau'+\gammau,\xu-\xu'\big]
\end{array}\!\!\!
\colon D^\# \stackrel{\cong}{\lra} {}^\#\!D
\ee
which can be checked to be equal to the canonical maps that are induced by the relation between left and right duals via braidings, cf.\ \eqref{eq:LeftRightDualX}. 
The maps~$\omega_D$ in particular mediate between left and right duals of 2-morphisms~$\varphi$ by conjugation, or equivalently 
\be
\omega_{D'}\cdot\varphi^\#={}^\#\!\varphi\cdot\omega_D\,.
\ee

\medskip 

We now turn to the interaction between $\#$-duals and $\dagger$-adjoints. 
First we note that it follows from the definitions that they are compatible for 1-morphisms~$D$ in the sense that 
\be
(D^\#)^\dagger={}^\#(D^\dagger)\,.
\ee
Moreover, by using~\eqref{eq:CuspInverse} we find that in~$\bigra$ the 2-isomorphisms~\eqref{eq:Omega} and~\eqref{eq:OmegaTilde} that ``bend lines around the corner'' become
\begin{align}
	\Omega_D
	=
	\tikzzbox{%

	}
	&\colon
	(1_\zu \btimes D)\circ\coev_\xu \lra ({}^\#\!D \btimes 1_\zu) \circ\coev_\zu
	\, . 
	\label{eq:MFbending2}
\end{align}
Due to the relation~\eqref{eq:CuspInverse} between left and right cusp isomorphisms also 
$\Omega_D$ and $\Omega_D^\prime$ are related by adjunction:
\be\label{eq:omegavsomegap}
\Omega^\dagger_D=(\Omega^\prime_{D^\dagger})^{-1}\,.
\ee
This in particular implies a compatibility between adjunction data for $D$ and its $\#$-duals. 
For instance 
\vspace{-0.8cm}
\be
\tikzzbox{%
	%
}%
\ee
This equation relates $\ev_D$ and $\ev_{{}^\#\!D}$ via $\Omega_{D^\dagger}$ and $\Omega'_D$ and  allows to pull 1-morphisms across the upper hemisphere.
An analogous relation holds for the coevaluations which allows to pull 1-morphisms across the lower hemisphere.

\medskip 

It turns out that for our applications to Rozansky--Witten models it is convenient to work with an equivalent description of duals of 1-morphisms, which we denote by ``$*$'' instead of ``$\#$'': for  $D=(\alphau;W(\alphau,\xu,\zu))\colon\xu\lra\zu$ we define
\be
D^*={}^*\!D
:=
\big(\alphau;W(\alphau,\xu,\zu)\big)\colon \zu\lra\xu\,.
\ee
That means $D^*$ is represented by the same data $(\alphau;W(\alphau,\xu,\zu))$ as~$D$, but viewed as a 1-morphism in the opposite direction. 
The expression for~$D^*$ is isomorphic to both $D^\#$ and ${}^\#\!D$, and it is simpler than the ones for~$D^\#$ and~${}^\#\!D$, compare e.\,g.\ \eqref{eq:ComplicatedDdual}. 
Moreover, it is convenient that left and right $*$-duals are equal, while in general $D^\# \neq {}^\#\!D$. 

To switch between these two models of duality, we fix 2-isomorphisms $D^\#\lra D^*$ given by 

\be
\label{eq:defclD}
c_{\textrm{l}}^D=
\tikzzbox{%

}
\;
:=
\begin{array}{cl}
	&\big[\alphau^\prime-\alphau,\Deltau W(\vect{\alphau^\prime}{\alphau},\xu^\prime,\zu^\prime)\big]\\[0.05cm]
	\otimes&\big[\xu-\xu^\prime,\gammau+\Deltau W(\alphau^\prime,\vect{\xu}{\xu^\prime},\zu)\big]\\[0.05cm]
	\otimes &\big[\zu-\zu^\prime,\betau+\Deltau W(\alphau^\prime,\xu^\prime,\vect{\zu}{\zu^\prime})\big]\{-\Phi_\xu-\Phi_\zu\} 
\end{array}%
\ee
and for analogous 2-isomorphisms $c_{\textrm{r}}^D\colon{}^\#\!D\lra D^*$ we choose
\be\label{eq:defcrD}
c_{\textrm{r}}^D:=
\Big( \big(c_{\textrm{l}}^{D^\dagger}\big)^\dagger\Big)^{-1}\,, \quad
\big(c_{\textrm{r}}^D\big)^{-1}
=
\big(c_{\textrm{l}}^{D^\dagger}\big)^\dagger\,,
\ee
where we used that $(D^*)^\dagger=(D^\dagger)^*$.
Note that for the special case $D=1_\xu$, the isomorphism $c_{\textrm{l}}^D$ reduces to the ordinary cusp isomorphism $c_{\textrm{l}}^\xu$ composed with a unitor. 

Importantly, due to the choice~(\ref{eq:defcrD}), the isomorphisms $c_{\textrm{l}}^D$ and $c_{\textrm{r}}^D$ intertwine the adjunction data for the $\#$- and $*$-duals of 1-morphisms. For instance, it implies the identity
\be\label{eq:hashstarcompatibility}
\tikzzbox{%

}
\ee
which relates $\ev_{D^*}$ and $\ev_{D^\#}$. Analogous relations hold for the evaluation of ${}^\#\!D$ and also for the respective coevaluations.

Beyond the simplification of duals for 1-morphisms, also $\omega_D\colon D^\#\lra {}^\#\!D$ from~\eqref{eq:omegaD} is rendered trivial when working with the $*$-version of duality. 
More precisely, it is straightforward to check that 
\be
\big(c_{\textrm{l}}^D\big)^{-1}\cdot\omega_D\cdot c_{\textrm{r}}^D=1_{D^*}\,.
\ee
Another advantage of $*$-duality is that its action on 2-morphisms is also quite simple: 
let $D^\prime
=(\alphau^\prime;W(\alphau^\prime,\xu,\zu))\colon \xu\lra\zu$ be another 1-morphism, and let
$\varphi\colon D\lra D^\prime$ be a 2-morphism represented by a matrix factorisation~$P$ of $W-W^\prime$. To obtain the $*$-dual of $\varphi$, one has to compose the $\#$-dual $\varphi^\#$ as defined in general in~(\ref{eq:varphiDual}) with 
the isomorphisms $c_{\textrm{l}}^{D^\prime}$ and $(c_{\textrm{l}}^D)^{-1}$:
\be
\varphi^*
:= 
c_{\textrm{l}}^{D^\prime}\cdot \varphi^\#\cdot \big(c_{\textrm{l}}^D\big)^{-1} \colon D^*\lra (D^\prime)^*\,.
\ee 
It is straightforward to verify that~$P$ also represents~$\varphi^*$. 
Thus, as for 1-morphisms, also for 2-morphisms, the $*$-dual is represented by the same data as the original 2-morphism, but with source and target swapped.

Also the ``bending isomorphisms'', which in the $*$-formulation we denote by $\widetilde{\Omega}_D$ and $\widetilde{\Omega}^\prime_D$, can be obtained from the ones in the $\#$-dual formulation (recall~\eqref{eq:MFbending1} and~\eqref{eq:MFbending2}) by composition with the respective isomorphisms $c_{\textrm{l}}^D$ and $(c_{\textrm{l}}^D)^{-1}$. 
For instance, by combining the inverse 
\be
\big(c_{\textrm{r}}^\xu\big)^{-1}
=
\tikzzbox{%

}
\,
=
\big[\gammau+\muu,\xu_1-\xu_3\big]\otimes\big[\xu_2-\xu_3,\deltau-\muu\big]
\ee
of~\eqref{eq:defcr} with the expression~\eqref{eq:defclD}, we obtain 
\be
\widetilde{\Omega}_D=
\tikzzbox{%
	
}
\;\;
= \;
\begin{array}{cl}
	&\big[\xu_1-\xu,\muu+\Deltau W(\alphau,\vect{\xu_1}{\xu},\zu)\big]\\[0.05cm]
	\otimes &\big[\alphau_1-\alphau,\Deltau W(\vect{\alphau_1}{\alphau},\xu_1,\zu)\big]\\[0.05cm]
	\otimes &\big[\deltau+\muu,\xu-\xu^\prime\big]\\[0.05cm]
	\otimes &\big[\xu_1-\xu^\prime,\betau-\muu\big]\\[0.05cm]
	\otimes &\big[\alphau^\prime-\alphau_1,\Deltau W(\vect{\alphau^\prime}{\alphau_1},\xu_1,\zu)\big]\\[0.05cm]
	\otimes &\big[\xu^\prime-\xu_1,\betau+\Deltau W(\alphau^\prime,\vect{\xu^\prime}{\xu_1},\zu^\prime)\big]\\[0.05cm]
	\otimes & \big[\zu^\prime-\zu,\gammau+\Deltau W(\alphau^\prime,\xu_1,\vect{\zu^\prime}{\zu})\big]\{-\Phi_\xu-\Phi_\zu\} \,.
\end{array}
\ee
Using the relations from Section~\ref{subsec:targetcategory} to simplify this, and proceeding analogously for the inverse, we get 
\begin{align}\label{eq:defomegap}
\widetilde{\Omega}_D = 
\tikzzbox{%
	
}  
&
\;\;=\;
\begin{array}{cl}
&\big[\alphau^\prime-\alphau,\Deltau W(\vect{\alphau^\prime}{\alphau},\xu,\zu)\big]\{-\Phi_\zu\}\\[0.05cm]
\otimes & \big[\xu^\prime-\xu,-\deltau+\Deltau W(\alphau^\prime,\vect{\xu^\prime}{\xu},\zu^\prime)\big]\\[0.05cm]
\otimes &\big[\zu^\prime-\zu,\gammau+\Deltau W(\alphau^\prime,{\xu},\vect{\zu^\prime}{\zu})\big]\,,
\end{array}%
\end{align}
Similarly, the relation (\ref{eq:omegavsomegap}) between $\Omega_D^\prime$ and $\Omega_{D^\dagger}$ as well as the relation (\ref{eq:defcrD}) between $c_{\textrm{r}}^D$ and $c_{\textrm{l}}^{D^\dagger}$ lead to 
\begin{align}
	\label{eq:defomega}
	\widetilde{\Omega}_D^\prime =
	\tikzzbox{%
		
	}  
	&=\;
	\begin{array}{cl}
		&\big[\alphau^\prime-\alphau,\Deltau W(\vect{\alphau^\prime}{\alphau},\xu,\zu)\big]\{-\Phi_\xu\}\\[0.05cm]
		\otimes & \big[\xu^\prime-\xu,-\deltau+\Deltau W(\alphau^\prime,\vect{\xu^\prime}{\xu},\zu)\big]\\[0.05cm]
		\otimes &\big[\zu^\prime-\zu,\gammau+\Deltau W(\alphau^\prime,{\xu^\prime},\vect{\zu^\prime}{\zu})\big]\,,
	\end{array}%
\end{align}
These maps inherit properties from the original bending isomorphisms. 
Since the latter as well as~$c_{\textrm{l}}^D$ and~$c_{\textrm{r}}^D$ are compatible with adjunction in the sense of~(\ref{eq:omegavsomegap}) and~(\ref{eq:defcrD}), respectively, it follows that $\widetilde{\Omega}_D$ and $\widetilde{\Omega}_D'$ intertwine the adjunction 2-morphisms of $D$ and $D^*$. In particular, 
the following identity is satisfied, cf.~(\ref{eq:hashstarcompatibility}):
\be\label{eq:hemdef}
\tikzzbox{%
	%
}
\,.
\ee
An analogous relation holds for coevaluations. 
These identities allow us to ``pull 1-morphisms across the poles of hemispheres''.

\subsection{The extended TQFT}
\label{subsec:ExtendedRWTQFT}

In this section we identify the data internal to the Rozansky--Witten 2-category~$\bigra$ to apply the cobordism hypothesis with defects (reviewed in Section~\ref{sec:ExtendedDefects}) to construct extended TQFTs. 
This is illustrated by explicit computations of quantum dimensions as well as state spaces for tori with defect networks. 
Further example computations for specific classes of defects (symmetry defects and boundary conditions) are carried out in more detail in Section~\ref{sec:Examples}.

\subsubsection{The bulk TQFT}
\label{subsubsec:extendedtqft}

Since all objects $\xu\in\bigra$ are fully dualisable (as established in \cite{BCR} and above in Section~\ref{subsubsec:fulldualisability}), according to the general theory outlined in 
Section~\ref{subsec:CHwithoutDfects} each~$\xu$ gives rise to a 2-dimensional extended (bulk) TQFT. 
In this section we provide an explicit choice of coherent full duality data for every $\xu\in\bigra$ (and hence explicit constructions of framed TQFTs valued in~$\bigra$) as well as all trivialisations of the Serre automorphisms~$S_\xu$ (and hence oriented extended TQFTs).

\medskip 

We claim that the full duality data 
\be 
\label{eq:CoherentFullDualityDataInC}
\Big( 
\xu,\xu,\tev_\xu,\tcoev_\xu,1_u, 1_u, c^\xu_{\textrm{l}},c^\xu_{\textrm{r}}, \ev_{\tev_\xu}, \coev_{\tev_\xu}, \ev_{\tcoev_\xu}, \coev_{\tcoev_\xu}, \lambda_{1_\xu}, \lambda_{1_\xu} 
\Big) 
\ee 
provided in Section~\ref{subsec:adjunctions} are already coherent. 
To prove this, we have to check that the swallowtail identity~\eqref{eq:Swallowtail} and the cusp-counit identity~\eqref{eq:CuspCounit} hold. 

Indeed, we already mentioned that the relation~\eqref{eq:CuspInverse} between our choices for the left and right cusp isomorphisms ensures the cusp-counit identity. 
Nevertheless, we give explicit calculations here. 
Using the evaluation maps (\ref{eq:adjevcoev}) and the left unitor (\ref{eq:deflambda}) we find that in our context its left-hand side is 
\begin{eqnarray}
&&
\tikzzbox{%

}\nonumber\\
&=&
\begin{array}{rl}
&\big[\xu_1'-\xu_1,\muu_1+\alphau_1'\big]\otimes\big[\xu_1-\zu,-\muu+\alphau_1'\big]\\[0.05cm]
\otimes&\big[\alphau_1'-\alphau_1,\xu_1-\xu\big]\otimes \big[\alphau_2'-\alphau_3,\xu_1'-\yu\big]\{-2\Phi_\xu\}\\[0.05cm]
\otimes&\big[\xu_1'-\xu_1,-\muu_1+\alphau_2\big]\otimes\big[\alphau_3-\alphau_2,\xu_1'-\yu\big]\vspace{0.2cm}\\
\end{array}\nonumber\\
&\cong&
\begin{array}{rl}
&\big[\xu_1-\zu,-\muu+\alphau_1'\big]\otimes\big[\alphau_1+\alphau_2',\xu_1'-\xu_1\big]\\[0.05cm]
\otimes &\big[\alphau_1'-\alphau_1,\xu_1'-\xu\big]\otimes\big[\alphau_2'-\alphau_2,\xu_1-\yu\big] \,.
\end{array}
\label{eq:CC1var}
\end{eqnarray}
On the other hand, with the cusp isomorphisms (\ref{eq:defcl}) and (\ref{eq:defcr}) and the left unitor (\ref{eq:deflambda}), the right-hand side of~\eqref{eq:CuspCounit} evaluates to
\begin{eqnarray}
&&
\tikzzbox{%
	
}\nonumber\\
&=&
\begin{array}{rl}
&\big[\muu_1+\alphau_2,\yu-\xu_1\big]\otimes\big[\xu-\xu_1,-\muu_1+\alphau_1\big]\{-\Phi_\xu\}\\[0.05cm]
\otimes&\big[\muu_2+\alphau_2',\xu_1'-\yu\big]\otimes\big[\xu_1'-\zu,-\muu_2+\alphau_1'\big]\\[0.05cm]
\otimes &\big[\zu-\yu,-\muu_2+\muu_1\big]\otimes\big[-\muu+\muu_1,\xu-\zu\big]\{-\Phi_\xu\}
\end{array}\nonumber\\
&\cong&
\begin{array}{rl}
&\big[-\alphau_1'-\alphau_2,\xu_1-\yu\big]\otimes\big[\xu-\xu_1,-\alphau_1'+\alphau_1\big]\\[0.05cm]
\otimes &\big[\alphau_1'+\alphau_2',\xu_1'-\yu\big]\otimes\big[-\muu+\alphau_1',\xu-\zu\big]\,.
\end{array}
\label{eq:CC2var}
\end{eqnarray}
Here we used variable elimination (\ref{eq:varelimination}) together with the shifting property of Koszul matrix factorisations (\ref{eq:Koszulshift}). 
It is now straightforward to check that upon further shifts and application of the identity (\ref{eq:mfmanipulation}), the two expressions (\ref{eq:CC1var}) and (\ref{eq:CC2var}) represent the same 2-morphism in~$\bigra$.

The swallowtail relation can also be checked very explicitly. 
In our situation it reads
\be\label{eq:swallowtail}
\tikzzbox{%
	%
}  
\, . 
\ee
Plugging in the formulas (\ref{eq:defcl}) and (\ref{eq:defcr}) for the cusp isomorphisms on the left-hand side yields
\begin{align}
&\big[\muu_2+\gammau,\yu'-\yu\big]\otimes\big[\zu-\yu,-\muu_2+\alphau\big] \otimes\big[\muu_1+\betau,\xu-\yu'\big]\otimes\big[\yu-\yu',-\muu_1+\gammau\big]\{-\Phi_\xu\}\nonumber
\\
& \cong
\big[\muu_1+\muu_2,\yu'-\yu\big]\otimes\big[\zu-\yu,-\muu_2+\alphau\big]\otimes\big[\muu_1+\betau,\xu-\yu'\big]\,.
\label{eq:st1}
\end{align}
On the other hand, inserting the formulas (\ref{eq:defrho}) and (\ref{eq:rhoinv}) for the right unitors on the right-hand side, we obtain 
\begin{align}
&\big[\xu-\yu,\muu_1+\alphau\big]\otimes\big[\gammau-\alphau,\xu-\zu\big]\otimes\big[\yu'-\zu,\muu_2-\gammau\big]\otimes\big[\betau-\gammau,\xu-\yu'\big]\{-\Phi_\xu\}\nonumber
\\
&\cong
\big[\yu'-\yu,\muu_1+\muu_2\big]\otimes\big[\zu-\yu,-\muu_2+\alphau\big]\otimes\big[\muu_1+\betau,\xu-\yu'\big]\{-\Phi_\xu\} \,.
\label{eq:st2}
\end{align}
But now we observe that the matrix factorisations (\ref{eq:st1}) and (\ref{eq:st2}) are isomorphic via the shift relation (\ref{eq:Koszulshift}). 
Hence the swallowtail identity~(\ref{eq:swallowtail}) holds. 

\medskip 

Since the coherence conditions are satisfied, each object $\xu=(x_1,\ldots,x_n)$ corresponding to a Rozansky--Witten model with target space $T^*\C^n$ defines a framed extended TQFT ${\mathcal Z}^{\textrm{fr}}_n\colon \Bordfr \lra \bigra$. 
As reviewed in Section~\ref{subsec:CHwithoutDfects}, the coherent full duality data~\eqref{eq:CoherentFullDualityDataInC} completely determine~$\zz^{\textrm{fr}}_n$ via the cobordism hypothesis, in particular: 
\begin{equation}
\begin{array}{rcl}
\mathcal{Z}_n^{\textrm{fr}}\colon \quad \Bordfr & \lra & \bigra
\\
+&\lmt& \xu\\
-&\lmt&\xu^\#=\xu\\
{\ev}_\pm&\lmt&\ev_\xu\\
{\coev}_\pm&\lmt&\coev_\xu\\
\ev_{\ev_\pm}&\lmt&\ev_{\ev_\xu}\\
\coev_{\ev_\pm}&\lmt&\coev_{\ev_\xu}\\
\widetilde{\ev}_{\ev_\pm}&\lmt&\widetilde{\ev}_{\ev_\xu}\\
\widetilde{\coev}_{\ev_\pm}&\lmt&\widetilde{\coev}_{\ev_\xu}\\
c_{\textrm{l}}&\lmt& c_{\textrm{l}}^\xu\\
c_{\textrm{r}}&\lmt& c_{\textrm{r}}^\xu \, . 
\end{array}
\end{equation}
Our explicit formulas for the morphisms on the right-hand side allow for a straightforward evaluation of all correlators in this extended TQFT.

\medskip 

For the construction of \textsl{oriented} extended TQFTs we have to choose in addition a 
trivialisation $S_\xu\lra 1_\xu$ of the Serre automorphism
\begin{align}
S_\xu&=(1_\xu\btimes\tev_\xu)\circ(b_{\xu,\xu}\btimes 1_{\xu^\#})\circ(1_\xu\btimes\tev_\xu^\dagger)\nonumber\\
&=
\tikzzbox{%
	%
}  
\nonumber
\\
&={\footnotesize \big(\alphau_1,\alphau_2,\betau_1,\betau_2,\xu_1,\xu_2,\xu_3;\alphau_1(\xu_1-\xu)+\alphau_2(\yu-\xu_3)+\betau_1(\xu_2-\xu_1)+\betau_2(\xu_3-\xu_2)\big)} \, . 
\end{align}
One such trivialisation is given by the 2-isomorphism
\begin{align}
\lambda_\xu & =
\tikzzbox{%
	
}
\nonumber
\\
&
=
\begin{array}{rl}&\big[-\alphau+\alphau_1,\xu-\xu_1\big]\otimes\big[-\alphau+\betau_1,\xu_1-\xu_2\big]\\[0.05cm]
\otimes&\big[-\alphau+\betau_2,\xu_2-\xu_3\big]\otimes\big[-\alphau+\alphau_2,\xu_3-\yu\big]\,.
\end{array}
\end{align}
As shown in \cite{BCR}, the trivialisation is unique up to grade shift. 
Hence any trivialisation of the Serre automorphism is of the form $\lambda_\xu\{\Phi\}$ for some shift $\Phi$ of $(\Z_2\times\Q\times\Q)$-gradings, cf.\ \eqref{eq:defPhi}. 
For any such choice, the 2-dimensional oriented cobordism hypothesis reviewed in Section~\ref{subsec:CHwithoutDfects} produces a TQFT 
\be 
\label{eq:BulkTQFTn}
\zz_n \colon \Bordor \lra \bigra \, . 
\ee 
For example, what~$\zz_n$ associates to a closed surface of genus~$g$ can be read off of~\eqref{eq:ZStateSpacesGeneral}, where we use our choice of $\lambda_u := \lambda_\xu\{\Phi\}$ in the formula~\eqref{eq:Lambda} for~$\Lambda_u$.

\subsubsection{The extended defect TQFT}
\label{subsec:samplecalculations}

Since the 2-category~$\bigra$ is pivotal (as demonstrated in Section~\ref{subsubsec:Pivotality}), the bulk TQFTs~\eqref{eq:BulkTQFTn} in fact ``glue'' to a single extended defect TQFT 
\be 
\label{eq:ExtendedRWDefectTQFT}
\zz \colon \Bord_{2,1,0}^{\textrm{def}}(\mathds{D}^{\bigra}) \lra \bigra \, ,
\ee 
where the defect label sets ${D}^\bigra_2, {D}^\bigra_1$ and~${D}^\bigra_0$ (see Section~\ref{subsec:ExtendedDefectBordisms} for the notation) are precisely the objects, 1- and 2-morphisms of~$\bigra$, respectively. 
From the 3-dimensional perspective, these describe bulk theories, surface and line defects, which according to Section~\ref{subsec:targetcategory} are basically given by lists of variables, polynomials and matrix factorisations. 

Since the 2-category~$\bigra$ is under very explicit control, the cobordism hypothesis with defects as formulated in Section~\ref{subsec:ExtendedDefectTQFTs} can be applied very explicitly to surfaces with defects of arbitrary complexity. 
In the remainder of this section we do so for a number of illustrative examples. 
Further applications are discussed in Section~\ref{sec:Examples} below.

\subsubsection*{Quantum dimensions}

As a warm-up consideration, we ask what the TQFT~$\zz$ assigns to a single loop 1-stratum in a local patch of a bordism in $\Bord_{2,1,0}^{\textrm{def}}(\mathds{D}^{\bigra})$. 
If the loop is labelled by a 1-morphism $D=(\gammau;W(\gammau,\xu,\zu))\colon\xu\lra \zu$ in $\bigra$, then by construction~$\zz$ assigns to it the (say, left) quantum dimension
\be\label{eq:qdim}
\dim_{\textrm{l}}(D) = 
\tikzzbox{%
	\begin{tikzpicture}[DefaultSettings,baseline = 0cm]
	\coordinate (b1) at (-2,-2);
	\coordinate (b2) at (+2,-2);
	\coordinate (m1) at (-2,0);
	\coordinate (m2) at (+2,0);
	\coordinate (t1) at (-2,2);
	\coordinate (t2) at (+2,2);
	\coordinate (bm) at ($0.5*(b1)+0.5*(b2)$);
	\coordinate (tm) at ($0.5*(t1)+0.5*(t2)$);
	\coordinate (mm) at ($0.5*(m1)+0.5*(m2)$);
	\coordinate (mq1) at ($0.5*(m1)+0.5*(mm)$);
	\coordinate (mq2) at ($0.5*(mm)+0.5*(m2)$);
	\coordinate (bq1) at ($0.5*(b1)+0.5*(bm)$);
	\coordinate (tq1) at ($0.5*(t1)+0.5*(tm)$);
	\coordinate (bq2) at ($0.5*(bm)+0.5*(b2)$);
	\coordinate (tq2) at ($0.5*(tm)+0.5*(t2)$);
	\coordinate (m) at (0,1);
	\filldraw [color=black, fill= BackSurfaceColor, thin] (b1) -- (b2) -- (t2) -- (t1) -- (b1); 
	%
	\draw[IdentityLine] (m) -- (tm);
	\draw[IdentityLine] ($-1*(m)$) -- (bm);
	\draw[costring, LineDefect] (mq1) .. controls +(0,1.3) and +(0,1.3) .. (mq2);
	\draw[costring, LineDefect] (mq2) .. controls +(0,-1.3) and +(0,-1.3) .. (mq1);
	%
	\fill[DefectColor] (mq1) circle (0pt) node[left] {\tiny $\gammau$};
	\fill[DefectColor] (mq2) circle (0pt) node[right] {\tiny $\betau$};
	\fill[DefectColor] ($(mm)+(1,1)$) circle (0pt) node {\small $D$};
	\fill[IdentityColor] (mm) circle (0pt) node {\scriptsize $\zu$};
	\fill[IdentityColor] ($(b1)+(0.5,0.5)$) circle (0pt) node {\scriptsize $\yu$};
	\fill[IdentityColor] ($(b2)+(-0.5,0.5)$) circle (0pt) node {\scriptsize $\xu$};
	\fill[IdentityLineColor] (bm) circle (0pt) node[above,xshift=0.2cm] {\tiny $\alphau$};
	\fill[IdentityLineColor] (tm) circle (0pt) node[below,xshift=0.2cm] {\tiny $\alphau^\prime$};
	\end{tikzpicture}
}
=\ev_D\cdot \coev_{D^\dagger}\,.
\ee
Inserting the explicit formulas (\ref{eq:defev}) and (\ref{eq:defcoev}) for the evaluation and coevaluation one obtains
\begin{align}
\nonumber
\dim_{\textrm{l}}(D) &=
\big[\yu-\xu,\alphau^\prime+\Deltau W(\gammau,\vect{\yu}{\xu},\zu)\big]
\otimes\big[\gammau-\betau,\Deltau W(\vect{\gammau}{\betau},\xu,\zu)\big]\{-\Phi_\zu\}\\
&\qquad \otimes\big[\yu-\xu,-\alphau-\Deltau W(\gammau,\vect{\yu}{\xu},\zu)\big]
\otimes\big[\gammau-\betau,-\Deltau W(\vect{\gammau}{\betau},\xu,\zu)\big]\nonumber\\
&\cong
\big[-\alphau^\prime+\alphau,\xu-\yu\big]\otimes
\big[0,\partial_\gammau W(\gammau,\xu,\zu)\big] 
\nonumber 
\\
&\qquad 
\otimes 
\big[0,\alphau^\prime+\Deltau W(\gammau,\vect{\yu}{\xu},\zu)\big]\{\Phi_\xu-\Phi_\zu\}\,.
\label{eq:qdim-general}
\end{align}
Note that the first tensor factor is indeed the matrix factorisation representing the 2-morphism $1_{1_\xu}$, and that $\zu$ and $\gammau$ are still internal variables. 

As a concrete example we calculate the quantum dimension of a defect incorporating the transformations of bulk variables by an invertible linear transformation $A\in\text{GL}(n;\C)$, i.\,e.\ $z_i=\sum_j A_{ij}x_j$. 
(See Section~\ref{subsec:gldefects} for more details about such symmetry defects.)
The potential corresponding to such a defect is given by
\be\label{eq:symmetrydefect}
W(\gammau,\xu,\zu)=\sum_i \gamma_i(z_i-A_{ij}x_j)\,.
\ee
Hence $\partial_\gammau W(\gammau,\xu,\zu)$ is given by $(z_i-\sum_j A_{ij} x_j)_{i}$, and
$\Deltau W(\gammau,\vect{\yu}{\xu},\zu)$ is given by $(-\sum_i\gamma_iA_{ij})_j$. Thus, the second tensor factor in (\ref{eq:qdim-general}) becomes $\bigotimes_i[0,z_i-A_{ij}x_j]$ and (together with the shift $\{-\Phi_\zu\}$) can be used to eliminate the variables $z_i$. Similarly, the third tensor factor in 
(\ref{eq:qdim-general}) becomes $\bigotimes_j[0,\sum_i\gamma_iA_{ij}-\alpha^\prime_j]$. Since the matrix $A$ is assumed to be invertible, this tensor factor 
(together with the shift $\{\Phi_\xu\}$) can be used to eliminate the variables $\gamma_i$. Hence, 
the quantum dimension for these defects is the identity, 
\be
\dim_{\textrm{l}}\big( \gammau; \, \gammau\cdot (\zu - A\xu)\big) 
=1_{1_\xu}\,.
\ee

Another important example is that of a quantum dimension of a boundary condition, i.\,e.\ a defect with the trivial theory on one side. 
In this case the (left) quantum dimension corresponds to a disc correlator. 
So assume that $\xu=\varnothing$, describing the trivial bulk theory. 
Then there are no variables $x_i,y_i,\alpha_i,\alpha^\prime_i$, the potential is a polynomial $W=W(\gammau,\zu)$ in the variables~$\gamma_i$ and~$z_i$ only, and the formula (\ref{eq:qdim-general}) reduces to
\be
\label{eq:DiscSpace}
\big[0,\partial_\gammau W(\gammau,\xu,\zu)\big]\{-\Phi_\zu\}\,.
\ee
So for instance for the boundary condition with potential $W(\gammau,\zu)=\gammau\cdot\zu$, this becomes $[0,\zu]\{-\Phi_\zu\}$. 
Eliminating the variables $z_i$, one ends up with 
\be 
\label{eq:SpecialDiscSpace}
\zz \Bigg( 
\tikzzbox
{
 
} 
\Bigg) 
\cong \C[\gammau]\{-2\Phi_\zu\} \,. 
\ee 

For later use we will also state the more general formula for the left trace (see \eqref{eq:traces})
of a 2-endomorphism $\phi\colon D\lra D$, represented by a matrix factorisation $P(\gammau,\gammau^\prime,\xu,\zu)$ of $W(\gammau^\prime,\xu,\zu)-W(\gammau,\xu,\zu)$. Composing $\phi$ with the respective adjunction 2-morphisms and subsequent simplification yields the formula
\be \label{eq:left_trace}
\tikzzbox{%
	
}
=\begin{array}{ll}
&\phi(\beta, \beta, \xu, \zu) \otimes 
\big[ \alphau^\prime - \alphau, \yu - \xu\big] 
\\[0.05cm]
\otimes& 
\big[ 0, \alphau^\prime + \Deltau W(\betau,\vect{\yu}{\xu},\zu)\big] \{\Phi_\xu - \Phi_\zu \} \,.
\end{array}
\ee
The variables~$\betau$ are the only internal variables left here.

\subsubsection*{Defects on a torus}

The evaluation of the TQFT~$\zz$ on the torus with a defect labelled by a 1-morphism $D=(\deltau;W(\deltau,\xu,\yu))\colon \xu\lra \yu$, wrapping the poloidal cycle, is also a special case of the formula for the quantum dimension~(\ref{eq:qdim}). 
Indeed, it can be expressed as the quantum dimension 
\be
\tikzzbox{%
	%
	}
	\nonumber
	\\
	&=\big(\alphau,\betau,\deltau,\xu_1,\xu_2,\yu; \;\alphau(\xu_2-\yu)+\betau(\yu-\xu_1)+W(\deltau,\xu_1,\xu_2)\big) \,.
\end{align}
Applying the general formula (\ref{eq:qdim}) and then simplifying the result yields
\be\label{eq:toruswithdefect}
\tikzzbox{%
	%
}
	=
	\big[\underline{0},\partial_{\deltau} W(\deltau,\xu,\xu)\big] \otimes \big[\underline{0},\partial_\xu W(\deltau,\xu,\xu)\big]
\ee
where $\xu$ and $\deltau$ are the only remaining variables. 

Note that if~$D$ is the invisible defect, $D=1_\xu$, then we have $W(\deltau,\xu_1,\xu_2)=\deltau\cdot(\xu_1-\xu_2)$, and hence $W(\deltau,\xu,\xu)=0$. 
In this case~\eqref{eq:toruswithdefect} reproduces the partition function of the torus without defects\footnote{The notation $[\underline{0},\underline{0}]$ is not really well-defined. One has to remember the associated gradings from formula~(\ref{eq:toruswithdefect}).} (cf.\ \cite[Prop.\,3.5]{BCR})
\be
\label{eq:0000}
\big[\underline{0},\underline{0}\big] \otimes \big[\underline{0},\underline{0}\big]
	\cong \big(
	\left(\C\oplus\C\{1,0,-1\}\right)
	\otimes_\C
	\left(\C\oplus\C\{1,0,1\}\right)
	\big)^{\otimes n}\C[\xu,\deltau]\,.
\ee
The shifted components $\C\{1,0,\pm 1\}$ correspond to fermionic generators, one for each of the variables $x_i$ and $\delta_i$. 

If~$D$ is a symmetry defect given by a potential as in~(\ref{eq:symmetrydefect}), one arrives at a result similar to~\eqref{eq:0000}, where now however one has $\dim_\C(\text{ker}(1-A))$-many variables~$x_i$ and~$\delta_i$ instead of $n$. This will be explained in more detail in Section~\ref{subsubsec:statespaces} below. 
We interpret this as obtaining as many boson-fermion pairs as the dimension of the eigenspace  of~$A$ for eigenvalue~1.

\subsubsection*{Intersecting defects on a torus}

Let $D_1$ and $D_2$ be two endomorphisms of $\xu\in\bigra$. 
By an intersection of such defects we mean a 2-morphism $\varphi\colon D_1\circ D_2\lra D_2\circ D_1$. Thus, if $D_1=(\alphau;W_1(\alphau,\xu,\yu)) \colon \xu\lra\yu$ and $D_2=(\betau;W_2(\betau,\xu,\yu))\colon \xu\lra\yu$, the map~$\varphi$ is represented by a matrix factorisation
\be\label{eq:intersectionmf}
\tikzzbox{%

}
=P(\alphau_1,\alphau_2,\betau_1,\betau_2,\xu_1,\xu_2,\xu_3,\xu_4)
\ee
of $W_1(\alphau_2,\xu_3,\xu_4)+W_2(\betau_2,\xu_2,\xu_3)-W_2(\betau_1,\xu_4,\xu_1)-W_1(\alphau_1,\xu_2,\xu_1)$. 
For any such intersection, one can now straightforwardly compute what the TQFT~$\zz$ associates to a torus with intersecting defects such as 
\be
\label{eq:DefectTorusDecomposed}
\tikzzbox{%
	%
}
\, . 
\ee
This can be done by decomposing it as
\be
\tikzzbox{%
	%
}\,.
\ee
Writing $X := \ev_\xu\circ(1_\xu^*\btimes D_1)\circ\coev_\xu$, here the upper and lower pieces just represent $\ev_X$ and $\coev_{X^\dagger}$, respectively. 
These adjunction maps can be obtained from the general formulas (\ref{eq:defev}) and (\ref{eq:defcoev}). 
The piece in the middle of~\eqref{eq:DefectTorusDecomposed} can be built out of unitors, $\ev_{D_2}$, $\coev_{D_2^\dagger}$, $\widetilde{\Omega}^\prime_{D_2^\dagger}$, $\widetilde{\Omega}^{-1}_{D_2^\dagger}$ and the intersection 2-morphism~$\varphi$. 
Putting everything together, one obtains a lengthy expression which simplifies to
\be\label{eq:torusintersection}
\tikzzbox{%
	\begin{tikzpicture}[DefaultSettings, scale = 0.75, baseline = 0cm]
		\coordinate (l1) at (0,0);
		\coordinate (l2) at (2,0);
		\coordinate (r2) at (6,0);
		\coordinate (r1) at (8,0);
		\coordinate (ld1) at (1,0);
		\coordinate (ld2) at (7,0);
		
		\fill[BackSurface] (l1) .. controls +(0,3) and +(0,3) .. (r1) .. controls +(0,-3) and +(0,-3) .. (l1) ;
		%
		\draw[blue!80!white, very thick] (r2) .. controls +(0,-0.5) and +(0,-0.5) .. (r1);
		\fill[Surface] (l1) .. controls +(0,3) and +(0,3) .. (r1) .. controls +(0,-3) and +(0,-3) .. (l1) ;
		\draw[black, thick] (l1) .. controls +(0,3) and +(0,3) .. (r1) .. controls +(0,-3) and +(0,-3) .. (l1) ;
		\fill [white] (l2) .. controls +(1,1) and +(-1,1) .. (r2) .. controls +(-1,-1) and +(1,-1) .. (l2) ;
		\draw[black, thick] (l2) .. controls +(1,1) and +(-1,1) .. (r2) .. controls +(-1,-1) and +(1,-1) .. (l2) ;
		\draw[black, thick] (l2) -- ($(l2) + (-0.25, 0.25)$) ;
		\draw[black, thick] (r2) -- ($(r2) + (0.25, 0.25)$) ;
		%
		\draw[costring, LineDefect] (ld1) .. controls +(0,2) and +(0,2) .. (ld2);
		\draw[LineDefect] (ld2) .. controls +(0,-2) and +(0,-2) .. (ld1);
		\draw[oostring, blue!80!white, very thick] (r1) .. controls +(0,0.5) and +(0,0.5) .. (r2);
		\fill[blue!80!white] (7.15,1.0) circle (0pt) node {\small $D_2$};
		\fill[DefectColor] (4.6,1.85) circle (0pt) node {\small $D_1$};
		\fill[black] (6.91,0.39) circle (4pt) ;
	\end{tikzpicture}
}
	\cong 
	P(\alphau,\alphau,\betau,\betau,\xu,\xu,\xu,\xu)\,.
\ee
This is the matrix factorisation of $0$ in the variables $\alphau,\betau$ and $\xu$, which is obtained by setting  $\alphau_1=\alphau_2=\alphau$, $\betau_1=\betau_2=\betau$, $\xu_1=\xu_2=\xu_3=\xu_4=\xu$ in the matrix factorisation (\ref{eq:intersectionmf}) describing the intersection of $D_1$ and $D_2$.

Note that if one chooses $D_2=1_\xu$ and~$\varphi$ to be the 2-morphism obtained by composing the respective unitors, (\ref{eq:torusintersection}) reproduces the expression (\ref{eq:toruswithdefect}) for a torus with a single defect.

\medskip 

If $D_2$ is a more general symmetry defect with potential (\ref{eq:symmetrydefect}), and if $D_1$ is covariant in the sense that $D_1\circ D_2\cong D_2\circ D_1$, such an isomorphism gives a natural junction of~$D_1$ and~$D_2$. 
If the symmetry also acts linearly on the defect variables $\alphau$ on $D_1$, then (\ref{eq:DefectTorusDecomposed}) evaluates to 
\be\label{eq:torusWithDefect_invResult}
\big[\underline{0},\partial_{\alphau^\text{inv}} W^\text{inv}(\alphau^\text{inv},\xu^\text{inv},\xu^\text{inv})\big]
\otimes \big[\underline{0},\partial_{\xu^\text{inv}} W^\text{inv}(\alphau^\text{inv},\xu^\text{inv},\xu^\text{inv})\big]\,.
\ee
Here $\xu^\text{inv}$ and $\alphau^\text{inv}$ are the variables fixed under the symmetry, i.\,e.~there are $\text{dim}(\text{ker}(1-A))$-many variables~$\xu^\text{inv}$ and $\text{dim}(\text{ker}(1-T_A))$-many variables $\alphau^\text{inv}$, where $T_A$ is the chosen representation of~$A$ on~$\alphau$, and $W^\text{inv}$ is obtained by setting all the non-invariant variables in $W$ to zero. 
We will prove this in Section~\ref{subsubsec:statespaces} below.

\section{Examples}
\label{sec:Examples}

In this section we apply our general results from Section~\ref{sec:DefectsRW} to two special classes of defects: symmetry defects (leading to a description of background gauge fields in the formalism we adopted) and boundary conditions (leading to a categorifications of simple intersection pairings and of a Hirzebruch--Riemann--Roch-type theorem).

\subsection{Symmetry defects}
\label{subsec:gldefects}

Symmetry defects incorporate symmetries of the respective bulk theories. We introduce natural trivalent junctions of such symmetry defects which can be used to assemble special symmetry defect networks, whose insertion corresponds to the introduction of non-trivial flat background gauge fields.
We then compute state spaces 
associated to surfaces decorated with such networks, see for instance \eqref{eq:sigmagdefspace} for the result of closed genus-$g$ surfaces.
We also show the independence of the state spaces under certain local changes of the network which correspond to Pachner moves between triangulations. 

Finally we compute the category of line operators in the twisted sectors of the theory. These are line operators which exhibit non-trivial holonomy on cycles around them. 
From the perspective of extended TQFT taken in this paper, these categories arise from circles with marked points that are labelled by group elements.

\subsubsection{Symmetries} 
\label{subsubsec:symmetries}

Rozansky--Witten models with target spaces $T^*\C^n$ exhibit an $\text{Sp}_{2n}(\C)$-symmetry, acting linearly on the target manifold.
Recall that $\text{Sp}_{2n}(\C)$ is generated by the three subgroups
\begin{align}
D&:=\left\{\begin{pmatrix}
A&0\\
0&(A^\dagger)^{-1}
\end{pmatrix}
\,\bigg|\, A\in\text{GL}_n(\C)\right\} ,
\\
N&:=\left\{\begin{pmatrix}
\1_n&B\\
0&\1_n
\end{pmatrix}
\,\bigg|\, B^\dagger=B\right\} ,
\\
L&:=\left\{\1_{2n},\,\Omega:=\begin{pmatrix}
0&\1_n\\
-\1_n&0
\end{pmatrix}
\right\} .
\end{align}
Given objects $\xu,\yu \in \Cc$ of length~$n$, 
\be
I_A :=\Big( \alphau ; \alphau \cdot (\xu-A\yu) \Big)
	\colon \yu \lra \xu \,\;\text{for } A\in\text{GL}_n(\C)
\ee
are the 1-isomorphisms implementing the symmetry generators in the subgroup $D\subset\text{Sp}_{2n}(\C)$. 
Here we use $(A \xu)_i = \sum_j A_{ij} x_j$ to denote the left action of the matrix $A$ on a list of variables~$\xu$. 
Similarly, $(\alphau A)_i = \sum_j \alpha_j A_{ji}$ denotes the right action on~$\alphau$. 

On the other hand, elements in $N\subset\text{Sp}_{2n}(\C)$ give rise to 1-isomorphisms
\be
N_B := \big(\alphau;\,\alphau\cdot(\yu-\xu)+\alphau\cdot B\alphau\big) \colon \xu\lra\yu
\ee
for Hermitian $(n\times n)$-matrices~$B$, while the 1-isomorphism corresponding to~$\Omega$ is the ``Legendre transformation'' 1-morphism, defined in \cite[Sect.\,2.3]{KR0909.3643}:
\be
J:=(\varnothing;\xu\cdot\yu)\colon \xu\lra\yu \, . 
\ee

It is straightforward to verify that the composition of the respective 1-isomorphisms indeed satisfies the required relations:
$I_{A_1}\circ I_{A_2}\cong I_{A_1A_2}$, $N_{B_1}\circ N_{B_2}=N_{B_1+B_2}$ and 
$J\circ J\cong D_{-\1_n}$. 
By composing 1-isomorphisms $I_A$, $N_B$ and $J$, we therefore obtain 1-isomorphisms $D(g)$ for any $g\in\text{Sp}_{2n}(\C)$, which in particular satisfy $D(g_1)\circ D(g_2)\cong D(g_1g_2)$. 

We note however that $N_B$ (for $B\neq 0$) and~$J$ are not compatible with the U(1)-flavour symmetry. 
As discussed at the beginning of Section \ref{sec:DefectsRW}, compatibility with this symmetry requires the polynomials $W$ defining 1-morphisms in $\mathcal C$ to be homogeneous of flavour charge $0$. 
Since we assign U(1)-flavour charge $q_{x_i} = -1$ to all the bulk variables $x_i$, the potentials appearing in the definition of $N_B$ for $B\neq 0$ and $J$ cannot have flavour charge $0$. 
Thus, only the subgroup $D\subset\text{Sp}_{2n}(\C)$ is realized in $\bigra$. The full symmetry group $\text{Sp}_{2n}(\C)$ can only be realised if we give up on the U(1)-flavour symmetry.\footnote{This is of course not surprising. After all the flavour symmetry acts with opposite signs on base respectively fibre variables of $T^*\C^n$. Hence only those $\text{Sp}_{2n}(\C)$-transformations which do not mix base and fibre variables can be compatible with the flavour symmetry. And these are exactly the ones in the subgroup $D\subset\text{Sp}_{2n}(\C)$.}

In the following, we shall preserve the flavour symmetry and only consider symmetry defects $I_A$, $A\in\text{GL}_n(\C)$. 
We represent them graphically as 
\begin{equation}
	\tikzzbox{%
		I_A:=

	} \, . 
\end{equation}
They indeed satisfy  $I_A\circ I_B\cong I_{AB}$ for all $A, B \in \text{GL}_n(\C)$, 
\begin{equation}
	\tikzzbox{%
		%
	}
\, , 
\end{equation}
as witnessed by the 2-isomorphisms
\begin{align}
	\label{eq:multiplication}
	\tikzzbox{%
		%
	}
	&= \Delta_{A,B} \coloneqq 
	\big[\yu - B\zu, \betau - \gammau A \big] \otimes \big[\alphau - \gammau, \xu - A \yu \big] .
\end{align}
These isomorphisms are related to structure morphisms of~$\bigra$. 
To explain this we first introduce some notation. 
Consider any 1-morphism $D\colon \xu \lra \yu$ given by $D = \left(\alpha; W (\alphau, \xu, \yu) \right)$, where the length of $\xu$ is $n$. Then for any $B \in \text{GL}_n(\C)$ we define the \textsl{right twist} of~$D$ by~$B$ as
\be
(D)_B 
	:=
	\big(\alpha; W (\alpha, B\xu, \yu ) \big)\,.
\ee
This twist also acts on 2-morphisms. 
Namely, let $E = (\alpha; V(\betau, \xu, \yu))\colon \xu \lra \yu$ be another 1-morphism and $\phi\colon D\lra E$ a 2-morphism represented by a matrix factorisation $P(\alphau, \betau, \xu, \yu)$ of
$V(\betau, \xu, \yu)-W(\alpha, \xu, \yu)$. 
Then the right twist $(\phi)_B\colon(D)_B \lra (E)_B$ is represented by $P(\alphau, \betau, B\xu, \yu)$.

We can check that $(I_A)_B=I_{AB}$. 
Indeed, applying $(-)_B$ to the right unitor (cf.~\eqref{eq:defrho})
\be
\rho_{I_A} =
\tikzzbox{%

}
= \big[\zu - \yu, \betau - \alphau A\big] \otimes \big[\gammau - \alphau, \xu - A \zu \big] \{-\Phi_{\yu}\} 
\ee
yields a 2-morphism $(\rho_{I_A})_B \colon I_A\circ I_B\lra I_{AB}$, which coincides with the 2-isomorphism $\mu_{A,B}$:
\be
(\rho_{I_A})_B = \big[ B\zu - \yu, \betau - \alphau A\big] \otimes \big[\gammau - \alphau, \xu - AB \zu \big] \{-\Phi_{\yu}\} = \mu_{A, B} .
\ee
Similarly, twisting the inverse $(\rho_{I_A})^{-1}$ of the right unitor (cf.~\eqref{eq:rhoinv})
\be
(\rho_{I_A})^{-1} =
\tikzzbox{%
	%
}
=\big[\yu - \zu, \betau - \gammau A \big] \otimes \big[\alphau - \gammau, \xu - A \yu \big] \, ,
\ee
one obtains the 2-morphism $(\rho_{I_A}^{-1})_B \colon I_{AB} \lra  I_A\circ I_B$
\be
(\rho_{I_A}^{-1})_B = \big[\yu - B\zu, \betau - \gammau A \big] \otimes \big[\alphau - \gammau, \xu - A \yu \big] = \Delta_{A,B} 
\ee
which coincides with $\Delta_{A,B}$.\footnote{The maps $\mu_{A,B}$ and $\Delta_{A,B}$ can equally be obtained by left-twisting the left unitors $\lambda_{I_B}$ and its inverse $(\lambda_{I_B})^{-1}$ by $A$, respectively.}

The isomorphisms $\mu_{A,B}$ and $\Delta_{A,B}$ are also related to the adjunction 2-morphisms of $I_A$.
For this we need the fact that $I_{A^{-1}} \cong (I_A)^\dagger$. Namely, 
for $I_A = (\alphau; \alphau \cdot (\xu - A\yu))\colon \yu \lra \xu$, we get from \eqref{eq:defdual} that 
\be
(I_A)^\dagger 
	= 
	\big( \alphau; \, - \alphau \cdot (\xu - A\yu) \big) 
	= 
	\big( \alphau; \, \alphau \cdot A (\yu - A^{-1} \xu) \big) \colon \xu \lra \yu \, .
\ee
This is isomorphic to $I_{A^{-1}} = ( \alphau^\prime;  \alphau^\prime \cdot (\yu - A^{-1} \xu))\colon \xu \lra \yu$ by means of the 
 isomorphism $\chi_A\colon (I_A)^\dagger \lra I_{A^{-1}}$ given by the Koszul factorisation
\be
\chi_A = \big[ \alphau^\prime - \alphau A, \, \yu - A^{-1} \xu \big] \,.
\ee
Using this isomorphism one finds
\be \label{eq:mu_ev_iso}
\mu_{A,A^{-1}} = \tev_{I_A} \cdot ( 1_{I_A} \circ \chi_A^{-1})\,, 
\quad 
\Delta_{A,A^{-1}} =  (1_{I_A} \circ \chi_A) \cdot \coev_{I_A} \,.
\ee
Indeed, using the adjunction 2-morphisms 
\begin{align}
	\tev_{I_A} &=
	\tikzzbox{%

	}
	= \big[\xu - \zu, -\gammau + \alphau \big]\otimes
	\big[\alphau-\betau, \zu - A \yu \big]
\end{align}
from Section \ref{sectdualisability}, as well as the properties of matrix factorisations \eqref{eq:defPhi}--\eqref{eq:varelimination}, we compute
\begin{align}
	\begin{split}
		\tev_{I_A} \cdot ( 1_{I_A} \circ \chi_A^{-1}) 
		=&
		\big[\xu-\zu,\gammau - \alphau \big] \otimes \big[ \alphau -\betau, -(\zu - A \yu)] 
		\\
		& \otimes \big[ \betau A - \betau^\prime , \yu - A^{-1} \zu \big] \{-\Phi_{\yu}\}
		\\
		\cong &
		\big[\xu-\zu,\gammau - \alphau \big] \otimes \big[ \alphau -\betau, -(\zu - A \yu)]
		\\
		& \otimes \big[ \betau - \betau^\prime A^{-1}, A \yu -  \zu \big] \{-\Phi_{\yu}\}
		\\
		\cong &
		\big[\xu-\zu,\gammau - \alphau \big] \otimes \big[ \alphau - \betau^\prime A^{-1}, -(\zu - A \yu)] \{-\Phi_{\yu}\}
		\\
		\cong & \big[\gammau - \alphau, \xu-\zu\big]  \otimes \big[ A^{-1}\zu - \yu, \betau^\prime -\alphau A] \{-\Phi_{\yu}\}
		\\
		= & \mu_{A,A^{-1}} \, .
	\end{split}
\end{align}
This yields the first relation in \eqref{eq:mu_ev_iso}. The second one can be checked in a similar fashion.

\subsubsection{State spaces}
\label{subsubsec:statespaces}

In this section we calculate the state spaces associated to genus-$g$ closed surfaces~$\Sigma_g$ with networks of symmetry defects~$I_A$. 
The vertices of the networks are given by (or built out of) the trivalent junctions $\mu$ and $\Delta$, cf.~\eqref{eq:multiplication} and~\eqref{eq:comultiplication}, respectively. 
Inserting such a fixed choice of network can be interpreted as turning on a flat background gauge field, where a defect $I_A$ on the surface introduces a gluing of the respective flat gauge connections on either side of the defect via~$A$. 
Note that there are many different defect networks which in this way describe the same gauge connection. 
We will argue in Section~\ref{subsubsec:junctions} below that the state spaces of two different defect networks associated to the same background gauge field are indeed isomorphic. 

\medskip 

We first compute state spaces on surfaces with symmetry defects wrapped along non-contractible cycles. 
Let us start with a torus with defect $I_A$ wrapped around the $a$-cycle
(corresponding to a background gauge field with holonomy~$A$ around the $b$-cycle): 
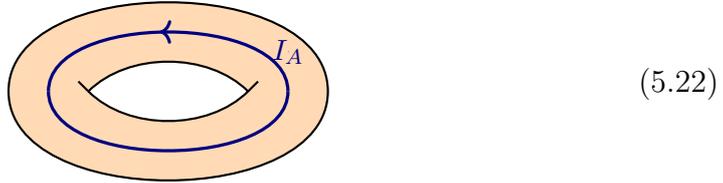
\begin{equation}
	\tikzzbox{%

	}
\end{equation}
As noted in Section \ref{subsec:samplecalculations}, the state space associated to this surface corresponds to the quantum dimension $\dim_{\textrm{l}}(X_A)$ of the 1-morphism\footnote{In this diagram we only label the variables remaining in \eqref{eq:toruswithdefect} after removing intermediate variables. Likewise, in the diagrams in the rest of the section, we only show the names of the internal variables when they are necessary for some computation.}
\begin{equation}
\label{eq:XA}
	X_A = 
	\tikzzbox{%
		%
	}
	\,.
\end{equation}
Applying formula \eqref{eq:toruswithdefect} for $W(\alphau,\xu,\yu) = \alphau \cdot (\xu - A\yu)$ and using the fact that the shifts $\Phi_\xu$, $\Phi_\alphau$ defined in \eqref{eq:defPhi} satisfy $\Phi_\xu + \Phi_\alphau = 0$, together with the property of matrix factorisations \eqref{eq:Koszulshift}, we arrive at
\begin{equation} 
\label{eq:decorated_torus_state_space_1}
	\dim_{\textrm{l}}(X_A) 
	= 
	\big[(\1 - A)\xu, \, \zeru\big] \otimes \big[ \alphau(\1-A) , \, \zeru \big]\,.
\end{equation}
This matrix factorisation can be simplified. 
The matrix $(\1-A)$ acts linearly on~$\xu$ and~$\alphau$. 
Hence via \eqref{eq:mfmanipulation} all the linear combinations of variables $x_i$ and $\alpha_i$ in the image of $(\1-A)$ and $(\1-A)^{\textrm{T}}$, respectively, can be eliminated by setting them to zero. 
This uses up all the matrix factorisations with non-zero differentials. 
Only $\dim(\ker(\1-A))$-many pairs of rank-1 matrix factorisations with zero differential remain.
This leaves $\dim(\ker(\1-A))$-many pairs of $x$- and $\alpha$-variables not in the image of $(\1-A)$, respectively $(\1-A)^{\textrm{T}}$, which we identify with the variables invariant under the $A$-action. 
We denote these invariant variables by $\xu^\text{inv}$ and $\alphau^\text{inv}$, respectively. 
Since $\dim_{\textrm{l}}(X_A)$ is a matrix factorisation of~$0$, it is equal in~$\bigra$ to its cohomology, which can be written as
\begin{equation}
\label{decorated torus state space final}
	\dim_{\textrm{l}} (X_A) 
	= 
	\big((\C \oplus \C\{1,0,-1\})\otimes(\C\oplus \C\{1,0,1\})\big)^{\dim \ker (\1-A)} \otimes_\C \C[ \xu^{\text{inv}} , \alphau^{\text{inv}} ] \, . 
\end{equation}
Note that for a generic $A\in\text{GL}_n(\C)$ for which $\ker(\1-A)=0$, one obtains the trivial state space
\begin{equation} \label{eq:state_space_neq1}
	\dim_{\textrm{l}} (X_A) = \C,
\end{equation}
while for $A=\1$ one recovers the state space 
\begin{equation}\label{eq:torus_trivial_element}
	\dim_{\textrm{l}} (X_A) 
	= 
	\big((\C \oplus \C\{1,0,-1\})\otimes(\C\oplus\C\{1,0,1\})\big)^{n} \otimes_{\C} \C[\xu,\alphau]
\end{equation}
of the torus without defect insertion, cf.~\eqref{eq:0000}.

\medskip 

We can do a similar computation to evaluate the state space of a torus with defect wrapping the other cycle:
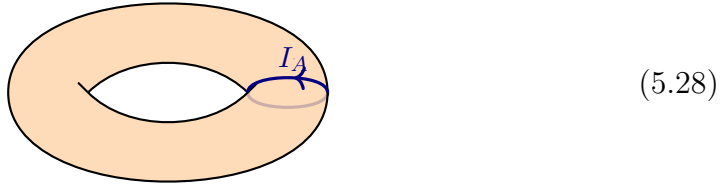
\be
\tikzzbox{%

}
\ee
Since the tori are related by a modular transformation, they belong to the same bordism class. 
Hence this computation should give the same result as~\eqref{eq:decorated_torus_state_space_1}. 
However, in the graphical calculus, this is a completely different computation and it will provide us with a consistency check of our construction. We compute this state space as a left trace $\tr_{\textrm{l}}(\phi)$ (cf.~\eqref{eq:traces}) of the 2-morphism 
\begin{align} \label{eq:cylinder_with_line}
	\phi&:= 
	\tikzzbox{%
		
	}
	= \begin{array}{ll} & \big[ 0, \xu - A \yu \big] 
	\otimes
	\big[ 0, \deltau^\prime - \gammau^\prime A\big]
	\\[0.05cm]
	\otimes &
	 \big[ \yu^\prime - \yu , \deltau - \deltau^\prime \big]
	 \\[0.05cm]
	\otimes &\big[ \xu - \xu^\prime , \gammau - \gammau^\prime \big] \{ -\Phi_\xu \} \, .
	\end{array} 
\end{align}
To arrive at the expression on the right-hand side, the bending isomorphisms \eqref{eq:defomegap} and \eqref{eq:defomega} were used. 
Inserting~$\phi$ into the general formula \eqref{eq:left_trace} for the left trace, we obtain
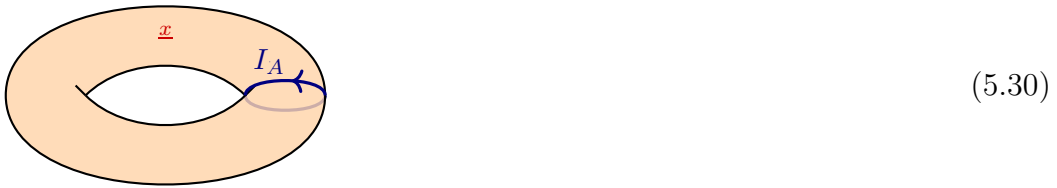
\begin{align} \label{eq:torus_bcycle}
		&
		\tikzzbox{%
			%
		}
		\\
		& \cong 
		\big[ \zeru, \xu - A \xu \big] 
		\otimes 
		\big[ \zeru, \gammau^\prime - \gammau^\prime A \big]
		\otimes \big[ \xu^\prime - \xu , \gammau - \gammau^\prime \big]
		\otimes \big[ \xu - \xu^\prime , \gammau - \gammau^\prime \big] \{ -\Phi_\xu \}
		\nonumber\\
		&\cong
		\big[ (\1-A)\xu  , \zeru \big] 
		\otimes 
		\big[ \gammau (\1- A), \zeru \big] \, . 
		\nonumber
\end{align}
In the last step we used the two rightmost matrix factorisations to eliminate the variables $\xu^\prime$ and $\gammau^\prime$.
Hence, we indeed arrive at the same result as \eqref{eq:decorated_torus_state_space_1}.

\medskip 

Next, let us wrap symmetry defects on both cycles of the torus:
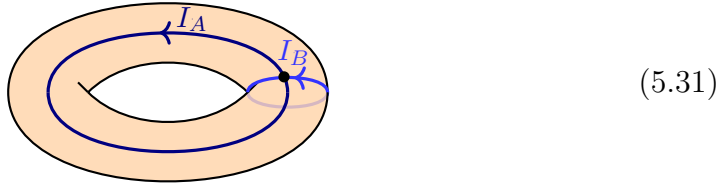
\be
\tikzzbox{%

}
\ee
We choose $A,B\in\text{GL}_n(\C)$ such that $AB=BA$. This allows us in particular to define the junction by resolving it into trivalent junctions of the type \eqref{eq:multiplication} and \eqref{eq:comultiplication}, i.\,e.\ by defining the 4-valent junction to be 
\be\label{eq:symjunction}
\tikzzbox{%
	%
}
\cong \big[ \widetilde \yu - A \zu , \widetilde \alphau - \widetilde \betau B \big]
\otimes \big[ B \zu - \yu ,  \betau - \alphau A \big]
\otimes \big[ \widetilde \betau - \alphau ,  \xu - AB \zu \big] \{ - \Phi_\yu \} \, . 
\ee
Inserting this junction 2-morphism into the general formula \eqref{eq:torusintersection} for the intersection of defects on a torus, we obtain
\begin{align}
		&\big[ \xu - A \xu ,  \alphau - \betau B \big]
		\otimes \big[ B \xu - \xu ,  \betau - \alphau A \big]
		\otimes \big[ \betau - \alphau ,  \xu - AB \xu \big] \{ - \Phi_\xu \}
		\\
		\cong
		&\big[ (\1-A)\xu ,  \alphau (\1-B) \big]
		\otimes \big[ (B-\1) \xu  , \alphau (\1- A)\big] \{ - \Phi_\xu \} \, . 
\end{align}
As before, we can eleminate all but the variables $\xu^\text{inv}$ and $\alphau^\text{inv}$ which are invariant under both the $A$- and $B$-action, i.\,e.\ those which lie in the common kernel $V_{A,B}:=\ker(\1-A)\cap\ker(\1-B)$, and end up with
\be
\tikzzbox{%

}=
\big((\C 
	\oplus \C\{1,0,-1\})\otimes(\C\oplus \C\{1,0,1\})\big)^{\dim V_{A,B}} \otimes_\C \C[ \xu^{\text{inv}} , \alphau^{\text{inv}} ]
\, .
\ee

\medskip 

Having calculated the state space of a torus with symmetry defects wrapped around the two homology cycles, we now turn our attention to general closed surfaces $\Sigma_g$ of genus~$g$, with symmetry defects wrapping all the homology cycles:
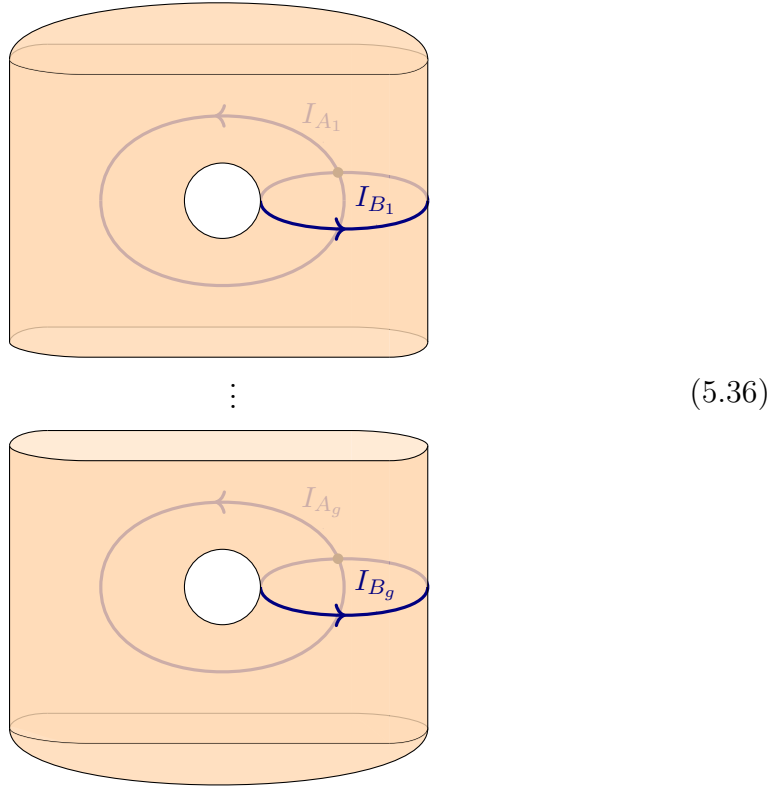
\begin{align}
\begin{split}
	& 
	\tikzzbox{%
		%
	}
\end{split}
\end{align}
As before we assume that $A_IB_I=B_IA_I$ for all $I \in \{1,\ldots, g\}$, and we use \eqref{eq:symjunction} as the respective junction fields. 
This amounts to turning on a non-trivial flat gauge background with holonomies $A_I,B_I$ around the respective cycles. 

In order to calculate the state space, we first evaluate the decorated cylinder 
\begin{equation}
\begin{split}
	& 
	\tikzzbox{%
		%
	}
	\\
	=& \;\big[\yu' - \yu , \deltau - \gammau B\big] \otimes \big[\yu'' - \yu' , \deltau' - \gammau B\big] \otimes \big[\xu' - \xu'' , \gammau' - \gammau \big] 
	\\
	& \; \otimes \big[\yu - A \yu'' , \alphau' - \gammau B \big] \otimes \big[B \yu'' - \xu'' , \gammau - \alphau A\big] \otimes \big[\gammau - \alphau , \xu - AB\yu''\big] \{ -2\Phi_\xu \} \, . 
\end{split}
\end{equation}
This can be used to evaluate the decorated handle element:
\begin{align}
\begin{split}
	& 
	\tikzzbox{%
	%
	}
	\\
	=& \; \big[\rhou - \sigmau , \xu' - \xu\big] \otimes  \big[ \deltau - \gammau , \xu - \xu' \big] 
	\\
	& \; \otimes \big[ (\1 - A)\xu , \gammau (\1- B)\big] \otimes \big[(B-\1) \xu, \gammau (\1- A)\big]  \{ 2\Phi_\xu\} \, .  
	\label{eq:decorated_handle}
\end{split}
\end{align}
Composing $g$ copies of the handle element and eliminating the internal variables yields
\begin{align}
	\big[\rhou - \sigmau , \xu' - \xu\big] 
		& \otimes  [ \deltau - \gammau , \xu - \xu' ] 
		\nonumber 
	\\
		& \otimes \bigotimes_{I=1}^g \big[ (\1- A_I)\xu , \gammau (\1- B_I)\big] \otimes \big[(B_I-\1) \xu , \gammau (\1- A_I)\big]  \{ 2g\Phi_\xu\} \,.
\end{align}
Pre- and post-composing with the cup and cap 2-morphisms from \eqref{eq:adjevcoev}, respectively, we arrive at 
\begin{equation}
	\bigotimes_{I=1}^g  \big[(\1 -  A_I)\xu , \gammau (\1- B_I)\big] \otimes \big[\gammau (\1- A_I) , (B_I-\1) \xu\big] \{ 3(g-1) \Phi_\xu \} \, .
\end{equation}
As before, we can eliminate all but the variables $\xu^\text{inv}$ and $\gammau^\text{inv}$ which are invariant under all $A_I$ and $B_I$, which then automatically sets the differentials in the remaining matrix factorisations to zero. One obtains the cohomology
\be\label{eq:sigmagdefspace}
\big((\C 
	\oplus \C\{1,0,-1\})\otimes(\C\oplus \C\{1,0,1\})\big)^{n(g-1)+\dim V_{A_\bullet,B_\bullet}} \otimes_\C \C[ \xu^{\text{inv}} , \alphau^{\text{inv}} ]\,,
\ee
where 
\be
V_{A_\bullet,B_\bullet}
	=
	\bigcap_{I=1}^g \big(\ker(\1-A_I)\cap\ker(\1-B_I)\big)
\ee
is the subspace of $\C^n$ which is invariant under all the $A_I$ and $B_I$, $I \in\{1,\ldots, g\}$. 

Note that if~$A_I$ and~$B_I$ are all trivial, no variables can be eliminated, and one recovers the state space associated to a genus-$g$ surface without defect network, as in \cite[Prop.\,3.5]{BCR}, which we recall for future reference: 
\be\label{eq:state_space_genus_g}
\big((\C 
	\oplus \C\{1,0,-1\})\otimes(\C\oplus \C\{1,0,1\})\big)^{ng} \otimes_\C \C[ \xu , \alphau ]\, .
\ee
The space is isomorphic as a vector space to the exterior algebra on $2ng$ fermions with coefficients in $\C[\xu, \gammau]$, as discussed in \cite[App.\,B.2]{BCR}. 
This in particular agrees with the original result for state spaces for free Rozansky--Witten models in \cite{RW1996} (see also \cite{BFK, CDGG} for a recent treatment).

If at least one of $A_I,B_I$ is non-trivial, variables can be eliminated and the matrix factorisation reduced. If for instance the $A_I,B_I$ are chosen such that $V_{A_\bullet,B_\bullet}=0$ (which for instance is the case if one of the $A_I,B_I$ has no eigenvalue~$1$), the state space \eqref{eq:sigmagdefspace} becomes
\be
\big((\C 
	\oplus \C\{1,0,-1\})\otimes(\C\oplus \C\{1,0,1\})\big)^{n(g-1)} \,.
\ee
It is the space generated by $2n(g-1)$ fermions but without bosonic degrees of freedom. 
This agrees with the state space of a theory of $n$ free hypermultiplets with corresponding non-trivial gauge background derived by other methods in \cite{CDGG}: turning on a generic non-trivial gauge connection modifies the BV differential  on the state space associated to $\Sigma_g$ by canonical quantisation in such a way that the $2n$ chiral fields corresponding to the $\xu$ and $\alphau$ become exact, and $2n$ of the fermions disappear from the kernel of the BV differential.

\subsubsection{Defect networks and triangulations} 
\label{subsubsec:junctions}

As alluded to above, insertion of a network of symmetry defects with 3-junctions defined in \eqref{eq:multiplication} and \eqref{eq:comultiplication} amounts to introducing a non-trivial flat background gauge field. 
In fact, any such background can be modelled by inserting such a network on the Poincar\'e dual of a triangulation of the surface. 
For this, one assigns the two 2-morphisms \eqref{eq:multiplication}, \eqref{eq:comultiplication} to the oppositely oriented triangles
\begin{equation}
\label{eq:triangles}
\tikzzbox{%

}
.
\end{equation}
Parallel transport along an edge of a triangle is given by the respective group element (or its inverse, depending on the relative orientation) corresponding to the symmetry defect it intersects. 
Note that the conditions on the vertices ($A_1=A_2A_3$ for $\Delta$ on the left, and $A_1A_2=A_3$ for $\mu$ on the right) ensure that holonomies around contractible cycles vanish. 

Of course the state space associated to a surface with gauge background should only depend on the isomorphism class of the respective bundle with connection, and not on a choice of triangulation. 
That this is in fact true in our construction follows from the following relations satisfied by the 3-junctions \eqref{eq:multiplication} and \eqref{eq:comultiplication}:
\begin{align} \label{eq:frobenius_relations}
\tikzzbox{%

}
\end{align}
in case $A_1 A_2 = A_3 A_4$, as well as 
\begin{equation}
\tikzzbox{%
	%
}
\,,
\end{equation}
its vertically reflected version involving~$\Delta$, and
\begin{equation} \label{eq:symmetry}
\tikzzbox{%
	%
}
\,.
\end{equation}
Thus, state spaces are invariant under the local changes \eqref{eq:frobenius_relations}--\eqref{eq:bubble} of the symmetry defect network. 
These local changes exactly correspond to the Pachner moves of triangulations, and hence, the state spaces do not depend on the triangulations.

As an aside we remark that relations \eqref{eq:frobenius_relations}--\eqref{eq:bubble} are ``component versions" of Frobenius, (co)associativity, symmetry and separability relations, respectively, which appear in the generalised orbifold construction (see \cite{ffrs0909.5013, cr1210.6363} for the 2-dimensional case of relevance here, and \cite{CRS1} for the basic theory in arbitrary dimension): given a $\Delta$-separable symmetric Frobenius algebra~$\A$ in some pivotal 2-category, one can insert $\A$-defect networks on any bordism and show that the corresponding correlators are independent of the choice of network (in the interior). 
Indeed, in case we restrict to a finite (sub)group, we can gauge the symmetry by summing over all group elements. 
A direct sum completion of the relations \eqref{eq:frobenius_relations}--\eqref{eq:bubble} can then be used to construct a $\Delta$-separable symmetric Frobenius algebra and thus apply the orbifold construction. 
See for instance \cite[Sect.\,7]{cr1210.6363} for the example of a Landau--Ginzburg orbifold. 
We do not perform the gauging in this paper, see however \cite{BFK} for results on state spaces for gauged theories obtained by other means.

In the following we prove the first identity of \eqref{eq:frobenius_relations}. 
The other relations follow in a similar way.
The leftmost diagram of \eqref{eq:frobenius_relations} evaluates to 
\begin{align}
		&\big[\widetilde \yu - A_4 \zu, \widetilde \betau - \gammau A_3 \big] \otimes \big[\widetilde \alphau - \gammau, \xu - A_3 \widetilde \yu \big] 
		\nonumber
		\\ 
		& \qquad \otimes \big[ A_2\zu - \yu, \betau - \alphau A_1\big] 
		\otimes \big[\gammau - \alphau, \xu - B_1 \zu \big] \{-\Phi_{\yu}\} 
		\nonumber
		\\
		\cong & \big[\widetilde \yu - A_4 \zu, \widetilde \betau - \alphau A_3 \big] \otimes \big[\widetilde \alphau - \alphau, \xu - A_3 \widetilde \yu \big] \otimes \big[ A_2\zu - \yu, \betau - \alphau A_1\big] \{-\Phi_{\yu}\} \,,
\end{align}
while the one in the middle yields
\begin{align}
		& \big[ A_2\zu - \yu, \betau - \gammau B_2\big] \otimes \big[\widetilde \betau - \gammau, \widetilde \yu - B_2 \yu \big] 
		\nonumber
		\\
		& \qquad \otimes \big[\widetilde \yu - B_2 \yu, \gammau - \alphau A_3 \big] \otimes \big[\widetilde \alphau - \alphau, \xu - A_3 \widetilde \yu \big] \{-\Phi_{\yu}\}
		\nonumber
		\\
		\cong
		& 
		\big[ A_2\zu - \yu, \betau - \widetilde \betau B_2\big] \otimes \big[\widetilde \yu - B_2 \yu, \widetilde \betau - \alphau A_3 \big] \otimes \big[\widetilde \alphau - \alphau, \xu - A_3 \widetilde \yu \big] \{-\Phi_{\yu}\} \,.
\end{align}
These two expressions are isomorphic if
\begin{equation}\label{eq:frob_iso}
	\big[\widetilde \yu - A_4 \zu, \widetilde \betau - \alphau A_3 \big] \otimes \big[ A_2\zu - \yu, \betau - \alphau A_1\big] \cong \big[ A_2\zu - \yu, \betau - \widetilde \betau B_2\big] \otimes \big[\widetilde \yu - B_2 \yu, \widetilde \betau - \alphau A_3 \big] 
\end{equation}
which in turn is a consequence of the general relation of matrix factorisations
\begin{equation}
	\begin{split}
		\bigotimes_i \big[p_1^{(i)}, p_0^{(i)} \big] \otimes \big[q_1^{(i)}, q_0^{(i)} \big] 
		\cong 
		\bigotimes_i \Big[p_1^{(i)} + \sum_j T_{ij} q_1^{(i)}, p_0^{(i)}\Big] \otimes \Big[q_1^{(i)}, q_0^{(i)} - \sum_j q_1^{(i)} T_{ji} \Big]
	\end{split}
\end{equation}
where $T_{ij}$ are arbitrary linear transformations. 
This is a direct generalisation of property~\eqref{eq:mfmanipulation}. 
Applying this formula to the left-hand side of \eqref{eq:frob_iso} for $T=B_2$, and using the conditions $A_1 = A_3 B_2$, $A_4 = B_2 A_2$, we obtain
\begin{align}
		& \big[\widetilde \yu - A_4 \zu, \widetilde \betau - \alphau A_3 \big] \otimes \big[ A_2\zu - \yu, \betau - \alphau A_1\big] 
		\nonumber
		\\
		\cong &
		\big[\widetilde \yu - B_2 A_2 \zu + B_2(A_2\zu - \yu), \widetilde \betau - \alphau A_3 \big] \otimes \big[ A_2\zu - \yu, \betau - \alphau A_3 B_2 - (\widetilde \betau - \alphau A_3) B_2 \big]
		\nonumber
		\\
		\cong &
		\big[\widetilde \yu  - B_2 \yu, \widetilde \betau - \alphau A_3 \big] \otimes \big[ A_2\zu - \yu, \betau - \widetilde \betau B_2 \big] \,,
\end{align}
which implies \eqref{eq:frob_iso}.

\subsubsection{Twisted sector line operators}


Given any extended $d$-dimensional TQFT~$\zz$ with values in some $d$-category~$\mathcal D$, one can extract information about its $k$-dimensional defects from what~$\zz$ assigns to the sphere $S^{d-k-1}$. 
More precisely, the (higher) category of $k$-dimensional defects is the (higher) Hom category of $(d-k-1)$-morphisms $\mathcal D(\zz(\varnothing_{d-k-1}),\zz(S^{d-k-1}))$ between what~$\zz$ associates to the $(d-k-1)$-dimensional empty set and what it assigns to~$S^{d-k-1}$, see e.\,g.\ \cite{Kapustin-review}. 

In our truncation of a 3-dimensional theory, taking values in the 2-category~$\bigra$, we do not see the full category of line operators, but only the vector space $\bigra(\zz(\varnothing_{1}),\zz(S^{1}))$ of its isomorphism classes. 
With this caveat we however continue to use the phrase ``category of line operators'', especially since~$\bigra$ is the homotopy 2-category of the 3-category~$\mathcal{RW}^{\textrm{aff}}$ of \cite{KR0909.3643}, and our extended TQFTs~$\zz_n$ are expected to lift to 3-dimensional ones valued in~$\mathcal{RW}^{\textrm{aff}}$ (over $\C[\xu,\alphau]$, not over~$\C$). 

As computed in \cite{BCR}, the image of the circle under~$\zz$ is
\begin{equation}\label{eq:lineops}
\tikzzbox{%
		\begin{tikzpicture}[DefaultSettings,baseline = -0.1cm]
			
			\coordinate (c1) at (0,0);
			\coordinate (c2) at (2,0);
			\coordinate (l1) at (1,0.8);
			\coordinate (l2) at (1,-0.8);
			
			\draw[IdentitySlice] (c1) .. controls +(0,0.8) and +(0,0.8) .. (c2);
			\draw[IdentitySlice] (c1) .. controls +(0,-0.8) and +(0,-0.8) .. (c2);
			\fill[IdentityColor] (c1) circle (0pt) node[left] {\scriptsize $\betau$};
			\fill[IdentityColor] (c2) circle (0pt) node[right] {\scriptsize $\alphau$};
			\fill[IdentityColor] (l1) circle (0pt) node {\scriptsize $\xu$};
			\fill[IdentityColor] (l2) circle (0pt) node {\scriptsize $\yu$};
		\end{tikzpicture}
	}
	=
	\big(\alphau,\betau,\xu,\yu;\; \betau\cdot(\yu-\xu) + \alphau\cdot( \xu - \yu)\big) \, ,
\end{equation}
and the image of $\varnothing_1$ is the zero potential. Hence the category of line defects is given by the homotopy category of matrix factorisations of the potential $W=(\betau-\alphau)\cdot(\yu-\xu)$. 
Indeed, in this way we recover the description of bulk line operators in the initial 3-dimensional target category of \cite{KR0909.3643}, where they arise as line operators on an invisible surface defect.  
As the identity surface defect corresponds to the 1-morphism $(\betau; \alphau\cdot(\xu-\yu))$, the latter are indeed given by matrix factorisations of $(\betau-\alphau)\cdot(\yu-\xu)$.
By Kn\"orrer periodicity, \eqref{eq:lineops} is equivalent to $(\alphau, \xu; 0)$, or the homotopy category of matrix factorisations of $0$, which in turn is equivalent to ${\Z}_2$-graded ${\C}[\xu, \alphau]$-modules. 
This agrees with \cite{r0112209,RobWill}.\footnote{Following \cite{KRS, Kapustin-review}, we stress that a priori and for general target there is only a ${\Z}_2$-grading on this category, which in our special case (due to the preserved symmetries) can be upgraded to a ${\Z}$-grading which is however not expected for general Rozansky--Witten models.}

\medskip 

In Sections \ref{subsubsec:statespaces}--\ref{subsubsec:junctions}, we considered closed 2-dimensional bordisms with arbitrary networks of symmetry defects, corresponding to fixed background gauge fields. 
Cutting discs out of such bordisms gives rise to circles dressed by group elements, marking the points where the symmetry defects end. 
The image of such circles under the extended TQFT encodes the information about line operators in a fixed gauge background with prescribed monodromy. 
This provides a description of ``twisted sector line operators", in close analogy to the well-known twisted sector point operators, often also refered to as disorder (point or line) operators, see \cite{DGGH} for a discussion in the context of gauge symmetries.

Note that circles~$X_A$ marked by group elements~$A$ (recall~\eqref{eq:XA}) can be connected by bordisms dressed with symmetry defects, where a defect  $I_A$ has to end on the marked point. 
For example, there is a pair-of-pants connecting two circles marked by~$A$ and~$B$, respectively, to one marked by the composition of the group elements $AB$, $X_A \circ X_B \lra X_{AB}$. 
Such a structure is expected physically, since different twisted line operators can be merged whenever this is compatible with the twist. 

Concretely, our extended TQFT associates the 1-morphism
\begin{equation}
	X_A=\tikzzbox{%
		\begin{tikzpicture}[DefaultSettings,baseline = -0.1cm]
			
			\coordinate (c1) at (0,0);
			\coordinate (c2) at (2,0);
			\coordinate (l1) at (1,0.8);
			\coordinate (l2) at (1,-0.8);
			
			\draw[IdentitySlice] (c1) .. controls +(0,0.8) and +(0,0.8) .. (c2);
			\draw[IdentitySlice] (c1) .. controls +(0,-0.8) and +(0,-0.8) .. (c2);
			\fill[IdentityColor] (c1) circle (0pt) node[left] {\scriptsize $\betau$};
			\fill[DefectColor] (c2) circle (3pt) node[right] {\small $(\alphau; A)$};
			\fill[IdentityColor] (l1) circle (0pt) node {\scriptsize $\xu$};
			\fill[IdentityColor] (l2) circle (0pt) node {\scriptsize $\yu$};
		\end{tikzpicture}
	}
	=
	\big(\alphau,\betau,\xu,\yu; \betau\cdot(\yu-\xu) + \alphau\cdot( \xu - A\yu)\big)
\end{equation}
to the circle with an insertion of the symmetry defect $I_A$. 
The category whose Grothendieck group is $\bigra(\zz(\varnothing_1),\zz(X_A))$ is then given by the homotopy category of (equivalence classes of) matrix factorisations 
\be\label{eq:cattwisted}
{\mathcal L}_A:=
\operatorname{hmf}\big(\C[\alphau,\betau,\xu,\yu],W_A\big)^\omega 
\ee 
of the potential
\be
W_A
	:=
	\betau\cdot(\yu-\xu) + \alphau\cdot( \xu - A\yu)
	=
	(\alphau-\betau)\cdot(\xu-\yu)+\alphau(\1-A)\yu\,.
\ee
It corresponds to the category of line defects in the sector twisted by the defect~$I_A$ of the underlying theory. 

Indeed, as in the untwisted case discussed above, this agrees with the perspective of the full 3-category~$\RW^{\textrm{aff}}$ of affine Rozansky--Witten models of \cite{KR0909.3643}. 
Here the line operators of the twisted sector correspond to line defects between the identity surface defect and the respective symmetry defect. 
The category of these line operators is given by $\RW^{\textrm{aff}}(I_\1,I_A)$ which is nothing but ${\mathcal L}_A$, where now the potential $W_A$ is obtained as the difference of the superpotential associated to~$I_A$ and the one associated to the identity defect.  

The category ${\mathcal L}_A$ can be further simplified:
substituting $\gammau = \alphau - \betau$ and $\zu = \xu - \yu$, one obtains $\operatorname{hmf}(\C[\alphau,\gammau,\yu,\zu],\cu\zu+\alphau\cdot(\1-A)\yu)^\omega$ which in turn, by virtue of Kn\"orrer periodicity, is equivalent to the category 
$\operatorname{hmf}(\C[\alphau,\yu],\alphau\cdot(\1-A)\yu)^\omega$.
In fact, Kn\"orrer periodicity can be further used to eliminate variables~$\alpha$ and~$y$ in the image of $\1-A$. 
This uses up the entire potential and only $\dim(\ker(\1-A))$-many variables $\yu^\text{inv}$ and $\alphau^\text{inv}$ remain. 
Thus ${\mathcal L}_A$ is equivalent to 
\be
\operatorname{hmf}\!\big(\C[\alphau^\text{inv},\yu^\text{inv}],0\big)
	\cong
	\modu^{\Z_2}\!\big(\C[\alphau^\text{inv},\yu^\text{inv}]\big)\,.
\ee
If $A$ is generic in the sense that $\ker(\1-A)=0$, then all variables can be eliminated and the category is equivalent to $\SVect$.
This agrees with the discussion of holonomy line defects in the theory of free hypermultiplets in \cite{CDGG}.

\subsection{Boundaries}
\label{subsec:boundaries}

In this section we fix a bulk theory and consider its boundary conditions. 
As discussed around \eqref{eq:boundarydatum}, boundary conditions can be regarded as defects with a trivial theory on one side. 
In affine Rozansky--Witten models, boundary conditions are described by 1-morphisms $(\alphau; W(\alphau, \xu))\colon \varnothing \lra \xu$ (or their adjoints). 

Restricting the functor~$\zz$ constructed in previous sections to such data, we can answer any question one might have on theories with boundaries in truncated affine Rozansky--Witten models. 
Below we exemplify this by computing state spaces associated to surfaces with boundaries. 
More generally, we explicitly construct the 2-dimensional open-closed TQFTs (satisfying the axioms formulated in \cite{l0010269, MooreSegal2dTQFT, lp0510664}) associated to truncated Rozansky--Witten models. 

Along the way, we provide a geometric interpretation of some boundary conditions as Lagrangian submanifolds in the target geometry $T^{*} \C^n$, and we construct a generalised intersection pairing that is manifestly invariant under Sp$_{2n}(\C)$. 
The open-closed TQFT construction directly implies a baby Hirzebruch--Riemann--Roch theorem that might be of interest in more involved target space geometries.

Our results also have an interpretation in a 3-dimensional theory of twisted hypermultiplets with 2-dimensional boundaries. 
The consistency conditions of the open-closed TQFT constrain operations involving line operators, as we explain below.

\subsubsection{State spaces}

As a first application, we compute state spaces on surfaces with boundary.
The simplest example is a disc. 
Regarding the boundary as a defect, the TQFT maps the disc to the quantum dimension~\eqref{eq:qdim-general}.
To treat surfaces of arbitrary genus, and with an arbitrary number of discs punched out, we utilise a cylinder $S^1 \times [0,1]$ with one disc removed and a boundary condition $D=(\alphau; W(\gammau,\xu))$ imposed on the new boundary. 
This serves  as a basic block for further calculations, as any surface with discs cut out can be obtained by combining this building block with handle operators. 
Evaluating the functor~$\zz$ yields
\begin{align}
	\tikzzbox{%

	}
	& =
	\begin{array}{cl}
		& \big[\zeru, \partial_{\gammau} W(\gammau, \xu) \big]
		\otimes  \big[\zeru, \partial_{\xu} W(\gammau, \xu) - \muu \big]
		\\[0.05cm]
		\otimes &\big[\rhou - \muu, \xu^\prime - \xu \big]\{\Phi_\xu\} \, .  
		\label{eq:cyl-hole}
	\end{array}
\end{align}
Since it does not matter where (on a given connected component) the discs are punched out, we subsequently evaluate the functor on a cylinder $C_N$ with $N$ discs removed, and then combine with handle operators, caps and cups in further steps. 
We impose boundary conditions $D_1, \dots, D_N$ on the different boundary components, where $D_I=(\gammau^I, W_I(\gammau^I,\xu))$. 
In this notation, the capital indices $I$ run over the discs, such that $\gamma^I_j$ is a variable living on the boundary of the $I$-th disc, and we write $C_N^{D_1,\dots,D_N}$ for the defect bordism with chosen boundary conditions. 
The partition function associated to $C_N^{D_1,\dots,D_N}$ can be obtained by composing $N$ copies of \eqref{eq:cyl-hole}. 
This results in
\begin{align}
	\zz\big(C_N^{D_1,\dots,D_N}\big) 
	= 
	\bigotimes_{I=1}^N   \Big( \big[0, \partial_{\gammau^I} W_I(\gamma^I, \xu)\big] 
	& 
	\otimes \big[0, \partial_\xu W_I(\gammau^I, \xu)-\rho^1\big] \Big) 
	\nonumber 
	\\
	& \otimes \big[{\rhou}^1-{\muu}^N, \xu'-\xu\big]\{N \Phi_\xu\}  \, .
\end{align}
Capping off $C_N^{D_1,\dots,D_N}$ with cap and cup we obtain a  defect sphere $S_N^{D_1,\dots,D_N}$, and using \eqref{eq:adjevcoev} we find 
\begin{align}  
	\zz\big(S_N^{D_1, \dots, D_N}\big)  = 
	\bigotimes_{I=1}^N 
	& 
	\big[0, \partial_{\gammau^I} W_I (\gammau^I,\xu)\big] 
	\nonumber 
	\\
	& 
	\otimes \bigotimes_{J=2}^{N} \big[0, \partial_\xu W_J(\gammau^J, \xu)-\partial_\xu W^1 (\gammau^1, \xu)\big]  \{(N-3) \Phi_\xu\} \, .
	\label{eq:minusNspheres}
\end{align}
Note that  for $N=1$, \eqref{eq:minusNspheres} reproduces our earlier result for the disc partition function \eqref{eq:DiscSpace}. 
$\zz\big(S_N^{D_1, \dots, D_N}\big)$ as in~\eqref{eq:minusNspheres} is a matrix factorisation of~$0$, and can therefore be represented by its cohomology
\be
\zz(S_N^{D_1, \dots, D_N} )
	= 
	\C[\gammau^I, \xu] / {\mathcal I} 
	\, , \quad 
	{\mathcal I}
	= 
	\big\langle \partial_{\gammau^I} W_I (\gammau^I, \xu), \ \partial_\xu W_I(\gammau^I, \xu)-\partial_\xu W_1 (\gammau^1, \xu) \big\rangle \, .
\ee

A special case that deserves some attention is $N=2$, the sphere with two holes, providing a pairing on the category of boundary conditions: 
\begin{align}
\mathcal Z\big(S_2^{D_1, D_2}\big)
&
=
\tikzzbox{%
	\begin{tikzpicture}[thick,scale=0.5,color=black, baseline=0cm]
	\coordinate (c1) at ($(-0.5,0)$);
	\coordinate (c2) at ($(-2.5,0)$);
	\coordinate (d1) at ($(0.5,0)$);
	\coordinate (d2) at ($(2.5,0)$);
	%
	%
	\fill[ball color=orange!40!white] (0,0) circle (4);
	%
	\fill [BackSurfaceColor]
	(d2) .. controls +(0,-1.3) and +(0,-1.3) .. (d1)
	-- (d1) .. controls +(0,1.3) and +(0,1.3) .. (d2)
	;
	\fill [BackSurfaceColor]
	(c2) .. controls +(0,-1.3) and +(0,-1.3) .. (c1)
	-- (c1) .. controls +(0,1.3) and +(0,1.3) .. (c2)
	;
	\draw[LineDefect] (d2) .. controls +(0,-1.3) and +(0,-1.3) .. (d1);
	\draw[costring, LineDefect] (d1) .. controls +(0,1.3) and +(0,1.3) .. (d2);
	\draw[LineDefect] (c1) .. controls +(0,-1.3) and +(0,-1.3) .. (c2);
	\draw[costring, LineDefect] (c2) .. controls +(0,1.3) and +(0,1.3) .. (c1);
	%
	\fill[DefectColor] (1.5,1.5) circle (0pt) node {{\small $W_2$}};
	\fill[DefectColor] (-1.5,1.5) circle (0pt) node {{\small $W_1$}};
	\fill[IdentityColor] (0,2.5) circle (0pt) node {{\scriptsize $\xu$}};
	%
	\end{tikzpicture}
}
\nonumber 
\\
&
= 
\big[0, \partial_{\gammau^1} W_1 (\gammau^1, \xu)\big] 
\otimes 
\big[0, \partial_{\gammau^2} W_2 (\gammau^2, \xu)\big] 
\nonumber 
\\ 
& \qquad 
\otimes \big[0, \partial_\xu W_2(\gammau^2, \xu)-\partial_\xu W_1 (\gammau^1, \xu)\big] \,.
\label{eq:pairing}
\end{align}
It follows from the construction that this pairing is invariant under the action of the symplectic group actions discussed in Section~\ref{subsubsec:symmetries}, i.\,e.
\be
\label{eq:invariant}
\zz\big(S_2^{D_1, D_2}\big)
	= 
	\zz\big(S_2^{D(g) \circ D_1, D(g) \circ D_2}\big) 
\ee
where $D(g)$ denotes any of the 1-morphisms $I_A, N_A$ or~$J$ in Section~\ref{subsubsec:symmetries}. 
To see this, we wrap the symmetry defect $D(g)$ around each of the two boundary components in~\eqref{eq:pairing}. 
This corresponds to the right-hand side of the above equation. 
Because of the topological nature of the theory and the coherence of the formalism, we may move these additional defects away from the two punctures and to a common line of longitude, 
partitioning the sphere into a left and right half. 
Then the two $D(g)$-labelled lines meet with opposite orientation. 
But since $D(g)^\dagger \cong D(g)^{-1} \cong D(g^{-1})$, the fusion of the two lines yields the invisible defect, leaving us with the left-hand side of~\eqref{eq:invariant}.

\medskip

The evaluation of~$\zz$ on surfaces of higher genus is a straightforward combination with the formulas for the handle operator \eqref{eq:decorated_handle}. 
The amplitude~\eqref{eq:state_space_genus_g} factorises into a piece that takes the form of the sphere partition function and another piece that depends on  the genus. Punching out holes only replaces the sphere partition function by \eqref{eq:minusNspheres}, leaving the other factor unchanged.

We spell out the results in simple examples, namely  $D_1=(\varnothing; 0)$ and $D_2=(\alphau; \xu\cdot\alphau)$. 
These two defects are related by the ``Legendre transformation" symmetry defect~$J$. 
Their disc partition functions are ${\C}[\xu]\{-\Phi_\xu \} $ and ${\C}[\alphau] \{-2\Phi_\xu\}$, respectively. 
The genus-$g$ amplitude with~$N$ insertions of $(\varnothing;0)$ is
\begin{align} 
\zz\big(\Sigma_{g,N}^{(\varnothing;0),\dots,(\varnothing;0)}\big)
	&
	=
	\Big((\C \oplus \C\{1,0,-1\})\otimes(\C\oplus\C\{1,0,1\})\Big)^{\otimes ng} 
	\nonumber
	\\
&\qquad\qquad\qquad\qquad\otimes_\C \C[\xu] \{ [3(g-1) +N]\Phi_\xu \} \, , 
\end{align}
while for $N$ insertions of $D_2$ one obtains
\begin{align} 
\zz\big(\Sigma_{g,N}^{(\alphau; \xu\cdot\alphau),\dots,(\alphau; \xu\cdot\alphau)}\big)
	&
	=
	\Big((\C \oplus \C\{1,0,-1\})\otimes(\C\oplus\C\{1,0,1\})\Big)^{\otimes ng} 
	\nonumber
	\\
	&\qquad\qquad\qquad\qquad\otimes_\C \C[\alphau] \{ [-3(g-1)]\Phi_\alphau \}  \,.
\end{align}
The pairing~\eqref{eq:pairing} between the two boundary conditions evaluates to
\be
\zz\big(S_2^{(\varnothing,0),(\alphau, \xu\alphau)}\big)
	=
	\big[0, \xu\big] \otimes \big[0,\alphau\big] \{ -\Phi_\xu \} 
	\cong 
	\C\{-\Phi_\xu\} \, ,
\ee 
while the pairing of either one with itself is isomorphic to the disc partition function with the respective boundary condition, up to a shift.

\medskip 

Geometrically, one expects the simplest boundary conditions to correspond to Lagrangian subspaces of the target space $T^*\C^n$, where the directions along the base correspond to Neumann boundary conditions on fields, while transversal directions correspond to Dirichlet conditions. 
Since variables $\xu$ correspond to the base, and $\alphau$ to the fibre, the two boundary conditions set either the fibre or base variables to a constant. 
This interpretation is further supported by the pairing between them, and it is also in agreement with the symmetries of the model.

In slightly more generality, we may describe a sublocus by a set of $m\leq n$ $\C$-linear equations
\be\label{eq:subspace}
\sum_i A_{ji} x_i =0 \,, \quad  j \in \{1, \dots, m \} \, .
\ee
For a generic matrix~$A$, these equations will have $n-m$ linearly independent solutions that we denote $\xi_1, \dots, \xi_{n-m}$.
To implement such equations, we introduce~$m$ ``Lagrange multipliers" $\gamma_j$ and relate the geometry to the 1-morphism
\be
D_A 
	:= 
	\bigbtimes_j \Big(\gamma_j; \, \sum_i \gamma_jA_{ji}x_i\Big) 
	= 
	\Big(\gamma_1, \dots, \gamma_m; \, \sum_{i,j} \gamma_j A_{ji} x_i\Big) \,  .
\ee
The fields $\gamma_j$ parametrise directions in the fibre. 
We thus obtain a {Lagrangian subspace} stretching over $n-m$ dimensions in the base and~$m$ dimensions in the fibre, generalising the special cases $(\varnothing; 0)$ and $(\alphau; \xu\cdot\alphau)$. 

For any constraints \eqref{eq:subspace}, the disc partition function  corresponding to $D_A$ is a polynomial ring in $n$ variables. 
As before, the image of the functor on the disc with boundary condition $D_A$ is obtained from \eqref{eq:DiscSpace}, 
\be 
\zz \Bigg( 
\tikzzbox
{
	\begin{tikzpicture}[very thick,scale=0.7,color=blue!50!black, baseline=-0.1cm]
		\fill[orange!40!white, opacity=0.7] (0,0) circle (0.95);
		\fill (0,0) circle (0pt) node[red!80!black] {{\scriptsize$\xu$}};
		\draw (0,0) circle (0.95);
		\fill (35:1.32) circle (0pt) node {{\small$\hspace{1em} D_A$}};
		\draw[<-, very thick] (0.100,-0.95) -- (-0.101,-0.95) node[above] {}; 
		\draw[<-, very thick] (-0.100,0.95) -- (0.101,0.95) node[below] {}; 
	\end{tikzpicture} 
} 
\Bigg) 
\cong \C[\xiu, \gammau]\{-\Phi_\xiu+2\Phi_\gammau\} \,. 
\ee 
Here, $\xiu$ denotes the list $(\xi_1, \dots, \xi_{n-n})$, and $\gammau$ stands for the list $(\gamma_1,\dots, \gamma_{m})$. 
Altogether, one obtains a polynomial ring in $n$ variables.
This supports the interpretation that~$n$ directions in the target geometry are subject to Dirichlet boundary conditions, and another~$n$ to Neumann boundary conditions.

The pairing between any two such boundary conditions $D_A, D_{A'}$  yields the polynomial ring on the geometric intersection, which is how one would describe intersections in the language of algebraic geometry. 
The ordinary geometric intersection is well-defined for subspaces that intersect transversally. Whenever the geometric transverse intersection is $1$, our result is~$\C$.  

To put these observations into perspective, we recall that a class of ordinary 2-dimensional TQFTs is provided by twisted non-linear sigma models, and that boundary conditions in the  A-twist correspond to middle-dimensional Lagrangian cycles. 
In this context, the open string Witten index computes intersection numbers of cycles in the target geometry. 
The corresponding TQFT diagram is a cylinder whose ends are labelled by boundary conditions $D_1, D_2$, which topologically is precisely $S_2^{D_1, D_2}$. 
Therefore, the functor~$\zz$ evaluated on $S_2^{D_1, D_2}$ provides an analogue of the Witten index in a new context. 
The above discussion suggests to regard it as a categorification of the ordinary intersection pairing, since in our construction the intersection number is replaced by a state space associated to the intersection.

To shed more light on these relations and to formulate a version of an index theorem, we next describe the complete open-closed subsector of our fully extended defect TQFT.

\subsubsection{Open-closed TQFT}
\label{subsubsec:open-closed}

To construct an open-closed TQFT from a fully extended TQFT with defects, we simply have to restrict the functor to the relevant bordism category. 
Its objects are 1-morphisms between the empty set and the empty set. 
This includes closed circles as well as intervals which are dressed with elements of $D_1^\partial$ at their boundaries, see the general discussion around \eqref{eq:boundarydatum} as well as \eqref{eq:Whistle}. 
Restricting the functor~$\zz$ constructed in Section~\ref{sec:DefectsRW} to these 1-morphisms as well as 2-morphisms between them, we obtain a consistent open-closed TQFT taking values in $\cat(\varnothing,\varnothing)$, i.\,e.\ the endomorphism category of the unit object in~$\bigra$. 
The axioms of \cite{l0010269, MooreSegal2dTQFT, lp0510664} simply follow from the properties of the underlying extended defect TQFT. 

The structure of the closed sector was computed in \cite{BCR}. 
We recall from there that to a circle, the functor associates
\be
\zz(S^1) 
	= 
	\big(\alphau,\deltau,\xu,\yu; \, (\alpha-\delta)\cdot(\xu-\yu)\big) \, .
\ee
The associated category of line operators is the homotopy category of matrix factorisations 
\be 
\hmf\!\big(\C[\alphau,\deltau,\xu,\yu], (\alpha-\delta)\cdot(\xu-\yu) \big)^\omega
\cong 
\bigra \big( 1_\varnothing, \zz(S^1) \big) 
\ee 
which in turn is isomorphic to the category of ${\Z}_2$-graded ${\C}[\xu, \alphau]$-modules.

In the following, we will identify the open-closed generators one by one. 
We start by computing the image of an interval
\be 
I_{D_1, D_2^\dagger} = 
\tikzzbox
{
	\begin{tikzpicture}[very thick,scale=1.0,color=blue!50!black, baseline=-0.1cm]
	\draw[red!80!black] (0,0) -- (1,0);
	\fill (0,0) circle (2.5pt) node[above] {{\small$D_2^\dagger$}};
	\fill (1,0) circle (2.5pt) node[above] {{\small$D_1 \vphantom{D_2^\dagger}$}};
	\end{tikzpicture} 
}
\, . 
\ee 
In our setting, $D_1=(\alphau; W_1(\alphau,\xu))\colon\varnothing\lra \xu$ and $D_2^\dagger=(\betau; -W_2(\betau,\xu))\colon\xu \lra \varnothing$, hence we find
\be
\label{eq:functor_on_interval}
\zz(I_{D_1, D_2^\dagger}) 
	= 
	\big(\alphau,\betau,\xu; \, W_1 (\alphau,\xu)-W_2(\betau,\xu)\big) \, . 
\ee
For example, for $D_1=(\alphau; \alphau\cdot\xu)$ and $D_2=D_1^\dagger$ one obtains
\be
\zz(I_{D_1, D_1^\dagger})
	= 
	\big(\alphau,\betau,\xu; \,(\alphau-\betau)\cdot\xu\big) \, .
\ee
Analogously to the bulk case one can associate to this the category of line operators between the respective boundary conditions, whose Grothendieck group is $\bigra(1_\varnothing,\zz(I_{D_1,D_2}))$. 
It is given by the homotopy category of matrix factorisations of $(\alphau-\betau)\cdot\xu$, which in turn is equivalent to ${\Z}_2$-graded ${\C}[\alphau]$-modules. 
Picking as before $D_1=(\alphau;\alphau \cdot \xu)$ but $D_2=(\varnothing,0)$, one is left with $\bigra(1_\varnothing,\zz(I_{D_1, D_2})) \cong \C$. 

Morphisms between circles which do not involve non-glueing boundaries are generated by the cap and the cup already given in~\eqref{eq:adjevcoev}, as well as the pair-of-pants and the upside-down pair-of-pants: 
\begin{align}
\tikzzbox{

}
&= 
\begin{array}{cl}
[\betau-\alphau, \yu-\yu'] \otimes [\xu-\yu,  \betau-\epsilonu] \otimes [\epsilonu'-\epsilonu, \xu'-\xu]\\[0.05cm]
\otimes\;  [\xu'-\yu', -\betau+\epsilonu]\otimes[\sigmau'-\sigmau, \xu-\xu'] \, ,
\end{array}%
\end{align}
Here, the dashed lines labelled $(\gammau, \gammau')$ denote invisible line defects on both the front ($\gammau$) and the back ($\gammau'$) of the surface. 

Morphisms between intervals are generated by the flat versions of cap, cup, and (upside-down) pair-of-pants, as well as ``whistles'' between circles and intervals. 
Upon evaluation with~$\zz$, one obtains
\begin{align}
\tikzzbox{%
	
}
&= 
\begin{array}{cl}
&\big[\tauu^\prime-\tauu,\Deltau W_1(\vect{\tauu^\prime}{\tauu},\yu)\big]
\\[0.05cm]
& \otimes \big[\yu-\xu,\Deltau W_1(\tauu^\prime,\vect{\yu}{\xu}) - \Deltau W_2(\betau,\vect{\yu}{\xu})\big]
\\[0.05cm]
& \otimes \big[\betau-\alphau,-\Deltau W_2(\vect{\betau}{\alphau}, \xu)\big] \,, \label{openpants}
\end{array}%
\end{align}
as well as the bulk-boundary and boundary-bulk maps 
\begin{align}
\tikzzbox{%
	
}
& =
\begin{array}{cl}
& \big[\widetilde \rhou-\rhou,\Deltau W(\omegau^\prime,  \vect{\widetilde \rhou}{\rhou})\big] \otimes \big[\zeru, - \deltau - \Deltau W(\rhou, \vect{\xu^\prime}{\xu})\big] 
\\[0.05cm]
& \otimes \big[\xu^\prime-\xu,\deltau^\prime-\deltau \big]\{-\Phi_\xu\} \, , 
\label{Bobu}
\end{array}%
\end{align}
In either bulk or boundary sector there are non-degenerate pairings given by gluing caps on to pairs-of-pants.

Any surface with corners can be built from the above building blocks. 
The decomposition  is not  unique, independence of the result follows from  a set of sewing relations. 
We refer to \cite{l0010269, MooreSegal2dTQFT, lp0510664} for a full list of these conditions, and to \cite[Sect.\,3]{2dDefectTQFTLectureNotes} for a review. 

We highlight the Cardy condition, which is one of the sewing conditions just mentioned. 
As an identity of 2-morphisms in~$\bigra$ it reads
\be
\label{eq:R16}
\tikzzbox
{

}  
\, ,
\ee
where the green and blue lines denote 1-morphisms $D_1\colon\varnothing\lra\xu$ and $D_2\colon\xu\lra\varnothing$, respectively. 
The right-hand side is obtained by gluing~\eqref{Bobu} to~\eqref{Bubo} along the common circle, resulting in 
\be
\tikzzbox
{
	
}  
= 
\begin{array}{l}
	\big[\widetilde \rhou-\rhou,\,\Deltau W_1(\omegau^\prime,  \vect{\widetilde \rhou}{\rhou})\big] \otimes \big[\widetilde \betau'-\rhou',\,\Deltau W_2(  \vect{\widetilde \betau}{\betau}, \xu')\big] 
	\\[0.05cm]
	\otimes \big[\zeru, \,\partial_{\xu'} W_1 (\widetilde \rhou, \xu') -  \partial_{\xu'} W_2 (\widetilde \betau', \xu') \big] \,. 
	\label{cardy}
\end{array}
\ee
The left-hand side of \eqref{eq:R16} indeed evaluates to the same expression after a standard calculation. 

\medskip 

Gluing to the diagram in  \eqref{cardy} a cup \eqref{opencup} at the bottom and a cap \eqref{opencap} at the top, one obtains a sphere with two holes punched out.  
As suggested before, this can be interpreted as a categorified Witten index, which 
 factorises into bulk-boundary and boundary-bulk maps closed off by cap and cup. 

On the other hand, gluing \eqref{opencap} to \eqref{Bobu} is basically the quantum dimension of the boundary condition~$D_2$. 
On the level of matrix factorisations, one can verify directly that the composition of  \eqref{opencap} and~\eqref{Bobu} yields \eqref{eq:qdim-general} (for $D=D_2$).  
To understand this relation, recall that quite generally $\bigra(1_\varnothing,\zz(S^1))\cong \End(1_{u})$, where~$u$ is the bulk theory label of the circle.\footnote{cf.~the discussion after~\eqref{eq:lineops}} 
For TQFTs valued in $\Vect_\C$, this is the familiar statement that the bulk state space agrees with the space of local operators on the invisible defect. 
The image of the quantum dimension of~$D_2$ under this equivalence is the boundary-bulk map evaluated on the ``identity'', by which we mean the neutral element with respect to the multiplication given by the flat pair-of-pants.\footnote{In the string theory context, this relation was referred to as ``tension is dimension" \cite{HKMS}.}

In summary, we arrived at a ``generalised index theorem'': 

\begin{center}
	\textsl{The gluing of quantum dimensions equals the categorified Witten index.}%
\end{center}
This is one way to read \eqref{cardy}, in the special case where the diagram is closed with cup and cap. 
We expect this interesting relation to hold also in more involved target geometries than $T^*\C^n$.

\subsubsection*{Towards an interpretation in three dimensions}

Further potentially interesting consequences of the sewing properties arise in the original, 3-dimensional setting from which the category $\cat$ was obtained by truncation. 
Recall that the objects $\xu$ have an interpretation in terms of  free fields in a 3-dimensional theory.
Surface defects of the 3-dimensional theory correspond to 1-morphisms $(\alphau; W(\alphau, \xu, \yu))$, line defects correspond to matrix factorisations, and point defects to morphisms of matrix factorisations up to homotopy. 
This is the setting of \cite{KR0909.3643}, see Section~\ref{subsec:targetcategory} for a summary of points relevant to our discussion. 
As explained there, in the setting of the current paper, only equivalence classes of matrix factorisations (hence line operators) play a role. 
In the following few paragraphs, we aspire to ``de-truncate'' and talk somewhat loosely about ``line operators'' and ``categories of line operators" instead of ``isomorphism classes of line operators" and ``Grothendieck rings". 
On the one hand, this simplifies formulations, on the other hand, in the initial 3-dimensional setting there is no need for a truncation and we expect some of our statements to hold in a stronger sense. 
Postponing a full analysis, we provide some interpretation of the ingredients of the 2-dimensional TQFT in the initial 3-dimensional setting. 

We start with the pair-of-pants. 
Our extended TQFT associates to intervals at the end of each leg categories $\bigra(1_\varnothing,\zz(I_{D_1, D_2}))$.
In the 3-dimensional interpretation, the latter contains for example line operators located on boundary surfaces, but also more general objects that geometrically correspond to configurations protruding into the bulk. 
To interpret the pair-of-pants, we test how~\eqref{openpants} acts on a pair of line operators. 
The latter are  described by matrix factorisations of differences of potentials; for example one could insert a matrix factorisation $P(\yu,\omegau,\betau)$ of $-W_3(\yu,\omegau)+W_2(\betau,\yu)$ at the left leg of~\eqref{openpants}, and $Q(\xu,\alphau,\tauu)$ of $-W_2(\alphau,\xu) +W_1(\tauu,\xu)$ at the right leg. 
Applying the pair-of-pants, one obtains the  matrix factorisation
\begin{align}
&\; \big[\tauu^\prime-\tauu,\Deltau W_1(\vect{\tauu^\prime}{\tauu},\yu)\big]
\otimes  \big[\yu-\xu,\Deltau W_1(\tauu^\prime,\vect{\yu}{\xu}) - \Deltau W_2(\betau,\vect{\yu}{\xu})\big] 
\nonumber 
\\
& \qquad 
\otimes \big[\betau-\alphau,-\Deltau W_2(\vect{\betau}{\alphau}, \xu)\big] \otimes P(\yu, \omegau, \betau) \otimes Q(\xu, \alphau,\tauu) 
\nonumber 
\\ 
\cong 
& 
\; P(\omegau,\betau,\yu) \otimes Q(\betau, \tauu, \yu) \, .
\end{align}
The result displayed in the last line is simply the fusion of line operators on a boundary surface. 
We will see a geometric picture for this in a moment.

There is also natural pairing on the category of line operators living on surface defects. In the framework of the 2-dimensional TQFT, a pairing is obtained by sewing a cap to the pair-of-pants. 
Applying this to matrix factorisations $P,Q$, we get the diagram on the left in~\eqref{eq:3dStuff} below. 
The right side provides a geometrical interpretation in the 3-dimensional setting: the pairing maps two line operators located on compatible surface defects to a cylinder with line operators extending transverally to~$S^1$:
\begin{equation}
\label{eq:3dStuff}
\tikzzbox{

}
\end{equation}

The above picture can also be used to visualise the action of the pair-of-pants without the cap: one simply has to cut the cylinder on the right-hand side of the picture along the vertical direction. 
We further note in passing that while in this section we consider only surface defects that are boundaries, the fusion of line defects on surface defects and their pairing in terms of defects rolled up on cylinders exists for any defect. 
A further relevant special case is for example the invisible surface defect. 
For that case, one obtains a pairing on the category of line operators in the bulk that is part of the structure investigated in \cite{BCR}.

Before we try to give a 3-dimensional interpretation to the upside-down pair-of-pants, we recall that for 2-dimensional TQFTs taking values in $\Vect_\C$ it provides the familiar comultiplication.
However, there is no independent cotensor product on the category of line operators: 
applying the upside-down pair-of-pants to a line operator represented by a matrix factorisation $P(\tauu,\omegau,\xu)$ of $W_1-W_3$ in \eqref{openinversepants} one obtains
\begin{align}
\big[\xu-\yu,\Deltau W_3(\omegau^\prime, \vect{\xu}{\yu}) + \Deltau W_2(\sigmau,\vect{\xu}{\yu})\big] 
\otimes &\big[\sigmau-\lambdau,\Deltau W_2(\vect{\sigmau}{\lambdau},\yu)\big] \otimes P(\omegau', \tau, \yu)
\end{align}
which in general is an object in $\zz(I_{D_1, D_2^\dagger})\btimes\zz(I_{D_2, D_3^\dagger})$ that cannot be factorised into a pair of matrix factorisations associated to the separate legs. 
Indeed, there are more general objects in $\zz(I_{D_1, D_2^\dagger})\btimes\zz(I_{D_2, D_3^\dagger})$, that geometrically correspond to for example line operators with surfaces protruding into the bulk attached to them. 
(Analogously, in the case of 2-dimensional TQFTs taking values in $\Vect_\C$, applying the comultiplication to a vector does not necessarily yield a ``pure state'' of the form $v\otimes w$, but a more general superposition, corresponding to an entangled state. 
What we see here is a somewhat related  statement for line operators.)

Finally, applying the boundary-bulk map \eqref{Bobu} to a line defect maps it to a line defect in the bulk  by rolling up the surface on a cylinder, thus giving rise to a line operator on the identity surface:
\begin{align}
\tikzzbox{%

}
\, .
\end{align}
Conversely, applying the bulk-boundary map to a line operator in the bulk moves the line operator to the boundary, turning it to a line operator on the boundary. 
In the picture, the bulk line operator is located on an invisible surface defect, shaded in grey, and the right-hand side illustrates how that line operator becomes a line operator at the boundary: 
\begin{align}
\tikzzbox{%
%
} 
\, . 
\end{align}

The sewing relations of the 2-dimensional TQFT correspond to sewing relations in three dimensions involving the building blocks described in this section. 
It is clear that such a correspondence should also exist more generally, for example in Rozansky--Witten models with other target spaces. In fact, we expect it to be relevant quite generally.

\end{document}